\documentclass[12pt]{article}
\usepackage{amsmath,amsfonts}
\usepackage{graphicx}
\input epsf

\makeatletter\@addtoreset{equation}{section}\makeatother

\def\be{\begin{equation}}
\def\ee{\end{equation}}
\def\bea{\begin{eqnarray}}
\def\eea{\end{eqnarray}}
\def\ie{\begin{equation}\begin{aligned}}
\def\fe{\end{aligned}\end{equation}}

\newcommand{\A}{{\alpha}}
\newcommand{\B}{{\beta}}
\newcommand{\C}{{\gamma}}

\newcommand{\da}{{\dot\alpha}}
\newcommand{\db}{{\dot\beta}}
\newcommand{\dc}{{\dot\gamma}}
\newcommand{\dd}{{\dot\delta}}

\makeatletter\@addtoreset{equation}{section}\makeatother

\renewcommand{\title}[1]{\vbox{\center\LARGE{#1}}\vspace{5mm}}
\renewcommand{\author}[1]{\vbox{\center#1}\vspace{5mm}}
\newcommand{\address}[1]{\vbox{\center\em#1}}
\newcommand{\email}[1]{\vbox{\center\tt#1}\vspace{5mm}}

\begin{document}
\begin{titlepage}
\begin{center}
\hfill \\
\hfill \\
\vskip 1cm

\title{Higher Spin Gauge Theory and Holography:\\ The Three-Point Functions}

\author{Simone Giombi$^{a}$ and
Xi Yin$^{b}$}

\address{
Center for the Fundamental Laws of Nature
\\
Jefferson Physical Laboratory, Harvard University,\\
Cambridge, MA 02138 USA}

\email{$^a$giombi@physics.harvard.edu,
$^b$xiyin@fas.harvard.edu}

\end{center}

\abstract{
In this paper we calculate the tree level three-point functions of
Vasiliev's higher spin gauge theory in $AdS_4$ and find agreement with the correlators of the
free field theory of $N$ massless scalars in three dimensions in the $O(N)$ singlet sector. This provides
substantial evidence that Vasiliev theory is dual to the free field theory, thus verifying a conjecture of Klebanov and Polyakov.
We also find agreement with the critical $O(N)$ vector model, when the bulk scalar field is subject to the
alternative boundary condition such that its dual operator has classical dimension 2.
}

\vfill

\end{titlepage}

\eject \tableofcontents

\section{Introduction and Summary of Results}

The AdS/CFT correspondence \cite{Maldacena:1997re,Gubser:1998bc,Witten:1998qj} has led to marvelous insights in quantum gravity and large $N$ gauge theories. Most progress has been made (see \cite{Aharony:1999ti} and references therein) relating
weakly coupled gravity or string theories in $AdS$ spaces of large radii to strongly coupled gauge theories.
On the other hand, a weakly coupled large $N$ gauge theory is expected to be dual to a weakly coupled string theory in $AdS$
space of small radius compared to the string length scale (see for instance \cite{Sundborg:2000wp, HaggiMani:2000ru,Aharony:2003sx,Beisert:2003te,Beisert:2004di,Gopakumar:2003ns,
Gopakumar:2004qb,Gopakumar:2005fx,David:2006qc,Berkovits:2008qc,Berkovits:2008ga}). In practice, while it is straightforward to
understand a large $N$ gauge theory at weak 't Hooft coupling perturbatively, the string theory side
involves a strongly coupled sigma model in the worldsheet description. It is difficult in general
to understand the string spectrum in the small radius limit, let alone the full string field theory in $AdS$.
In general, one expects the free limit of the boundary gauge theory to be dual to a higher spin gauge theory in 
the bulk. In this limit, the bulk strings become tensionless in $AdS$ units \cite{Isberg:1992ia,Isberg:1993av,Lindstrom:2003mg,Bonelli:2003zu,Bakas:2004jq,Sagnotti:2003qa}, and the string spectrum should
contain a tower of higher spin gauge fields. 

A remarkable conjecture made by Klebanov and Polyakov \cite{Klebanov:2002ja}, closely related to earlier ideas put forth in \cite{Sundborg:2000wp,HaggiMani:2000ru,Konstein:2000bi,Shaynkman:2001ip,Sezgin:2001zs,Vasiliev:2001zy,
Mikhailov:2002bp} and in particular \cite{Sezgin:2002rt}, has provided the first example
of a potential dual pair that involves a weakly coupled (possibly free) large $N$ gauge theory on one side,
and an explicitly known bulk theory on the other side. More precisely, the conjecture states that Vasiliev's minimal bosonic
higher spin gauge theory in $AdS_4$ \cite{Vasiliev:1995dn,Vasiliev:1999ba,Vasiliev:2003ev}, which contains gauge fields of all non-negative even integer spins, is
dual to either the three-dimensional free field theory of $N$ massless scalar fields, in its $O(N)$-singlet sector (we will refer to
this as the ``free $O(N)$ vector theory"), or the critical $O(N)$ vector model, depending on the choice of the boundary condition for the bulk scalar field.\footnote{To be a bit more precise, the restriction to the $O(N)$-singlet sector in the dual boundary CFT should be implemented by gauging the $O(N)$ global symmetry, at zero coupling. In order to preserve conformal invariance, we can couple the scalars to $O(N)$ Chern-Simons gauge fields at level $k$, and take the limit $k\to \infty$.} The bulk theory contains one scalar field, of mass square
$m^2=-2/R^2$, $R$ being the $AdS$ radius. Depending on the choice of the boundary condition for this scalar,
its dual operator has either dimension $\Delta=1$ or $\Delta=2$, classically.
We will refer to them as $\Delta=1$ and $\Delta=2$ boundary conditions, respectively.
The conjecture is that Vasiliev theory with $\Delta=1$ boundary condition is dual to the free $O(N)$ vector theory,
which contains a scalar operator of dimension 1,
and the Vasiliev theory with $\Delta=2$ boundary condition is dual to the critical $O(N)$ vector model, the
latter containing a scalar operator of classical dimension 2 (plus $1/N$ corrections).

Thus far there has been little evidence for the conjecture of \cite{Klebanov:2002ja} beyond
the $N=\infty$ limit, which involves free higher spin gauge theory in the bulk. The only nontrivial piece of evidence
we are aware of
that involves the detailed structure of Vasiliev theory has been the observation of \cite{Petkou:2003zz, Sezgin:2003pt}
that the cubic coupling of the scalar field in the bulk theory vanishes identically.
This implies that, with the choice of $\Delta=2$ boundary condition, the three-point function
of scalar operators in the leading $1/N$ expansion of the dual CFT vanishes. This is indeed the case
for the critical $O(N)$ model, that is special to dimension 3 (and is not the case if one works in dimension $2<d<4$
with $d\not=3$). One may then be puzzled by the $\Delta=1$ case, where the dual CFT is expected to be
the free $O(N)$ vector theory, in which the three-point function of scalar operators do not vanish.
A potential resolution to this, analogous to the ``extremal correlators" of \cite{D'Hoker:1999ea},\footnote{
The correlation function of the $\Delta=1$ scalar operator in the large $N$ limit may also be understood
in terms of the $\Delta=2$ case using the Legendre transform
\cite{Klebanov:1999tb,Gubser:2002vv,Diaz:2006nm}.}
is that the integration over the boundary-to-bulk propagators on $AdS_4$ is divergent for the $\Delta=1$ scalar,
hence even though the bulk interaction Lagrangian vanishes, a subtle regularization is needed
to compute the three-point function. Such a regularization is not previously known in Vasiliev theory.
It will be explained in section 4 and section 6 of this paper, as a byproduct of our results on
three-point functions involving more general spins.

It has been shown in \cite{Mikhailov:2002bp} that in a CFT with higher spin symmetry,
if the OPEs of the conserved higher spin currents (or equivalently, the three-point functions) have the same structure as in a free massless scalar field theory,
then all the $n$-point functions of the currents are determined by the higher spin symmetry
up to finitely many constants for each $n$. It is however far from obvious, a priori, that the three point functions
in Vasiliev theory are those of a free field CFT. One of the main goals of our paper is to establish
this.

In this paper, we will calculate tree level three-point functions of the scalar and
higher spin currents of Vasiliev theory in $AdS_4$. As we will review in section 2, this is highly nontrivial because,
while Vasiliev's theory is formulated in terms of nonlinear equations of motion, there
is no known Lagrangian from which these equations are derived (see for instance
\cite{Fotopoulos:2008ka} and references therein for works on the Lagrangian approach
to higher spin gauge theories). Further, in Vasiliev's formalism,
each physical degree of freedom is introduced along with infinitely many auxiliary fields,
which are determined in terms of the physical fields recursively and nonlinearly.
We will develop the tools for the computation of correlation functions in section 3 and 4. In particular,
we will derive the relevant boundary-to-bulk propagators in terms of Vasiliev's master fields
in section 3, and use the second order nonlinear fields in the perturbation theory to derive the
three-point functions. In some cases, there are superficial divergences due to the nonlocal nature
of Vasiliev theory, and suitable regularization in the bulk will be needed.

More concretely, our strategy is as follows. For the three-point function of currents $J_{s_1}, J_{s_2}, J_{s_3}$,
of spin $s_1, s_2$ and $s_3$, we choose two of them to be sources on the boundary. We will first solve
for the boundary-to-bulk propagators of the master fields sourced by the two currents, say $J_{s_1}$ and $J_{s_2}$.
Then we will solve for the master fields at second order in perturbation theory, using Vasiliev's equations
of motion. Finally, we examine the boundary expectation value of the spin-$s_3$ components of
this second order field, and read off the
coefficient of the three point function $\langle J_{s_1}J_{s_2}J_{s_3}\rangle$.
In fact, through this procedure, we can only determine the ratio
\ie\label{jjjone}
{\langle J_{s_1}J_{s_2}J_{s_3}\rangle\over \langle J_{s_3}J_{s_3}\rangle} \sim C(s_1,s_2;s_3),
\fe
with some a priori unknown normalization of $J_s$. In particular, the coupling constant of Vasiliev theory
must be put in by hand at the end, which multiplies all three-point functions.
The normalization of $J_s$ can be determined by comparing different computations of the same three-point function, grouping different pairs of currents as sources.

Note that the spatial and polarization dependence of the three point function $\langle J_{s_1}
J_{s_2} J_{s_3}\rangle$ is constrained by conformal symmetry and the conservation of the currents, to a linear combination of finitely many possible structures \cite{Osborn:1993cr}.\footnote{We thank J. Maldacena and I. Klebanov for discussions on this point.} All we need to
calculate is the coefficients, as a function of the three spins. Our $C(s_1,s_2;s_3)$ will be defined using
(\ref{jjjone}) in the limit where the first two currents, $J_{s_1}$ and $J_{s_2}$, approach one another. In other words, we will be computing the coefficient of $J_{s_3}$ in the OPE of $J_{s_1}$ with $J_{s_2}$.

Throughout this paper, we will take our default boundary condition for the bulk scalar field to be the $\Delta=1$
boundary condition. This is because classically, the higher spin currents $J_s$ have scaling dimension $s+1$;
with the choice of $\Delta=1$ boundary condition, the scalar field is treated on equal footing as the higher spin currents.
The ``standard" $\Delta=2$ will be considered separately.

In section 4 and section 6.1, we will explicitly calculate $C(s_1,s_2;0)$ and $C(0,s_1;s_2)$ for $s_1>s_2$.
In our normalization convention, which will be explained in section 3 and 4, we find
\ie
& C(s_1,s_2;0) = -{\sqrt\pi\over 2}\Gamma(s_1+s_2+{1\over 2}),
\fe
and that
\ie
& C(0,s_1;s_2) = -{\sqrt\pi\over 2} 2^{-s_2}{\Gamma(s_1+{1\over 2})\over s_2!},~~~~~s_1>s_2.
\fe
This are in fact precisely consistent with taking two different limits of the same three-point function
of conserved higher spin currents, which by itself is a nontrivial consistency check on Vasiliev's equations.
The results allow us to fix the relative normalization of $J_s$, and to determine the tree-level three-point
functions of the normalized currents, involving one scalar operator and two general spin operators,
as we show in section 6.1. Much more strikingly, we will find complete agreement
with the corresponding three-point functions in the free $O(N)$ vector theory.
We regard this as a substantial evidence for the duality between the two theories.

In section 5, we study the same tree level correlators in Vasiliev theory, but with $\Delta=2$ boundary condition
on the bulk scalar field. We will find that the three point function coefficient $C(s_1,s_2;0)$
in the $\Delta=2$ case is in precise agreement with that of the critical $O(N)$ vector model,
at the leading nontrivial order in the $1/N$ expansion.

Let us emphasize that from the perspective of the bulk higher spin gauge theory, the computations of, say
$C(s_1,s_2;0)$, $C(0,s_1;s_2)$ with $s_1>s_2$, and $C(0,s_1;s_2)$ with $s_1<s_2$, are very different. For instance, when the two spins coincide, $C(s,s;0)$
is naively identically zero from the nonlinear equations of motion. However, our result for $C(s_1,s_2;0)$
with general $s_1\not=s_2$ suggests that the seeming vanishing of $C(s,s;0)$ is an artifact due to
the highly nonlocal and singular nature of Vasiliev theory, and in fact a proper way to regularize the computation is to start
with different spins $s_1,s_2$, analytically continue in the result and take the limit $s_2\to s_1$.
In section 6, we also attempt to calculate $C(0,0;s)$, for $s>0$. Somewhat unexpectedly, this in fact involves a qualitatively different computation than the cases mentioned above. Our result on $C(0,0;s)$
appear to be inconsistent with the general properties of the three-point functions,
and we believe that this is because the computation is singular, similarly to the case of $C(s,s;0)$,
where the spins of the two sources coincide. We hope to revisit this and the more general $C(s_1,s_2;s_3)$
in the near future.

The story has a few important loose ends. First of all, Vasiliev's minimal bosonic higher spin gauge theory,
as a classical field theory in $AdS_4$, has an ambiguity in its interaction that involve quartic and higher order couplings
\cite{Sezgin:2002ru}.
This ambiguity is entirely captured by a single function of one complex variable. It does not affect
our computation of tree level three point functions, but will affect higher point correlation functions
as well as loop contributions. Presumably, this interaction ambiguity is uniquely determined by requiring that
the bulk theory is dual to the free $O(N)$ vector theory. Further, it is conceivable that this is the only
pure bosonic higher even-spin gauge theory that is consistent at the quantum level.

Secondly, there is an important missing ingredient in the case of Vasiliev theory with $\Delta=2$ boundary condition, which is expected to be dual to the critical $O(N)$ vector model. While higher spin symmetries
are symmetries of the $O(N)$ model in the $N=\infty$ limit, and hence at tree level in $1/N$ expansion,
they are not exact symmetries of the theory at finite $N$. The bulk Vasiliev theory, on the other
hand, has exact higher spin gauge symmetry. One possibility is that at loop level, an
effective Lagrangian is generated for the scalar field, such that the scalar field will condense in a new $AdS_4$ vacuum, and spontaneously break the higher spin gauge symmetries (see \cite{Girardello:2002pp,Gubser:2002zh}).
We will comment on these points in section 7, leaving the details to future works.

\section{General Structure of Vasiliev Theory}

In this section we shall review the construction of Vasiliev theory and set up the notations.
Throughout this paper we will be considering the minimal bosonic higher spin gauge theory in $AdS_4$,
which contains one spin-$s$ gauge field of each even spin $s=0,2,4,\cdots$. We will denote by $x^\mu=(\vec x,z)$ the
Poincar\'e coordinates of $AdS_4$, and write $x^2=\vec x^2+z^2$, ${\bf x}=x^\mu \sigma_\mu$, etc.
Our spinor convention is as follows.
\ie
& u^\A = \epsilon^{\A\B}u_\B,~~~~ u_\A = u^\B \epsilon_{\B\A},~~~~ \epsilon_{12}=\epsilon^{12}=1,
\fe
and the same for the dotted indices. When two spinor indexed matrices $M$ and $N$ are multiplied, it is understood that the indices are contracted as
$M_{\cdots}{}^\A N_{\A\cdots}$. ${\rm Tr}M = M_\A{}^\A$. We define $V_{\A\db} = V_\mu \sigma^\mu_{\A\db}$, and hence
$V_\mu = -{1\over 2}V_{\A\db} \sigma_\mu^{\A\db}$, $V_{\A\db}V^{\A\db} = -{\rm Tr}(V_\mu\sigma^\mu)^2=-2V^\mu V_\mu$.

Following Vasiliev we introduce the auxiliary variables $y_\A, \bar y_\da, z_\A, \bar z_\da$, where
$\bar y$ and $\bar z$ are complex conjugates of $y,z$. When there is possible confusion, we shall distinguish $z_\A$ from
the Poincar\'e radial coordinate by adding a hat, and write $\hat z_\A$ instead. While we will mostly be working with ordinary functions of
$y,\bar y, z,\bar z$, in writing down the equations of motion of Vasiliev theory we need to define a star product, $*$, through
\ie
f(y,z)*g(y,z) = \int d^2u d^2v e^{u^\A v_\A} f(y+u,z+u) g(y+v,z-v),
\fe
where the integral is normalized such that $f*1=f$, and similarly for the conjugate variables $\bar y, \bar z$. The star product
between functions of the unbarred variables and the barred variables is the same as the ordinary product. In particular, for $y$ and $\bar y$, we have
\ie
&y_\A * y_\B = y_\A y_\B + \epsilon_{\A\B},\\
&y^\A * y^\B = y^\A y^\B + \epsilon^{\A\B},\\
&\bar y_\da * \bar y_\db = \bar y_\da \bar y_\db + \epsilon_{\da\db},\\
&\bar y^\da * \bar y^\db = \bar y^\da \bar y^\db + \epsilon^{\da\db}.
\fe
whereas $z$ and $\bar z$ have similar $*$-contractions with opposite signs. Note that although
$z_\A$ and $y_\B$ $*$-commute, their $*$-product is not the same as the ordinary product, i.e. the $*$-contraction between
$z_\A$ and $y_\B$ is $\epsilon_{\A\B}$ rather than zero.

It will be useful to define the Kleinian of the star algebra, $K=e^{z^\A y_\A}$, and $\bar K = e^{\bar z^\da \bar y_\da}$. For convenience we will also define $K(t)=e^{t z^\A y_\A}$, and $\bar K(t) = e^{t\bar z^\da \bar y_\da}$. They have the property under $*$-product
\ie
&f(y)*K(t) = f((1-t)y-tz)K(t),\\
&K(t)*f(y) = f((1-t)y+tz)K(t),\\
&F(y,z)*K(t) = F((1-t)y-tz,(1-t)z-ty)K(t),\\
&K(t)*F(y,z) = F((1-t)y+tz,(1-t)z+ty)K(t).
\fe
In particular, $K$ $*$-anti-commutes with $y,z$, $\bar K$ $*$-anti-commutes with $\bar y,\bar z$, and
$K*K=\bar K*\bar K=1$.

We are now ready to introduce the master fields, $W=W_\mu(x|y,\bar y,z,\bar z)dx^\mu$,
$S=S_\A(x|y,\bar y,z,\bar z) dz^\A + S_\da(x|y,\bar y,z,\bar z) d\bar z^\da$, and $B=B(x|y,\bar y,z,\bar z)$. Here
$dz^\A$ and $d\bar z^\da$ behave as ordinary 1-forms under $*$-product. Our convention is slightly different from
Vasiliev's in that we will be writing $z_\A+S_\A$ for Vasiliev's $S_\A$, and similarly for $S_\da$.
We begin by presenting a fully covariant form of Vasiliev's equations of motion. To do this, we shall further define
\ie
&\hat {\cal A} = W + (z_\A+ S_\A) dz^\A + (\bar z_\da+S_\da) d\bar z^\da,\\
&{\cal A} = W + S_\A dz^\A +S_\da d\bar z^\da,\\
&\hat d=d_x + d_Z,~~~~d=d_x,\\
&\Psi = B*K,~~~~\bar\Psi = B*\bar K,\\
&\Theta = K dz^2 + \bar K d\bar z^2,~~~~R=K\bar K.
\fe
where $d_x$ is the exterior derivative in $x^\mu$ and $d_Z$ is the exterior derivative in $(z_\A, \bar z_\da)$,
$dz^2=dz^\A dz_\A$, $d\bar z^2=d\bar z^\da d\bar z_\da$.
The equation of motion of Vasiliev theory can be written as
\ie\label{veqna}
& d \hat {\cal A} + \hat{\cal A}*\hat{\cal A} = f(\Psi) dz^2 + \overline{f(\Psi)} d\bar z^2, \\
& \bar\Psi = \Psi*R,~~~~[R,W]_*=\{R,S\}_*=0.
\fe
where $f$ is a complex $*$-function of one variable, and ${\cal A}$ and $\Psi$ are understood here as otherwise unconstrained fields,
${\cal A}$ being a 1-form in $(x,\hat z,\hat{\bar z})$. For instance, $\bar\Psi=\Psi*R$ is just a rewriting of the statement
that both $\Psi$ and $\bar\Psi$ are related to the real field $B$.
(\ref{veqna}) the admits gauge symmetry
\ie
& \delta \hat{\cal A} = d\epsilon + [\hat{\cal A},\epsilon]_* = \hat d\epsilon + [{\cal A},\epsilon]_*,\\
& \delta\Psi = [\Psi,\epsilon]_*.
\fe

With field redefinitions of $S$ and $\Psi$, one can put $f(\Psi)$ in the form
$f(\Psi) = 1+\Psi + ic\Psi*\Psi*\Psi+\cdots$ where $c$ is a real constant and $\cdots$ are a remaining $*$-odd function in $\Psi$.
The $*$-cubic and higher order terms in $f(\Psi)$ will not affect the computation of tree level
three-point function, and may be ignored in most of this paper. We will comment on them later. It was observed in \cite{Sezgin:2003pt} that if one imposes parity invariance one can in fact 
fix\footnote{We thank Per Sundell for pointing this out to us.} $f(\Psi)=1+\Psi$. We will work with this choice in this paper, and refer to it as the ``minimal" Vasiliev theory \cite{Sezgin:2003pt}\footnote{We may assume that the scalar is even under parity. If the scalar is taken to be parity odd, the resulting bulk theory was proposed to be dual to 3d free $O(N)$ fermions/critical Gross-Neveu model \cite{Sezgin:2003pt,Leigh:2003gk}.}. 
In this case the equations of motion can be written simply as\footnote{This form of the equation of motion may appear similar to the string field theory equation of the form
$Q{\cal A}+{\cal A}*{\cal A}=0$ \cite{Witten:1985cc}. However, due to the RHS of (\ref{veqnb}), and the fact that $B$ field transforms
in the twisted adjoint representation with respect to the star algebra, we do not see an obvious way to cast
the equation (\ref{veqnb}) in the form of a cubic string field theory equation with some BRST operator.}
\ie\label{veqnb}
{\cal F}_{\cal A}\equiv\hat d {\cal A} + {\cal A}*{\cal A} = B*\Theta.
\fe 
Note in particular it follows from (\ref{veqnb}) that $dB*\Theta + [W,B*\Theta]_*=0$ and $[z_\A+S_\A, B*\bar K]_*=0$.
In terms of $W, S, B$, in a more digestable form, the equations are
\ie
& d_x W+ W*W = 0,\\
& d_Z W + d_x S + \{W,S\}_* = 0,\\
& d_Z S + S*S = B*K dz^2 + B*\bar K d\bar z^2,\\
& d_x B + W*B-B*\pi(W) = 0,\\
& d_Z B + S*B-B*\pi(S) = 0.
\fe
Here $\pi$ and $\bar\pi$ is defined by
\ie
&\pi(f(y,\bar y,z,\bar z, dz,d\bar z))=
f(-y,\bar y,-z,\bar z, -dz,d\bar z),\\
&\bar\pi(f(y,\bar y,z,\bar z, dz,d\bar z))=
f(y,-\bar y,z,-\bar z, dz,-d\bar z).
\fe
Note that because of the constraints $[W,R]_*=\{S, R\}_*=[B,R]_*=0$,
$\pi$ and $\bar\pi$ in fact act the same way on $W, S$ and $B$.
The gauge symmetry is now written as
\ie
& \delta W = d\epsilon + [W,\epsilon]_*,\\
& \delta S = d_Z \epsilon + [S,\epsilon]_*,\\
& \delta B = B*\pi(\epsilon) -\epsilon*B,
\fe
for some $\epsilon(x|y,\bar y,z,\bar z)$.

Note that the overall coupling constant of Vasiliev theory is absent from the equations,
which will need to be put in by hand in computing correlation functions using the AdS/CFT
dictionary. While one may verify the consistency of the equations of motion,
we do not know the explicit form of the Lagrangian from which these equations can be derived.

The $AdS_4$ vacuum is given by
\ie
W= W_0, ~~~S=0,~~~B=0.
\fe
where $W_0 = \omega_0^L+e_0$ satisfies the equation $dW_0 + W_0*W_0=0$. Here $\omega_0^L$ and $e_0$
are the $AdS_4$ spin connection and vierbein written in terms of the $*$-noncommutative variables $y$ and $\bar y$,
in Poincar\'e coordinates,
\ie
&\omega_0^L = {1\over 8} {dx^i\over z} \left[ (\sigma^{iz})_{\A\B}y^\A y^\B
+(\sigma^{iz})_{\da\db}\bar y^\da \bar y^\db \right],\\
& e_0 = {1\over 4} {dx_\mu\over z} \sigma^\mu_{\A\db}y^\A \bar y^\db.
\fe
We will often use the notations
\ie
& d_L = d+ [\omega_0^L,\,\cdot\, ]_*,\\
& D_0 = d+ [W_0,\,\cdot\, ]_*,\\
& \tilde D_0 = d + W_0*\,\cdot\, - \,\cdot\,*\pi(W).
\fe
Writing $W=W_0+\hat W$, we can write the equations of motion in a perturbative form as
\ie\label{perteqn}
& D_0 \hat W = -\hat W*\hat W,\\
& d_Z\hat W + D_0 S = -\{\hat W,S\}_*,\\
& d_Z S - B*\Theta = -S*S,\\
& \tilde D_0 B = -\hat W*B+B* \pi(\hat W),\\
& d_Z B = -S*B+B*\pi(S).
\fe
The linearized equations are simply obtained from (\ref{perteqn}) by setting the RHS to zero.
The strategy to solving the equations perturbations is as follows. First, using the last line of (\ref{perteqn})
we solve for the $\hat z$-dependence of $B$. Then using the third equation of (\ref{perteqn})
we solve for the $\hat z$-dependence of $S$ in terms of $B$. One can always gauge away the $\hat z$-independent part of $S$.
Using the second equation, one solves for the $\hat z$-dependence of $\hat W$ in terms of $B$.
We shall write $\hat W=\Omega+ W'$, where $\Omega=\hat W|_{\hat z=\bar{\hat z}=0}$, and $W'$ contains
the $\hat z$-dependent part of $\hat W$. The first equation will now give a relation between $\Omega$ and $B$,
either one will contain all the physical degrees of freedom (except the scalar, which is only contained in $B$).
Finally one can recover the equation of motion for the physical higher spin fields from either the fourth equation (which is often easier) or the
first equation in (\ref{perteqn}).

We will defer a discussion on the explicit relation between the linearized fields and the ``physical" symmetric traceless $s$-tensor gauge
fields to the next section, where we will solve for the boundary-to-bulk propagator for the master fields both using the ``conventional"
symmetric traceless tensor field and directly using Vasiliev's equations for the master fields. For now, let us point out
that that the physical degrees of freedom are entirely contained in $\hat W$ and $B$ restricted to $z_\A=\bar z_\da=0$. In fact, writing
the Taylor expansion
\ie
& \Omega = \hat W|_{z=\bar z=0} = \sum_{n,m=0}^\infty \Omega^{(n,m)}_{\A_1\cdots\A_n\db_1\cdots\db_m} y^{\A_1}\cdots y^{\A_n} \bar y^{\db_1}\cdots
\bar y^{\db_m},\\
& B|_{z=\bar z=0} = \sum_{n,m=0}^\infty B^{(n,m)}_{\A_1\cdots\A_n\db_1\cdots\db_m} y^{\A_1}\cdots y^{\A_n} \bar y^{\db_1}\cdots
\bar y^{\db_m},
\fe
the spin-$s$ degrees of freedom are entirely contained in $\Omega^{(s-1+n,s-1-n)}$ ($|n|\leq s-1$), $B^{(2s+m,m)}$ and $B^{(m,2s+m)}$
($m\geq 0$). In particular, $\Omega^{(s-1,s-1)}$ will be the symmetric $s$-tensor field, and $B^{(2s,0)}$, related to up to $s$ spacetime
derivatives of $\Omega^{(s-1,s-1)}$, plays the role of
the higher spin analog of Weyl curvature tensor.

\section{The Boundary-to-Bulk Propagator}

The goal of this section is to derive the boundary-to-bulk propagator for the Vasiliev master fields corresponding
to a spin-$s$ current in the boundary CFT. In the first subsection, we will derive the boundary-to-bulk propagator for
a free higher spin gauge field described by a traceless symmetric tensor. We will then recover the same result in the linearized
Vasiliev theory, while providing explicit formulae for the propagator of the master fields as well.

\subsection{The spin-$s$ traceless symmetric tensor field}

Let us consider a traceless symmetric $s$-tensor gauge field $\varphi_{\mu_1\cdots\mu_s}$ in $AdS_{d+1}$.
The equation of motion is given by
\ie\label{linten}
-(\Box-m^2)\varphi_{\mu_1\cdots\mu_s} + s\nabla_{(\mu_1}\nabla^\nu \varphi_{\mu_2\cdots\mu_s)\nu}
-{s(s-1)\over 2(d+2s-3)} g_{(\mu_1\mu_2}\nabla^{\nu_1}\nabla^{\nu_2}\varphi_{\mu_3\cdots\mu_s)\nu_1\nu_2} = 0,
\fe
where $m^2=(s-2)(d+s-3)-2$. This equation can be derived using the linearized form of Vasiliev's equation in $AdS_{d+1}$
for general $d$ in the $Sp(2)$-invariant formalism \cite{Vasiliev:2003ev,Bekaert:2005vh} (see also
\cite{Mikhailov:2002bp}). In this paper we will not use this formalism. Instead we will directly recover the result of
this section by starting with Vasiliev's master equations in $AdS_4$ in the next subsections.

Under the gauge condition $\nabla^\nu \varphi_{\nu \mu_1\cdots\mu_{s-1}}=0$, (\ref{linten}) simplifies to
\begin{equation}
(\Box-m^2)\varphi_{\mu_1\cdots\mu_s}=0.
\end{equation}
A solution to this equation has the boundary behavior as $z\to 0$,
\begin{equation}
\varphi_{i_1\cdots i_s}(\vec x,z)\sim z^\delta,~~~~(\delta+s)(\delta+s-d)-s=m^2.
\end{equation}
where the indices $i_k$ are along the boundary directions, running from $0$ to $d-1$.
From this we read off the dimension of the dual operator, a spin-$s$ current $J_{i_1\cdots i_s}$,
\begin{equation}
\Delta=d-\delta-s = {d\over 2} + \sqrt{m^2 +s+ \left({d\over 2}\right)^2} = d-2+s
\end{equation}
This scaling dimension also follows from the conformal algebra under the assumption that $J_{i_1\cdots i_s}$ is a conserved current
and a primary operator. In particular, in a free scalar field theory in $d$ dimensions, the currents of the
form $\phi \partial_{i_1}\cdots\partial_{i_s}\phi+\cdots$ have dimension $\Delta=d-2+s$.

Now let us study the boundary-to-bulk propagator for $\varphi_{\mu_1\cdots\mu_s}$.
Using the traceless condition on $\varphi$, the gauge condition $\nabla^\nu \varphi_{\nu \mu_1\cdots\mu_{s-1}}=0$
can be written in Poincar\'e coordinates explicitly as
\begin{equation}\label{expgauge}
\begin{aligned}
& (\partial_z-{d-1\over z}) \varphi_{z\mu_1\cdots \mu_{s-1}} + \partial_i \varphi_{i\mu_1\cdots \mu_{s-1}}=0.
\end{aligned}
\end{equation}
where the index $i$ is summed over $0,\cdots,d-1$, while $\mu_k$ runs through $0,\cdots,d$.

The operator $\Box=\nabla^a\nabla_a$ acts on $\varphi_{\mu_1\cdots\mu_s}$
as
\begin{equation}
\begin{aligned}
\Box \varphi_{\mu_1\cdots\mu_s} & =\left[z^2 (\partial_z+{s-d+1\over z})(\partial_z+{s\over z})
+z^2\partial_i \partial_i -s \right]\varphi_{\mu_1\cdots\mu_s}\\
&~~~-2sz\partial_{(\mu_1}\varphi_{\mu_2\cdots\mu_s)z} + s(s-1)\eta_{(\mu_1\mu_2} \varphi_{\mu_3\cdots\mu_s)zz} \\
&~~~-s(d+2s-3)\delta_{z(\mu_1} \varphi_{\mu_2\cdots\mu_s)z} + 2sz \partial_\rho \delta_{z(\mu_1} \varphi_{\mu_2
\cdots\mu_s)\rho}\\
& =\left[z^2 (\partial_z+{s-d+1\over z})(\partial_z+{s\over z})
+z^2\partial_i \partial_i -s \right]\varphi_{\mu_1\cdots\mu_s}\\
&~~~-2sz\partial_{(\mu_1}\varphi_{\mu_2\cdots\mu_s)z} + s(s-1)\eta_{(\mu_1\mu_2} \varphi_{\mu_3\cdots\mu_s)zz} +s(d-2s+1)\delta_{z(\mu_1} \varphi_{\mu_2\cdots\mu_s)z}.
\end{aligned}
\end{equation}
where in the second step we used the gauge condition. Now splitting the indices according to
boundary and radial directions, $(\mu_1\cdots\mu_s)=(i_1\cdots i_r z\cdots z)$, $0\leq r\leq s$, we have
\begin{equation}
\begin{aligned}
\Box \varphi_{i_1\cdots i_r z\cdots z}
& =\left[z^2 (\partial_z+{s-d+1\over z})(\partial_z+{s\over z})
+z^2\partial_i \partial_i -2(s-r)z\partial_z\right.\\
&\left.+(s-r)(d-s-r)-s \right]\varphi_{i_1\cdots i_rz\cdots z}
-2rz\partial_{(i_1}\varphi_{i_2\cdots i_r)z\cdots z} + r(r-1)\eta_{(i_1i_2} \varphi_{i_3\cdots i_r)z\cdots z}.
\end{aligned}
\end{equation}
Define the generating function
\ie
\Phi_s(x,z|Y) &= z^s\sum \varphi_{\mu_1\cdots\mu_s}(x,z) Y^{\mu_1}\cdots Y^{\mu_s}\\
&= z^s \sum {s\choose r} \varphi_{i_1\cdots i_r z\cdots z}(x,z) Y^{i_1}\cdots Y^{i_r} (Y^z)^{s-r},
\fe
with auxiliary variables $Y^\mu$.
We can express the equation of motion for $\varphi$ in terms of the generating function $\Phi_s$ as
\ie
&\left[ z^2 (\partial_z+{s-d+1\over z})(\partial_z+{s\over z})
+z^2\partial_i \partial_i  - 2z Y^z \partial_z \partial_{Y^z}
+(d-2s+Y^z \partial_{Y^z})Y^z \partial_{Y^z}\right.
\\
&\left. -s -2z Y^\mu \partial_\mu \partial_{Y^z} + Y^2 \partial_{Y^z}^2-m^2 \right]z^{-s}\Phi_s = 0.
\fe
Now we perform a Fourier transform on the variables $(\vec x, Y^z)$, into $(\vec p, v)$,
and write the Fourier transformed generating function as $\tilde\Phi_s(\vec p,z|\vec Y,v)$. Then the equation of motion simplifies to
\ie\label{eomaaa}
&\left[ (z\partial_z + v\partial_v)^2 + (2s-d+2)(z\partial_z + v\partial_v)
-(z\vec p-v \vec Y)^2 \right.
\\
&~~~\left.+ s(s-d+1)+1-d-m^2 \right] z^{-s}\tilde \Phi_s (\vec p,z|\vec Y,v) = 0.
\fe
In solving this equation, we must take into the traceless condition and the gauge condition,
which are expressed in terms of $\tilde\Phi_s$ as
\ie\label{extracd}
& \left[ v(z\partial_z+1-d)+z \vec p \cdot \vec\partial_Y \right]z^{-s}\tilde\Phi_s=0,\\
& (\vec\partial_Y^2-v^2)\tilde \Phi_s=0.
\fe
(\ref{eomaaa}) is essentially the Bessel equation, solved by
\ie
\tilde\Phi_s = z^s \psi_s(|z\vec p-v \vec Y|)
f({v\over z},\vec Y-{z\over v}\vec p,\vec p)
\fe
for some arbitrary function $f$, where
\ie
\psi_s(t) = t^{{d\over 2}-s-1} K_{{d\over 2}+s-2}(t)
\fe
solves the equation
\ie
\left[(t\partial_t)^2+(2s-d+2)t\partial_t - t^2 + s(s-d+1)+1-d-m^2 \right] \psi_s(t)=0.
\fe
To solve the gauge condition (first equation of (\ref{extracd})), we may take $f$ to be of the form
\ie
f({v\over z},\vec Y-{z\over v}\vec p,\vec p) = ({v\over z})^{1-d} F(\vec Y-{z\over v}\vec p,\vec p)
\fe
Now we shall specializing to the case of $AdS_4$, i.e. $d=3$. Replace the variable $v$ by $u=v/z$. Fourier transforming back,
we can turn the integration over $u$
into a contour integral around $u=0$, and write
the generating function for the spin-$s$ field as
\ie
\Phi_s &= z^{s+1} \int d^3\vec p \oint {du \over u^2} e^{i\vec p\cdot \vec x+izuY^z}  \psi_s(z|\vec p- u\vec Y|) F(\vec Y-{\vec p\over u},\vec p)\\
&= z^{s+1} \int d^3\vec p \oint {du \over u^2} e^{i\vec p\cdot \vec x+iux^\mu Y_\mu} \psi_s(z|\vec p|) F(-{\vec p\over u},\vec p+u\vec Y) \\
&= \oint {du\over u^2} e^{iux^\mu Y_\mu} F({i\vec\partial\over u},-i\vec\partial+u\vec Y) |\partial|^{1-2s} \left( z\over x^2+z^2 \right)^{s+1}
\fe
The traceless condition, i.e. second line of (\ref{extracd}), can be expressed as a condition on $F(\vec q,\vec p)$,
\ie
\vec q\cdot\vec\partial_q F=(s-1)F,~~~~\vec\partial_q^2 F=0.
\fe
We can therefore write $F$ as
\ie
F(\vec q,\vec p) = |\vec q|^{s-1} G({\vec q\over |q|},\vec p)
\fe
and the generating function as
\ie
\Phi_s = \oint {du\over u^{s+1}} e^{iux^\mu Y_\mu} G(i{\vec\partial\over |\partial|},-i\vec\partial+u\vec Y) |\partial|^{-s} \left( z\over x^2+z^2 \right)^{s+1}
\fe
The traceless condition now says $G(\vec q/|q|,\vec p)$ is a (singular) spherical harmonic on
$S^2$ with spin $s-1$ (or 0 for $s=0$).

Without loss of generality, we only need to consider the boundary-to-bulk propagator corresponding to a spin-$s$ current
contracted with a null polarization vector $\varepsilon$.
It turns out that the solution that gives the desired boundary behavior is
\ie
G(\hat q,\vec p) =const\times {(\vec \varepsilon\cdot \vec p)^{2s}\over (\vec\varepsilon\cdot \hat q)^s}
\fe
and so
\ie
\Phi_s = \tilde N_s\left. e^{iux^\mu Y_\mu} {(\varepsilon\cdot (-i\vec\partial+u\vec Y))^{2s}\over (i\varepsilon\cdot \vec\partial)^s}\right|_{u^s} \left( z\over x^2+z^2 \right)^{s+1}
\fe
for some normalization constant $\tilde N_s$. Here $|_{u^s}$ means to pick out the coefficient of
$u^s$ in a series expansion in $u$. Near the boundary $z\to 0$, $\Phi_s(\vec x,z|Y)$ behaves as
\ie
\Phi_s(x,z|Y) &\to \tilde N_s\sum_{t=0}^s {2s\choose t}(\vec \varepsilon\cdot\vec Y)^t
(\vec x\cdot\vec Y)^{s-t}{(\vec \varepsilon\cdot\vec \partial)^t\over (s-t)!}
\left[ \pi^{3\over 2}{\Gamma(s-{1\over 2})\over s!} z^{2-s}(\varepsilon\cdot \vec Y)^s \delta^3(\vec x) \right]
\\
&= \tilde N_s \pi^{3\over 2}{\Gamma(s-{1\over 2})\over s!}\sum_{t=0}^s(-)^t{2s\choose t} z^{2-s}(\varepsilon\cdot \vec Y)^s \delta^3(\vec x)
\\
&= \tilde N_s \pi^{3\over 2}{\Gamma(s-{1\over 2})(2s)!\over 2 (s!)^3} z^{2-s}(\varepsilon\cdot \vec Y)^s \delta^3(\vec x)
\fe
where we have dropped terms of the form $\partial^n \left[(\vec x\cdot \vec Y)^n \left( z\over x^2+z^2 \right)^{s+1}\right]$
which vanish at order $z^{2-s}$ near the boundary.
$s$ will be assumed to be an even integer from now on. By requiring the coefficient of $z^{2-s}(\varepsilon\cdot\vec Y)^s\delta^3(\vec x)$ to
be 1, the normalization constant $\tilde N_s$ is determined to be
\ie
\tilde N_s = {2\pi^{-{3\over 2}} (s!)^3\over \Gamma(s-{1\over 2})(2s)!}.
\fe

It is sometimes convenient to work in light cone coordinates on the boundary $\vec x = (x^+, x^-, x_\perp)$,
with $\vec x^2=x^+x^-+x_\perp^2$ and $\vec\varepsilon\cdot\vec \partial
=\partial_+$, i.e. $\varepsilon^+=1,\varepsilon^-=0$. We can then write the boundary-to-bulk propagator for
$\Phi_s$ simply as
\ie
\Phi_s &= i^s\tilde N_s e^{iux^\mu Y_\mu} (-i\partial_++uY_+)^{2s} {1\over \partial_+^s} \left( z\over x^+x^-+x_\perp^2+z^2 \right)^{s+1}
\\
&=i^s{\tilde N_s\over s!} e^{iux^\mu Y_\mu} (-i\partial_++uY_+)^{2s} {1 \over x^+x^-+x_\perp^2+z^2} \\
&=\left. i^s{\tilde N_s z^{s+1}\over s!(x^-)^s} \partial_+^{2s}  {e^{iux^\mu Y_\mu} \over x^+x^-+x_\perp^2+z^2}\right|_{u^s} \\
&= \tilde N_s {z^{s+1}\over (s!)^2(x^-)^s} \partial_+^{2s}  {(x^\mu Y_\mu)^s \over \vec x^2+z^2}.
\fe

\subsection{The boundary-to-bulk propagator for the master field $B$}

In this subsection, we will begin with the linearized equation for $B$ in Vasiliev theory,
and derive its boundary-to-bulk propagator. Recall that $B$ contains the higher spin analogs of Weyl curvature.
One of the linearized equations, $d_ZB=0$, simply says that at the linearized order, $B=B(x|y,\bar y)$ is independent of
$z_\A$ and $\bar z_\da$.

The other linearized equation, $\tilde D_0 B=0$, can be written explicitly as
\ie \label{beqn}
&dB + [\omega_0^L, B]_* + \{e_0, B\}_*\\
&=dB -{dx^i\over 2z}\left[(\sigma^{iz})_\A{}^\B y^\A \partial_\B
+ (\sigma^{iz})_\da{}^\db \bar y^\da \partial_\db \right] B + {dx^\mu\over 2z}\sigma_\mu^{\A\db}
\left( y_\A \bar y_\db +\partial_\A\partial_\db \right) B \\
&=0,
\fe
or in components,
\ie\label{eomcomp}
{\partial\over \partial x^{\A\db}}B + {1\over 2z}\left[ (\sigma^z)^{\C}{}_{\db} y_\A\partial_\C + (\sigma^z)_{\A}{}^{\dc}\bar y_\db \partial_\dc \right] B - {\sigma^z_{\A\db}\over 4z} \left( y^\C\partial_\C + \bar y^\dc \partial_\dc \right) B
+{1\over 2z}(y_\A\bar y_\db +\partial_\A\partial_\db)B = 0 .
\fe
Recall our convention $dx^\mu = -{1\over 2}dx^{\A\db}\sigma^\mu_{\A\db}$. By contracting (\ref{eomcomp}) with $y^\A\bar y^\db$ or by acting on
(\ref{eomcomp}) with $\partial^\A \partial^\db$, we obtain
\ie
&y^\A\bar y^\db \partial_{\A\db} B - {\sigma^z_{\A\db}y^\A\bar y^\db\over 4z} \left( y^\C\partial_\C + \bar y^\dc \partial_\dc \right) B
+{1\over 2z}(y^\A\partial_\A)(\bar y^\db\partial_\db) B = 0,\\
& \partial^\A\partial^\db \partial_{\A\db} B + {\sigma^z_{\A\db}\partial^\A\partial^\db\over 4z} \left( y^\C\partial_\C + \bar y^\dc \partial_\dc +4\right) B
+ {1\over 2z}(y^\A \partial_\A+2)(\bar y^\db\partial_\db +2)B=0,
\fe
or rather, expanded in powers of $y$ and $\bar y$,
\ie\label{eomab}
&y^\A\bar y^\db \partial_{\A\db} B^{(n,m)}- {\sigma^z_{\A\db}y^\A\bar y^\db\over 4z} \left( n+m \right) B^{(n,m)}
+{1\over 2z}(n+1)(m+1) B^{(n+1,m+1)} = 0, \\
& \partial^\A\partial^\db \partial_{\A\db} B^{(n,m)} + {\sigma^z_{\A\db}\partial^\A\partial^\db\over 4z} \left( n+m +4\right) B^{(n,m)}
+ {1\over 2z}(n+1)(m+1)B^{(n-1,m-1)}=0.
\fe
The scalar field and its derivatives are contained in $B^{(n,n)}$. In particular, it follows from the first line of (\ref{eomab}) that
$B^{(1,1)} = -2z y^\A\bar y^\db \partial_{\A\db}B^{(0,0)}$, and from the second line of (\ref{eomab}) that
\ie
&\partial^\A\partial^\db \partial_{\A\db} B^{(1,1)} + {3\sigma^z_{\A\db}\partial^\A\partial^\db\over 2z}  B^{(1,1)}
+ {2\over z}B^{(0,0)}\\
&=\left(-2\partial^{\A\db}z\partial_{\A\db} -3 \sigma_z^{\A\db}\partial_{\A\db}+{2\over z}\right)B^{(0,0)} \\
&=\left(z\partial^\mu \partial_\mu -2\partial_z+{2\over z}\right)B^{(0,0)} \\
&=0.
\fe
This is solved by scalar boundary-to-bulk propagator $B^{(0,0)}= K(x,z)^\Delta$ for $\Delta=1$ or $\Delta=2$,
where $K(x,z)\equiv {z\over x^2+z^2}$. This verifies that the linearized equation for $B$ indeed produces the correct
boundary-to-bulk propagator for the scalar field $B^{(0,0)}$.

Further solving for the higher components $B^{(n,n)}$ using (\ref{eomab}), we recover the boundary-to-bulk propagator for
the scalar component of the master field $B$. The answer for the $\Delta=1$ scalar is
\ie\label{eomac}
B = K e^{-y(\sigma^z-2{\bf x}K)\bar y} = K e^{-y\Sigma \bar y}
\fe
where we recall the notation ${\bf x} \equiv x^\mu \sigma_\mu = x^i \sigma_i + z\sigma^z$.
We also defined $\Sigma = \sigma^z-{2z\over x^2}{\bf x}$.
It is straightforward to check that (\ref{beqn}) is indeed solved by (\ref{eomac}). The boundary-to-bulk propagator
for the scalar component of $B$ field in the $\Delta=2$ case will be given in section 5.

Now let us generalize to the spin $s$ components of $B$. Consider an ansatz to the linearized $B$-equation of motion of the form
\ie\label{eomad}
B = {1\over 2} K e^{-y\Sigma \bar y} T(y)^{s}+c.c.
\fe
where $T(y)$ is a quadratic function in $y$, so that (\ref{eomad}) indeed corresponds to the spin-$s$ degrees of freedom.
Our normalization convention is such that for $s=0$ (\ref{eomad}) agrees with the scalar component of the boundary-to-bulk
propagator (\ref{eomac}).
This ansatz solves (\ref{beqn}) if $T(y)$ obeys
\ie
dT-{dz\over z}T + {K\over z}y {\bf x}d{\bf x}\partial_y T = 0.
\fe
The solution is given by
\ie\label{eomae}
T = {K^2\over z} y{\bf x}\vec{\varepsilon}\cdot\vec\sigma \sigma^z {\bf x} y,
\fe
for an arbitrary polarization vector $\vec\varepsilon$ along the 3-dimensional boundary. We will verify in the
next subsection that this is indeed the master field corresponding to the boundary-to-bulk propagator for the spin-$s$ tensor
gauge field derived in the previous subsection, with polarization vector $\varepsilon$.
(\ref{eomad}) together with (\ref{eomae}) give the boundary-to-bulk propagator for $B$ of general spin.

Sometimes we will write $C(x|y) = B|_{\bar y=0}$. It is useful to invert this relation and
recover $B$ from $C$, using (\ref{eomab}). For the spin $s$ components,
\ie
B^{(2s+m,m)} &= -{1\over m(2s+m)} y\left[z{\slash\!\!\!\partial}+(s+m-1)\sigma^z\right]\bar y B^{(2s+m-1,m-1)}
\\
&=-{1\over m(2s+m)} z^{2-s-m}(y{\slash\!\!\!\partial}\bar y) z^{s+m-1} B^{(2s+m-1,m-1)}
\\
&=(-)^m{(2s)!\over m!(2s+m)!} z^{-s-m}(z^2y{\slash\!\!\!\partial}\bar y)^m z^{s} C(x|y)
\fe
where our convention for ${\slash\!\!\!\partial}$ is ${\slash\!\!\!\partial}_{\A\db}\equiv \sigma^\mu_{\A\db}\partial_\mu = -2\partial_{\A\db}$. The entire spin $s$ part of $B$ is then given by
\ie\label{bfromc}
B &= \sum_{m=0}^\infty (-)^m{(2s)!\over m!(2s+m)!} z^{-s-m}(z^2y{\slash\!\!\!\partial}\bar y)^m z^{s} C(x|y) + c.c.
\fe

\subsection{The master field $W$}

As discussed earlier, our strategy of solving Vasiliev's equations perturbatively is to solve for the master fields
and then restrict to $z_\A=\bar z_\da=0$ at the end to extract the physical degrees of freedom. Writing $\hat W$ for the fluctuation
of $W$ away from the vacuum configuration $W_0$, it will be useful to split it into two parts,
\ie
\hat W(x|y,\bar y,z,\bar z) = \Omega(x|y,\bar y) + W'(x|y,\bar y,z,\bar z),
\fe
where $\Omega=\hat W|_{z=\bar z=0}$, and $W'$ is the remaining $\hat z$-dependent part of $W$.
At the linearized level, $W|_{z=0}\equiv \Omega$ will be expressed in terms of the traceless symmetric $s$-tensor
gauge fields and their derivatives, whereas $W'$ is determined by $B$ through the equations of motion. Let us first consider $\Omega$.
Expanding in a power series in $y$ and $\bar y$, we will denote by $\Omega^{(n,m)}$ the part of $\Omega$ of degree $n$ in $y^\A$ and
degree $m$ in $\bar y^\da$. Recall the generating function $\Phi_s(x|Y)$ for which we derived the boundary-to-bulk propagator
in section 3.1. If we identify $Y^\mu = \sigma^\mu_{\A\db} y^\A\bar y^\db$, then the component $\Omega^{(s-1,s-1)}$ is related to
$\Phi_s$ by
\ie
\Omega^{(s-1,s-1)} \sim {dx^\mu\over z}{\partial\over\partial Y^\mu}\Phi_s
\fe
We will fix our convention for the relative normalization later. The linearized equation
$D_0\hat W=0$, or $D_0 \Omega = -D_0 W' = -D_0 W'|_{z=\bar z=0}$, relates the other spin-$s$ components of $\Omega$ to $\Omega^{(s-1,s-1)}$
as well as to $B$.

Let us start with the linearized field $B(x|y,\bar y)$. Using the linearized equation
$d_Z S = B*K dz^2+B*\bar K d\bar z^2$, we can solve for the $z$-dependence of the master field $S$ by
integrating $d_Z S$,
\ie\label{sbrel}
S &= - z_\A d z^\A \int_0^1 dt \,t (B*K)|_{\hat z\to t\hat z} + c.c. \\
&= - z_\A d z^\A \int_0^1 dt \,t B(-t\hat z,\bar y) K(t) + c.c.
\fe
where $K(t) = e^{t z^\A y_\A}$. Define
\ie
s(y,\bar y,z) = \int_0^1 dt\,t B(-tz,\bar y)K(t)
\fe
so that we can write $S=-z_\A s(y,\bar y,z) dz^\A+c.c.$ Note that although $S$ may a priori
have a $z$-independent part, it can be gauged away using a $z$-dependent gauge parameter
$\epsilon(x|y,\bar y,z,\bar z)$. Next, using $d_Z \hat W = -D_0 S$,
we can solve for $W'$ by integrating again in $z^\A$ and $\bar z^\da$,
\ie\label{wpr}
W' &=  z^\A \int_0^1 dt\, D_0S_\A|_{z\to tz} +c.c. \\
&= - \int_0^1 dt z^\A \left( [W_0, z_\A s]_*|_{z\to tz}\right) +c.c. \\
&= - \int_0^1 dt z^\A \left( \left[{\partial W_0\over \partial y^\A}, s\right]_*|_{z\to tz}\right) +c.c. \\
&= {z^\A\over 2z} \left( dx^i (\sigma^{iz})_\A{}^\B\partial_\B + dx_\mu (\sigma^\mu)_\A{}^\db \partial_\db \right)
\int_0^1 dt \,s(y,\bar y,tz) +c.c. \\
&= {z^\A \over 2z} \left[dx_\A{}^\db\left(  \partial_\db+(\sigma^z)^\C{}_\db\partial_\C \right)-dz\partial_\A\right]
\int_0^1 dt \,s(y,\bar y,tz) +c.c. \\
&= {z^\A \over 2z} \left[dx_\A{}^\db\left(  \partial_\db+(\sigma^z)^\C{}_\db\partial_\C \right)-dz\partial_\A\right]
\int_0^1 dt \,(1-t)B(-tz,\bar y)K(t) +c.c. \\
&= {z^\A dx_\A{}^\db\over 2z}
\int_0^1 dt \,(1-t)\left(  \partial_\db-t(\sigma^z)^\C{}_\db z_\C \right)B(-tz,\bar y)K(t) +c.c.
\fe
In the above we used the notation $\partial_\A\equiv {\partial\over\partial y^\A}$,
$\partial_\da\equiv {\partial\over \partial \bar y^\da}$. The relation (\ref{wpr}) between
the linearized fields $W'$ and $B$ will be repeatedly used throughout this paper.

Now, we can write
\ie\label{ewaa}
D_0 \Omega &= - D_0 W'|_{z=\bar z=0} \\
&= \left.- \{ W_0, W'\}_*\right|_{z=\bar z=0} \\
&= -{1\over 2z}\left. \left[ W_0, z^\A\int_0^1 dt \,(1-t)\left(  \partial_\db-t(\sigma^z)^\C{}_\db z_\C \right)B(-tz,\bar y)K(t) \right]_*\right|_{z=\bar z=0} \wedge dx_\A{}^\db + c.c.\\
&= {1\over 2z}\left. \left[ {\partial W_0\over \partial y^\A}, \int_0^1 dt \,(1-t)\left(  \partial_\db-t(\sigma^z)^\C{}_\db z_\C \right)B(-tz,\bar y)K(t) \right]_*\right|_{z=\bar z=0} \wedge dx^{\A\db} + c.c.\\
\fe
Note that ${\partial_\A}W$ is linear in $y$ or $\bar y$; its $*$-commutator acts by taking a derivative on $y$
or $\bar y$. In the first term in the last line of (\ref{ewaa}), the $y$-derivative only acts on $K(t)$, and the result is zero after setting $z_\A=\bar z_\da=0$. So we have
\ie
D_0 \Omega &= {1\over 8z^2}\left. \left[ \bar y^\dc, \int_0^1 dt \,(1-t)\left(  \partial_\db-t(\sigma^z)^\C{}_\db z_\C \right)B(-tz,\bar y)K(t) \right]_*\right|_{\hat z=\bar{\hat z}=0} dx_{\A\dc}\wedge dx^{\A\db} + c.c.\\
&= {1\over 8z^2} \left[\partial_\db \partial_\dc B(0,\bar y) dx^{\A\db}\wedge dx_\A{}^{\dc} +
\partial_\B \partial_\C B(y,0) dx^{\B\da}\wedge dx^\C{}_\da\right]\\
\fe
In other words, the linearized equation for the spin-$s$ component of $\Omega$ takes the form
\ie\label{omge}
d_L \Omega = -{dx^{\A\db}\over 2z} \left( y_\A {\partial_\db}
+ \bar y_\db {\partial_\A}\right) \Omega+{\cal C}^{(2s-2,0)}+{\cal C}^{(0,2s-2)}.
\fe
where ${\cal C}^{(2s-2,0)}$ and ${\cal C}^{(0,2s-2)}$ are functions of only $y$
and only $\bar y$, respectively, of degree $2s-2$.
Expanding (\ref{omge}) in powers of $y$ and $\bar y$, we have
\ie\label{omgenew}
& (d_L\Omega^{(n,m)})_{\A\B} =-{1\over 2z}\left[ y_{(\A} \partial^\dc (\Omega^{(n-1,m+1)})_{\B)\dc}
+ \bar y^\dc \partial_{(\A} (\Omega^{(n+1,m-1)})_{\B)\dc}\right], \\
& (d_L\Omega^{(n,m)})_{\da\db} = -{1\over 2z}\left[\bar y_{(\da} \partial^\C (\Omega^{(n+1,m-1)})_{\C\db)}
+ y^\C \partial_{(\da} (\Omega^{(n-1,m+1)})_{\C\db)}\right] ,
\fe
for $n,m\geq 1$,
where $d_L\Omega$ can be explicitly written in Poincar\'e coordinates as
\ie
(d_L \Omega)_{\A\B}
&=\left. \left[ \partial_{\A}{}^{\dc} + {1\over 2z}\left( y_\A (\sigma^z\partial_y)^{\dc} + \bar y^\dc(\sigma^z\partial_{\bar y})_{\A} \right)  - {(\sigma^z)_\A{}^\dc\over 4z} \left( y\partial_y + \bar y \partial_{\bar y} \right)  \right] \Omega_{\B\dc}\right|_{(\A\B)}.
\fe
We will now solve for $\Omega^{(s-1+n,s-1-n)}$, for $n=1-s,\cdots,s-1$, $n\not=0$, in terms of
$\Omega^{(s-1,s-1)}$, or $\Phi_s$, using (\ref{omge}).
The following useful relations follow from (\ref{omgenew}),
\ie\label{usefla}
& y^\A y^\B (d_L\Omega^{(n,m)})_{\A\B} = - {n+1\over 2z} y^\A \bar y^\db \Omega^{(n+1,m-1)}_{\A\db},\\
& \partial^\da \partial^\db (d_L\Omega^{(n,m)})_{\da\db}
={m+1\over 2z}  \partial^\A \partial^\db \Omega^{(n+1,m-1)}_{\A\db},\\
& y^\A \partial^\B (d_L\Omega^{(n,m)})_{\A\B} = {n+2\over 4z} y^\A \partial^\db \Omega^{(n-1,m+1)}_{\A\db} - {n\over 4z} \bar y^\db \partial^\A \Omega^{(n+1,m-1)}_{\A\db},\\
& \bar y^\da \partial^\db (d_L\Omega^{(n,m)})_{\da\db}
= {m+2\over 4z} \bar y^\db \partial^\A \Omega^{(n+1,m-1)}_{\A\db}
-{m\over 4z} y^\A \partial^\db \Omega^{(n-1,m+1)}_{\A\db}.
\fe
For now we will restrict ourselves to the spin-$s$ sector.
Define the shorthand notation $\Omega^n=\Omega^{(s-1+n,s-1-n)}$. We will split $\Omega^n_{\A\db}$ into
four terms, $\Omega_{\pm\pm}^n$, defined as
\ie
&\Omega_{++}^n = y^\A \bar y^\db \Omega^{n}_{\A\db},\\
&\Omega_{--}^n = \partial^\A \partial^\db \Omega^{n}_{\A\db}, \\
&\Omega_{-+}^n = \bar y^\db  \partial^\A \Omega^{n}_{\A\db},\\
& \Omega_{+-}^n = y^\A \partial^\db \Omega^{n}_{\A\db},\\
& \Omega^n_{\A\db} = {1\over s^2-n^2}\left( \partial_\A\partial_\db \Omega_{++}^n
- \bar y_\db \partial_\A \Omega_{+-}^n - y_\A \partial_\db \Omega_{-+}^n + y_\A \bar y_\db \Omega_{--}^n \right).
\fe
We can now invert (\ref{usefla}) and express $\Omega^n_{\pm\pm}$ in terms of $d_L$ acting on
$\Omega^{n+1}$ or $\Omega^{n-1}$ as
\ie
& \Omega^n_{++} = -{2z\over s+n-1}y^\A y^\B (d_L\Omega^{n-1})_{\A\B} ,\\
& \Omega^{n}_{--} = {2z\over s-n+1}
\partial^\da \partial^\db (d_L\Omega^{n-1})_{\da\db},\\
& \Omega^{n}_{-+}
={z\over s}\left[ (s-n) y^\A \partial^\B (d_L\Omega^{n-1})_{\A\B}
+(s+n) \bar y^\da \partial^\db (d_L\Omega^{n-1})_{\da\db} \right],\\
& \Omega^n_{+-} = {z\over s} \left[ (s-n) y^\A \partial^\B (d_L\Omega^{n+1})_{\A\B}
+(s+n) \bar y^\da \partial^\db (d_L\Omega^{n+1})_{\da\db} \right].
\fe
These relations allow us to raise or lower the index $n$, hence relating different components of $\Omega$,
all of which containing the spin-$s$ field.
To proceed we must now fix some gauge degrees of freedom.
The gauge transformations with a $\hat z$-independent parameter $\varepsilon(x|y,\bar y)$ act on $\Omega$ as
\ie
&\delta \Omega^n = d_L \varepsilon^n + dx^{\A\db} (y_\A \bar\partial_\db \varepsilon^{n-1}
+ \bar y_\db \partial_\A \varepsilon^{n+1}), \\
& \delta \Omega_{+-}^n = y^\A\partial^\db (d_L\varepsilon^n)_{\A\db}
+(s^2-n^2) \varepsilon^{n+1},\\
& \delta \Omega_{-+}^n = \bar y^\db\partial^\A (d_L\varepsilon^n)_{\A\db}
+(s^2-n^2) \varepsilon^{n-1}.
\fe
where we used the notation $\varepsilon^n\equiv \varepsilon^{(s-1+n,s-1-n)}$, analogously to $\Omega^n$.
We can use $\varepsilon^1,\cdots,\varepsilon^{s-1}$ to gauge away $\Omega^n_{+-}$ for $n\geq 0$,
and use $\varepsilon^{-1},\cdots,\varepsilon^{1-s}$ to gauge away $\Omega^n_{-+}$ for $n\leq 0$.
In the $n=0$ case, this is simply the statement that we can gauge away the trace part of the symmetric $s$-tensor
field obtained from $\Omega^{(s-1,s-1)}$, which is a priori double traceless rather than traceless.
This allows us to fix all $\Omega^n$'s in terms of $\Omega^0=\Omega^{(s-1,s-1)}$, and hence in terms of
$\Phi_s$.
Schematically, these relations take the form
\ie
& \Omega^n = \hat T_+ \Omega^{n-1},~~~~~n>0,\\
& \Omega^n = \hat T_- \Omega^{n+1},~~~~~n<0,\\
&\Omega = \left( 1+\sum_{n=1}^{s-1}\hat T_+^n+\sum_{n=1}^{s-1}\hat T_-^n\right) \Omega^0.
\fe
for some raising and lowering operators $\hat T_\pm$.
More explicitly, for $n>0$, we have
\ie
\Omega_{++}^n & = -{2z\over s+n-1} y^\A y^\B (d_L\Omega^{n-1})_{\A\B}
\\
& = -{2z\over s+n-1} y^\A y^\B \left[\partial_\A{}^\dc + {1\over 2z} y_\A (\sigma^z \partial_y)^\dc
+{1\over 2z}\bar y^\dc (\sigma^z\partial_{\bar y})_\A -{(\sigma^z)_\A{}^\dc
\over 4z}(2s-2)\right] \Omega^{n-1}_{\B\dc}
\\
& = -{2z\over s+n-1} y^\A \left[\partial_\A{}^\dc
+{1\over 2z}\bar y^\dc (\sigma^z\partial_{\bar y})_\A
- {s-1\over 2z}(\sigma^z)_\A{}^\dc\right] \Omega^{n-1}_{+\dc}
\\
& = -{2z\over s^2-(n-1)^2} y^\A \left[\partial_\A{}^\dc
+{1-n\over 2z}(\sigma^z)_\A{}^\dc \right] \partial_\dc \Omega^{n-1}_{++}
\\
&= {z\over s^2-(n-1)^2} y \left( {\slash\!\!\! \partial}
+{n-1\over z}\sigma^z \right)\partial_{\bar y} \Omega_{++}^{n-1}
\\
&\equiv \hat L_{++} \Omega_{++}^{n-1}
\fe
Recall that ${\slash\!\!\!\partial}_{\A\db}\equiv \sigma^\mu_{\A\db}\partial_\mu = -2\partial_{\A\db}$. The operator $\hat L_{++}$ can also be written as
\ie
\hat L_{++}\Omega_{++}^n = {1\over s^2-n^2}z^{1-n}(y{\slash\!\!\!\partial} \,\partial_{\bar y}) z^n \Omega_{++}^n.
\fe
Analogously, we can write down recursive formulae relating $\Omega_{-+}^n$ and $\Omega_{--}^n$
to those of index $n-1$, for $n>0$,
\ie
\Omega_{--}^n &= {2z\over s-n+1}\partial^\da \partial^\db (d_L\Omega^{n-1})_{\da\db}
\\
&={2z\over s-n+1}\partial^\da \partial^\db \left[\partial^\C{}_\da + {1\over 2z} y^\C (\sigma^z \partial_y)_\da
+{1\over 2z}\bar y_\da (\sigma^z\partial_{\bar y})^\C -{(\sigma^z)^\C{}_\da
\over 4z}(2s-2)\right] \Omega^{n-1}_{\C\db}
\\
&= {2z\over s-n+1}\partial^\da \left[\partial^\C{}_\da
+{1-n\over 2z} (\sigma^z)^\C{}_\da \right] \Omega^{n-1}_{\C-}
\\
&= {z\over s^2-(n-1)^2}y \left( {\slash\!\!\!\partial}
+{n-1\over z} \sigma^z \right) \partial_{\bar y} \Omega^{n-1}_{--}\\
&=\hat L_{++}\Omega^{n-1}_{--},
\fe
and
\begin{equation}\nonumber\begin{aligned}
&\Omega_{-+}^n = {z\over s} \left[ (s-n) y^\A \partial^\B (d_L\Omega^{n-1})_{\A\B}
+(s+n) \bar y^\da \partial^\db (d_L\Omega^{n-1})_{\da\db} \right]
\\
&= {z\over 2s} \left\{ (s-n) (\partial^\A y^\B+\partial^\B y^\A) \left[\partial_\A{}^\dc
+ {1\over 2z} y_\A (\sigma^z \partial_y)^\dc
+{1\over 2z}\bar y^\dc (\sigma^z\partial_{\bar y})_\A -{s-1\over 2z}(\sigma^z)_\A{}^\dc\right] \Omega^{n-1}_{\B\dc}
\right.\\
&~~~\left.+(s+n) (\partial^\da \bar y^\db+\partial^\db \bar y^\da) \left[\partial^\C{}_\da + {1\over 2z} y^\C (\sigma^z \partial_y)_\da +{1\over 2z}\bar y_\da (\sigma^z\partial_{\bar y})^\C
-{s-1\over 2z}(\sigma^z)^\C{}_\da \right] \Omega^{n-1}_{\C\db} \right\}
\end{aligned}\end{equation}
\ie
&= {z\over 2s} \left\{ (s-n)  \left[\partial^{\B\dc}
+{1\over 2z}\bar y^\dc (\sigma^z\partial_{\bar y})^\B -{s-1\over 2z}(\sigma^z)^{\B\dc}\right] \Omega^{n-1}_{\B\dc}
\right.\\
&~~~+(s-n) y^\A \left[\partial_\A{}^\dc
+{1\over 2z}\bar y^\dc (\sigma^z\partial_{\bar y})_\A -{s-1\over 2z}(\sigma^z)_\A{}^\dc\right] \Omega^{n-1}_{-\dc}
\\
&~~~+(s-n) \partial^\A\left[\partial_\A{}^\dc
+{1\over 2z}\bar y^\dc (\sigma^z\partial_{\bar y})_\A -{s-1\over 2z}(\sigma^z)_\A{}^\dc\right]
\Omega_{+\dc}^{n-1} + {s^2-n^2\over 2z}\left[ (\sigma^z)^{\A\db}\Omega^{n-1}_{\A\db}
-(\sigma^z\partial_y)^\dc \Omega^{n-1}_{+\dc}\right]
\\
&~~~+(s+n) \left[\partial^{\C\db} + {1\over 2z} y^\C (\sigma^z \partial_y)^\db
-{s-1\over 2z}(\sigma^z)^{\C\db} \right] \Omega^{n-1}_{\C\db}
\\
&~~~+(s+n) \bar y^\da \left[\partial^\C{}_\da + {1\over 2z} y^\C (\sigma^z \partial_y)_\da
-{s-1\over 2z}(\sigma^z)^\C{}_\da \right] \Omega^{n-1}_{\C-}
\\
&~~~
+(s+n) \partial^\da \left[\partial^\C{}_\da + {1\over 2z} y^\C (\sigma^z \partial_y)_\da -{s-1\over 2z}(\sigma^z)^\C{}_\da \right] \Omega^{n-1}_{\C+}\\
&~~~
\left.+{(s+n)(s-n+2)\over 2z}\left[ (\sigma^z)^{\A\db}\Omega^{n-1}_{\A\db}
- (\sigma^z\partial_{\bar y})^\C \Omega^{n-1}_{\C+} \right]
\right\}
\\
&= {z\over 2s} \left\{ 2s  \left[\partial^{\A\db}
-{s\over 2z}(\sigma^z)^{\A\db}\right] \Omega^{n-1}_{\A\db} + {s-n\over 2z} (\sigma^z\partial_{\bar y})^\B \Omega^{n-1}_{\B+}
+ {s+n\over 2z} (\sigma^z\partial_y)^\db \Omega^{n-1}_{+\db}
\right.
\\
&~~~+(s-n) y^\A \left[\partial_\A{}^\dc
-{s\over 2z}(\sigma^z)_\A{}^\dc\right] \Omega^{n-1}_{-\dc}
+{s-n\over 2z}y\sigma^z\partial_{\bar y} \Omega^{n-1}_{-+}
+ (s+n) \bar y^\da \left[\partial^\C{}_\da -{s\over 2z}(\sigma^z)^\C{}_\da \right] \Omega^{n-1}_{\C-}\\
&~~~
+(s-n) \partial^\A\left[ \partial_\A{}^\dc +{n\over 2z}(\sigma^z)_\A{}^\dc \right] \Omega_{+\dc}^{n-1}
+(s+n)\partial^\da \left[ \partial^\C{}_\da -{n-2\over 2z}(\sigma^z)^\C{}_\da \right] \Omega_{\C+}^{n-1}
\\
&~~~\left.
+{s\over z}\partial_y \sigma^z\partial_{\bar y} \Omega_{++}^{n-1}
+{(s+n)(s-n+1)\over z} (\sigma^z)^{\A\db}\Omega_{\A\db}^{n-1}
\right\}
\\
&= {z\over 2s} \left\{ 2s  \left[\partial^{\A\db}
-{s\over 2z}(\sigma^z)^{\A\db}\right] \Omega^{n-1}_{\A\db} - {s-n\over 2(s+n-1)z}
(\partial_y\sigma^z\partial_{\bar y}\Omega^{n-1}_{++}-y\sigma^z\partial_{\bar y} \Omega^{n-1}_{-+})
\right.
\\
&~~~- {s+n\over 2(s-n+1)z} \partial_y\sigma^z\partial_{\bar y} \Omega^{n-1}_{++}
-{s-n\over 2(s-n+1)} y \left( {\slash\!\!\!\partial}
+{s\over z}\sigma^z\right) (\partial_{\bar y} \Omega^{n-1}_{-+} - \bar y \Omega^{n-1}_{--})
\\
&~~~\left. +{s-n\over 2z}y\sigma^z\partial_{\bar y} \Omega^{n-1}_{-+} + {s+n\over 2(s+n-1)} y \left({\slash\!\!\!\partial}
+{s\over z}\sigma^z \right) {\bar y} \Omega^{n-1}_{--} \right.
\\
&~~~
-{s-n\over 2(s-n+1)} \partial_y \left( {\slash\!\!\!\partial} -{n\over z}\sigma^z \right)\partial_{\bar y} \Omega_{++}^{n-1}
-{s+n\over 2(s+n-1)}\partial_{\bar y} \left( {\slash\!\!\!\partial} +{n-2\over z}\sigma^z \right)
(\partial_y \Omega_{++}^{n-1}-y\Omega_{-+}^{n-1})
\\
&~~~\left.
+{s\over z}\partial_y \sigma^z\partial_{\bar y} \Omega_{++}^{n-1}
+{(s+n)(s-n+1)\over z} (\sigma^z)^{\A\db}\Omega_{\A\db}^{n-1}
\right\}
\fe
where we have used $\Omega^{n-1}_{+-}=0$. Finally, we arrive at recursive formula for
$\Omega^n_{-+}$,
\ie
\Omega^n_{-+}&= -{(s-n) z\over 2s(s-n+1)} \partial_y\left({\slash\!\!\!\partial}-{s+1\over z}
\sigma^z\right)\partial_{\bar y}\Omega^{n-1}_{++} + {(s+n)z\over 2s(s+n-1)}
y\left({\slash\!\!\!\partial}+{s-1\over z}\sigma^z\right){\bar y} \Omega^{n-1}_{--}
\fe
In the case $\Omega^0_{\A\db}\sim \partial_\A\partial_\db\Phi$, $\Omega^n_{--}=0$ for all $n\geq 0$ (and by the complex conjugate relations, for $n\leq 0$ as well), and $\Omega^0_{-+}=0$.
Therefore, to solve for $\Omega^n$ with $n>0$ we only need the recursive relations
\ie\label{recursive}
\Omega^n_{++} &= {z\over s^2-(n-1)^2} y \left( {\slash\!\!\! \partial}
+{n-1\over z}\sigma^z \right)\partial_{\bar y} \Omega_{++}^{n-1},\\
\Omega^n_{-+} &=  -{(s-n) z\over 2s(s-n+1)} \partial_y\left({\slash\!\!\!\partial}-{s+1\over z}
\sigma^z\right)\partial_{\bar y}\Omega^{n-1}_{++}.
\fe
Similarly, to solve for $\Omega^n$ with $n<0$, we only need the analogous relations for $\Omega_{++}$
and $\Omega_{+-}$.

Now using the $(2s-2,0)$ component of (\ref{omge}), $C(x|y)=B|_{\bar y=0}$ is related to $\Omega$ by
\ie
&(d_L\Omega^{s-1})_{\A\B} + {1\over 2z}y_{(\A} \partial^\dc \Omega^{s-2}_{\B)\dc} =
(d_L\Omega^{s-1})_{\A\B} - {1\over (2s-2)2z} y_\A y_\B \Omega_{--}^{s-2}\\
&= {1\over 4z^2}\partial_\A\partial_\B  C(x,z|y)
\fe
and so
\ie
C(x,z|y) &= {2z^2\over s(2s-1)}y^\A y^\B (d_L\Omega^{s-1})_{\A\B}\\
&= {2z^2\over s(2s-1)}y^\A y^\B \left[\partial_\A{}^\dc + {1\over 2z} y_\A (\sigma^z \partial_y)^\dc
+{1\over 2z}\bar y^\dc (\sigma^z\partial_{\bar y})_\A -{(\sigma^z)_\A{}^\dc
\over 4z}(2s-2)\right]\Omega^{s-1}_{\B\dc}\\
&= {2z^2\over s(2s-1)}y^\A \left[\partial_\A{}^\dc -
{s-1\over 2z}(\sigma^z)_\A{}^\dc\right]\Omega^{s-1}_{+\dc}\\
&={z^2\over s(2s-1)}y ({\slash\!\!\!\partial}+{s-1\over z}\sigma^z)\partial_{\bar y} \Omega_{++}^{s-1}.
\fe
We will choose a normalization convention for $\Omega^{(s-1,s-1)}$ in terms of $\Phi_s$, such that
the boundary-to-bulk propagator for $C(x|y)$ takes the simple form in the previous section, $C(x|y)=KT(y)^s$.
This is given by
\ie\label{omnorm}
&\Omega_{\A\db}^{(s-1,s-1)} = {(s!)^2\over 2\tilde N_s (2s)!}{1\over sz}\partial_\A\partial_\db \Phi_s, \\
&\Omega_{++}^0 = {(s!)^2\over 2\tilde N_s (2s)!}{s\over z}\Phi_s\\
&~~~~~~= {s z^{s}\over 2 (2s)! (x^-)^s}\partial_+^{2s}{(y{\bf x}\bar y)^s\over x^2}.
\fe
We can then express the generalized Weyl curvature $C(x|y)$ in terms of $\Phi_s$,
\ie
C(x|y) &=  {z\over s}\hat L_{++}\Omega_{++}^{s-1}
\\
&= {(s!)^2\over 2\tilde N_s (2s)!} z\hat L_{++}^s z^{-1}\Phi_s \\
&= {(s!)^2\over\tilde N_s (2s)!} {1\over (2s)!} z^{1-s}(z^2y{\slash\!\!\!\partial} \,\partial_{\bar y})^s z^{-1} \Phi_s \\
&=\left. {(s!)^2\over\tilde N_s (2s)!} { s!\over (2s)!} z^{1-s}e^{z^2y{\slash\!\!\!\partial} \,\partial_{\bar y}} z^{-1} \Phi_s\right|_{\bar y=0}.
\fe
Using the boundary-to-bulk propagator for $\Phi_s$ derived in the
first subsection, we have
\ie\label{caa}
C(x|y) &= \left. {s!\over ((2s)!)^2} z^{1-s}e^{z^2y{\slash\!\!\!\partial} \,\partial_{\bar y}} z^{-1}
{z^{s+1}\over (x^-)^s}\partial_+^{2s}{(y{\bf x} \bar y)^s\over x^2}
\right|_{\bar y=0}
\\
&= \left. {1\over ((2s)!)^2} z^{1-s}\partial_+^{2s} (z^2y{\slash\!\!\!\partial} \,\partial_{\bar y})^s (\bar y{\bf x} y)^s {z^s\over (x^-)^s}{1\over x^2}
\right|_{\bar y=0}
\\
&= { s!\over ((2s)!)^2} z^{1-s} \partial_+^{2s}
\left[ {1\over x^2 }(z^2y {\bf x}{\slash\!\!\!\partial} y)^s
{z^s\over (x^-)^s}\right]
\fe
Recall that we are now working in the light cone coordinate, with the polarization vector
given by $\varepsilon^+=1,\varepsilon^-=\varepsilon^\perp=0$.
To proceed, observe that
\ie
(z^2 y {\bf x}{\slash\!\!\!\partial} y) {z^n\over (x^-)^n} = n (y{\bf x}(x^-\sigma^z-z\sigma^-)y){z^{n+1}\over (x^-)^{n+1}}
=n Q {z^{n+1}\over (x^-)^{n+1}}
\fe
where $Q$ is defined by
\ie
Q &\equiv y{\bf x}(x^-\sigma^z-z\sigma^-)y\\
&={1\over 2}(x^2 y\sigma^{-z}y - y{\bf x}\sigma^{-z}{\bf x}y)
\fe
We shall also make use of the property $\left[z^2 y {\bf x}{\slash\!\!\!\partial} y, Q\right] = 0$.
Continuing on (\ref{caa}), we can write 
\ie
C(x,z|y) &= {1\over 2(2s)!} {z^{s+1}\over (x^-)^{2s}} \partial_+^{2s}
{Q^s\over x^2 }\\
&= {2^{-s-1}\over (2s)!} {z^{s+1}\over (x^-)^{2s}} \partial_+^{2s}
{(y{\bf x}\sigma^{-z}{\bf x}y)^s\over x^2 }
\\
&= {2^{-s-1}\over (2s)!} {z^{s+1}\over (x^-)^{2s}}(y{\bf x}\sigma^{-z}{\bf x}y)^s \partial_+^{2s}
{1\over x^2 }
\\
&=  2^{-s-1} (y{\bf x}\sigma^{-z}{\bf x}y)^s {z^{s+1}\over (x^2)^{2s+1} }\\
&= {1\over 2}K \left[{z\over 2(x^2)^2}y{\bf x}\sigma^{-z}{\bf x}y \right]^s
= {1\over 2} K T(y)^s
\fe
where $T(y)$ is defined as in section 3.1.
We can then recover the entire spin-$s$ part of the linearized master field $B$,
\ie
B(x|y,\bar y) &= (2s)! \sum_{n=0}^\infty {1\over n!(n+2s)!}{1\over z^{n+s}}(-z^2 y{\slash\!\!\!\partial}\bar y)^n z^s C(x|y) + c.c. \\
&= {1\over 2}(2s)! \sum_{n=0}^\infty {2^n\over n!(n+2s)!}{({1\over 2}y{\bf x}\sigma^{-z}{\bf x}y)^s \over z^{n+s}}
(-z^2 y{\slash\!\!\!\partial}\bar y)^n (
K^{2s+1})  + c.c.
\fe
The following relations are useful,
\ie
&(-z^2 y{\slash\!\!\!\partial}\bar y) K(x) = (-z y\Sigma \bar y) K(x),\\
&(-z^2 y{\slash\!\!\!\partial}\bar y)^n K(x)^m = {(n+m-1)!\over (m-1)!} (-z y\Sigma \bar y)^n K(x)^m,
\fe
where we used the fact $(-z^2 y{\slash\!\!\!\partial}\bar y)(-z y\Sigma \bar y)=(-z y\Sigma \bar y)^2$.
Now we arrive at the expression
\ie
B(x,z|y,\bar y)
&=  {1\over 2}\sum_{n=0}^\infty {1\over n!}{({1\over 2}y{\bf x}\sigma^{-z}{\bf x}y)^s \over z^{n+s}}(-z y\Sigma \bar y)^n
K^{2s+1}  + c.c. \\
&= {1\over 2}\left({y{\bf x}\sigma^{-z}{\bf x}y\over 2z}\right)^s
K^{2s+1} e^{-y\Sigma\bar y}  + c.c.
\label{B-btb}
\fe
This is the result we claimed in the previous subsection, the boundary-to-bulk propagator for $B$.

Finally, let us derive the formulae for the boundary to bulk propagator of $\Omega^n$
for $n=1,\cdots,s-1$.
Using (\ref{omnorm}), we obtain
\ie
\Omega_{++}^n &= {(s-n)!\over s(s+n-1)!} z^{-n}(z^2 y{\slash\!\!\!\partial}\partial_{\bar y})^n \Omega_{++}^0
\\
&= {(s-n)!\over 2(2s)!(s+n-1)!} z^{-n}\partial_+^{2s} (z^2 y{\slash\!\!\!\partial}\partial_{\bar y})^n(y{\bf x}\bar y)^s{z^s\over (x^-)^s} {1\over x^2}
\\
&= {(s-n)!\over 2(2s)!(s+n-1)!}{(-)^n s!\over (s-n)!} z^{-n} \partial_+^{2s} (y{\bf x}\bar y)^{s-n}(z^2 y{\bf x}{\slash\!\!\!\partial}y)^n{z^s\over (x^-)^s} {1\over x^2}
\\
&= {(-)^n\over 4(2s-1)!} {z^s\over (x^-)^{s+n}} \partial_+^{2s} \left[(y{\bf x}\bar y)^{s-n} { Q^n\over x^2}\right]
\\
&= {2^{-n-2}\over (2s-1)!} {z^s\over (x^-)^{s+n}} (y{\bf x}\sigma^{-z}{\bf x} y)^n
\partial_+^{2s} {(y{\bf x}\bar y)^{s-n}\over x^2}
\fe
Recall $Q = {1\over 2}(x^2y\sigma^{-z}y-y{\bf x}\sigma^{-z}{\bf x}y)$. On the other hand, for $\Omega_{-+}^n$,
\ie
\Omega^n_{-+} &= {(s-n)z\over 2s(s-n+1)} \partial_y ({\slash\!\!\!\partial}-{s+1\over z}\sigma^z)\partial_{\bar y} \Omega_{++}^{n-1}
\\
&= {2^{-n}(s-n)\over (2s)!(s-n+1)} \partial_+^{2s}
z^{s+2} (\partial_y {\slash\!\!\!\partial}\partial_{\bar y})
(y{\bf x}\bar y)^{s-n+1} (y{\bf x}\sigma^{-z}{\bf x} y)^{n-1}
{z^{-1}\over (x^-)^{s+n-1} x^2}
\\
&= -{2^{-n-1}(s-n)\over (2s)!} \partial_+^{2s}
z^{s+2} (\partial_y {\slash\!\!\!\partial}{\bf x}y)
(y{\bf x}\bar y)^{s-n} (y{\bf x}\sigma^{-z}{\bf x} y)^{n-1}
{z^{-1}\over (x^-)^{s+n-1} x^2}
\\
&= {2^{-n}(s-n)\over (2s)!} \partial_+^{2s}
z^{s+2} \partial_\mu \left[ x^\mu
(y{\bf x}\bar y)^{s-n} (y{\bf x}\sigma^{-z}{\bf x} y)^{n-1}
{z^{-1}\over (x^-)^{s+n-1} x^2} \right]
\\
&=0.
\fe
So in fact the boundary-to-bulk propagator for $\Omega_{-+}^n$ vanishes identically for all $n$.
We can therefore recover the boundary-to-bulk propagator for $\Omega$ entirely from $\Omega_{++}^n$,

\section{Three Point Functions}

In this section we will study three point functions of currents dual to higher spin gauge fields in $AdS_4$,
at tree level in Vasiliev theory. While we do not know the explicit Lagrangian
of Vasiliev theory, we can compute the correlation functions directly using the equation of motion, up to certain
normalization factors. In general, an $n$-point function $\langle J_1(\vec x_1) \cdots J_n(\vec x_n)\rangle$ can be computed by solving for the expectation value of the field dual to $J_n$, $\varphi_n(\vec x, z)$, at $\vec x=\vec x_n$ near the boundary $z\to 0$, sourced by the currents $J_1(\vec x_1),\cdots, J_{n-1}(\vec x_{n-1})$.
Strictly speaking, this computation gives the $n$-point function up to a normalization factor that depends
only on the field $\varphi_n$.

\begin{figure}
\begin{center}
\includegraphics[width=40mm]{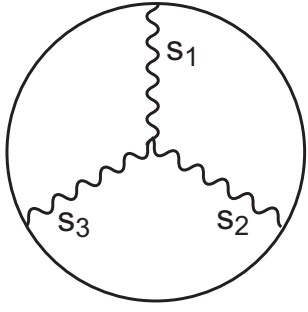}
\parbox{13cm}{
\caption{$C(s_1,s_2;s_3)$ will be computed by sewing two boundary-to-bulk propagators, corresponding to
sources of currents of spin $s_1$ and $s_2$, into a spin-$s_3$ field via the nonlinear equations of motion.  }}
\end{center}
\end{figure}

Let us analyze this more closely. Suppose a boundary operator $J(\vec x)$ is dual to a bulk field $\varphi$. We can express the AdS/CFT dictionary in a Schwinger-Dyson form
\ie
\left\langle J(\vec x_0) e^{\int d^3\vec x J(\vec x)\phi(\vec x)} \right\rangle = \int d^3\vec x \langle J(\vec x_0)J(\vec x) \rangle_{free} \phi(\vec x)-\int D\varphi|_\phi e^{-S} \int d^4x\sqrt{g}\, K_\varphi(x;\vec x_0) {\delta S_{int}\over \delta\varphi(x)}
\fe
where $(\cdots)|_\phi$ refers to the boundary condition $\varphi(\vec x,z\to 0)\to z^{\delta_-}\phi(\vec x)$,
$\delta_-$ being the appropriate scaling exponent associated to the boundary source of the field $\varphi$.
$K_\varphi(x;\vec x_0)$ is the boundary-to-bulk propagator for $\varphi$. We have separated the bulk action $S$ into a free action for $\varphi$ and the interaction part $S_{int}$; $\langle J(\vec x_0)J(\vec x)\rangle_{free}$ stands for the two-point function computed from the free action in the bulk. Here we assume that $\phi(\vec x)$ is supported away from $\vec x_0$. On the other hand, the expectation value of $\varphi$ near the boundary point $(\vec x_0,z=0)$ is given by
\ie
\left\langle \varphi(\vec x_0,z) \right\rangle_\phi = \int d^3\vec x' K_\varphi(\vec x_0,z;\vec x') \phi(\vec x')- \int D\varphi|_\phi e^{-S} \int d^4x'\sqrt{g}\, G_\varphi(\vec x_0,z;x') {\delta S_{int}\over \delta\varphi(x')}
\fe
where $G_\varphi(x;x')$ is the bulk propagator for $\varphi$. The boundary-to-bulk propagator is related by
\ie
G_\varphi(\vec x,z\to 0;\vec x',z')\to z^{\delta_+} K_\varphi(\vec x',z';\vec x).
\fe
Therefore, we have
\ie
\left\langle \varphi(\vec x_0,z\to 0) \right\rangle_\phi \to z^{\delta_+} \left\langle J(\vec x_0) e^{\int d^3\vec x J(\vec x)\phi(\vec x)} \right\rangle
\fe
In Vasiliev theory, however, we do not know a priori the normalization of the kinetic terms of
the spin-$s$ gauge fields, in terms of components of the master fields. Each spin-$s$ field $\varphi_s$ is
dual to the current $J_s$ in the boundary CFT with a certain normalization constant $a_s$. Here the currents
$J_s$ are understood to have appropriately normalized two-point functions. Furthermore, we have chosen an arbitrary
normalization for the boundary-to-bulk propagator for $\varphi_s$. So the boundary expectation value of $\varphi_s$
in the presence of sources is related to the correlation function of the currents by
\ie
\left\langle \varphi_s(\vec x_0,z\to 0) \right\rangle_\phi \to z^{s+1} {\cal C}_s \left\langle J_s(\vec x_0) e^{\sum a_i\int d^3\vec x J_i(\vec x)\phi_i(\vec x)} \right\rangle_{CFT},
\fe
or for the $n$-point function,
\ie
\left\langle \varphi_s(\vec x_0,z\to 0) \right\rangle_{(s_i;\vec x_i),~i=1,\cdots,n-1} \to z^{s+1} \left({\cal C}_s \prod_{i=1}^{n-1}a_{s_i}\right) \left\langle J_s(\vec x_0) J_{s_1}(\vec x_1)\cdots J_{s_{n-1}}(\vec x_{n-1}) \right\rangle_{CFT}.
\fe
Here ${\cal C}_s$ is an undetermined normalization constant that depends entirely on the normalization of
the field $\varphi_s$.
By comparing with the boundary-to-bulk propagator $K_{\varphi_s}$, one deduces
\ie\label{ratiosc}
{\left\langle \varphi_s(\vec x,z\to 0) \right\rangle_{(s_i;\vec x_i),~i=1,\cdots,n-1}\over
\left\langle \varphi_s(\vec x,z\to 0) \right\rangle_{(s;\vec x')}}\to {\prod_{i=1}^{n-1}a_{s_i}\over a_s} {\left\langle J_s(\vec x) J_{s_1}(\vec x_1)\cdots J_{s_{n-1}}(\vec x_{n-1}) \right\rangle_{CFT}\over
\left\langle J_s(\vec x) J_s(\vec x') \right\rangle_{CFT}}.
\fe
Combining various expectation values of $\varphi$ with sources, we can determine all the normalized correlation 
functions of the currents up to an overall constant, which may be identified with the coupling constant of Vasiliev theory.

The spatial and polarization dependence of the three point functions of the form
$\langle J_{s_1}(\vec x_1,\varepsilon_1)J_{s_2}(\vec x_2,\varepsilon_2)J_{s_3}(\vec x_3,\varepsilon_3) \rangle$
is fixed by conformal symmetry and the conservation of the currents up to a linear combination of finitely many possible structures. The coefficients characterize Vasiliev theory,
and we would like to compute them and compare with free and critical $O(N)$ vector models.
In the current paper, as a first step toward verifying the conjectured duality, we will assume that the spatial and polarization dependence
of $\langle J_{s_1}J_{s_2}J_{s_3}\rangle$ is proportional to that of free or critical $O(N)$ vector models, and compute the overall coefficient as a function of the three spins, which we denote by $C_{s_1s_2s_3}$. 
A general argument has been provided in \cite{Mikhailov:2002bp} stating that if
the three point functions of the currents (in other words, the OPEs of the currents)
have the same structure as in the free scalar field theory, then the structure of the $n$-point function of
the conserved higher spin currents $\langle J_{s_1}(\vec x_1)\cdots J_{s_n}(\vec x_n)\rangle$ is determined
in terms of
the corresponding currents $\phi\partial_{\mu_1}\cdots\partial_{\mu_s}\phi+\cdots$ in the free scalar field theory,
with the fields $\phi$ contracted in a cyclic order, and summed over permutations of these free field currents, with constant coefficients $A_\sigma$ that may depend on the particular permutation $\sigma\in S_n$.
It is far from obvious, a priori, that the assumption of \cite{Mikhailov:2002bp} that the three point functions
have the same structure as in free field theory, holds for the currents in Vasiliev theory. To demonstrate this is the main goal of this paper.

What we can compute using Vasiliev's equations of motion is the LHS of (\ref{ratiosc}). For three-point function
$\langle J_{s}(\vec x,\varepsilon)J_{s_1}(\vec x_1,\varepsilon_1)J_{s_2}(\vec x_2,\varepsilon_2) \rangle$,
where $\varepsilon, \varepsilon_1, \varepsilon_2$ are null polarization vectors, it suffices to consider the case
$\varepsilon_1=\varepsilon_2=\varepsilon$, and in the limit $\vec x_{12}\to 0$. The LHS
of (\ref{ratiosc}) in this limit, after stripping off the standard $\vec x$ and polarization dependence,
will be denoted $C(s_1,s_2;s)$. This is computed by the Witten diagram with two boundary-to-bulk propagators corresponding to spin $s_1$ and $s_2$ respectively, sewed together using the interaction terms in the equation of motion, and solving for the outcoming second order field of spin $s$ near the boundary. We will now carry out this computation explicitly.

\subsection{Some generalities}

We have seen that at the linearized level, $\Omega^{(s-1,s-1)}$ contains the symmetric traceless $s$-tensor gauge field, and $B^{(2s,0)}$ contains the generalized Weyl curvature. Either field can be used to extract
the correlation functions of the spin-$s$ current in the boundary CFT. It will be more convenient to
work with $B^{(2s,0)}$. Our strategy for computing $C(s_1,s_2;s_3)$ will be to compute the expectation value of $B^{(2s_3,0)}$ at the
second order in perturbation theory, with two sources on the boundary corresponding to the currents $J_{s_1}$
and $J_{s_2}$ respectively.

To do so, we make use of the equation of motion
\ie\label{bbeqn}
\tilde D_0 {\cal B} = -\hat W*B+B*\pi(\hat W),
\fe
While the linearized field $B$ does not depend on $z_\A, \bar z_\da$, at the second order the $B$ field in general does. From now on we will use the notation ${\cal B}$ to indicate the $\hat z$-dependence,
and write $B={\cal B}|_{z=\bar z=0}$. It suffices to consider (\ref{bbeqn}) restricted to
$z=\bar z=0$,
\ie
\tilde D_0 {\cal B}|_{z=\bar z=0} = \left.-\hat W*B+B*\pi(\hat W)\right|_{z=\bar z=0} \equiv J^Y
\fe
In order to solve for $B$ at the second order, we split ${\cal B}$ into $B$ and the $z$-dependent part
${\cal B}'$, $\tilde D_0{\cal B}_{z=\bar z=0}=\tilde D_0 B + \tilde D_0 {\cal B}'|_{z=\bar z=0}$. ${\cal B}'$ is solved from the equation
\ie\label{bseqn}
d_Z {\cal B}' = -S*B+B*\pi(\tilde S).
\fe
We will write $J^Z=-\tilde D_0 {\cal B}'|_{z=\bar z=0}$, and so
\ie
\tilde D_0 B = J^Y+ J^Z \equiv J = J_\mu dx^\mu.
\fe
This allows us to solve for $B(x|y,\bar y)$ from $J_\mu$.
More explicitly, in Poincar\'e coordinates,
\ie\label{eqnbb}
\left[\nabla_{\A\db} + {1\over 2z} (y_\A \bar y_\db + \partial_\A \partial_\db) \right] B(x|y,\bar y) = J_{\A\db}(x|y,\bar y)
\fe
where we have split $\tilde D_0$ into the Lorentz derivative $\nabla^L = d+[\omega_0^L,\,\cdots\,]_*$
and $\{e_0,\,\cdot\,\}_*$. We have previously encountered the homogeneous form of (\ref{eqnbb}) in solving
for the boundary-to-bulk propagator, but now with source $J$. By contracting (\ref{eqnbb}) with $y^\A \bar y^\db$,
and extracting the degree $(2s+1,1)$ term in the expansion in $y$ and $\bar y$, we obtain
\ie
y^\A \bar y^\db \nabla_{\A\db} B^{(2s,0)} + {2s+1\over 2z}B^{(2s+1,1)} = J^{(2s,0)}_{\A\db} y^\A y^\db
\fe
On the other hand, by acting on (\ref{eqnbb}) with $\partial^\A\partial^\db$, we have
\ie
\partial^\A \partial^\db \nabla_{\A\db} B^{(2s+1,1)} + {2(s+1)\over z}B^{(2s,0)} = \partial^\A\partial^\db J_{\A\db}^{(2s+1,1)}
\fe
Putting them together, we obtain a second order differential equation on $B^{(2s,0)}$ only,
\ie\label{eqncc}
&\left[\partial^\A \partial^\db \nabla_{\A\db} z y^\C \bar y^\dd \nabla_{\C\dd} - {(s+1)(2s+1)\over z}\right] B^{(2s,0)} \\
&=
\partial^\A \partial^\db \nabla_{\A\db} z J_{\C\dd}^{(2s,0)} y^\C \bar y^\dd - {2s+1\over 2}\partial^\A\partial^\db J_{\A\db}^{(2s+1,1)}  \equiv J(y).
\fe
The following formula for the first term on the RHS of (\ref{eqncc}) will be useful,
\ie\label{jysimplify}
&\partial^\A \partial^\db \nabla_{\A\db} z J_{\C\dd}^{(2s,0)} y^\C \bar y^\dd\\
&= \partial^\A \partial^\db \left[ \partial_{\A\db} + {1\over 2z}(\sigma^z)^\tau{}_\db y_\A\partial_\tau
+{1\over 2z}(\sigma^z)_\A{}^{\dot\tau} \bar y_\db \partial_{\dot\tau}
-{s+1\over 2z}\sigma^z_{\A\db}  \right] z J_{\C\dd}^{(2s,0)} y^\C \bar y^\dd \\
&=\left[ \partial^\A \partial^\db \partial_{\A\db} - {s+3\over 2z}(\partial_y \sigma^z\partial_{\bar y}) \right] z J_{\C\dd}^{(2s,0)} y^\C \bar y^\dd\\
&=-{z\over 2}\partial_y ({\slash\!\!\!\partial}\,-{s+2\over z}\sigma^z) {\slash\!\!\! J}^{(2s,0)} y
\fe
We will defer the solution to $B^{(2s,0)}$ from (\ref{eqncc}) to the next subsection. Now we will
consider the computation of $J(y)$
at the second order in perturbation theory, from the boundary-to-bulk propagators of the linearized fields.
As clear from (\ref{eqncc}) we only need to know $J_\mu|_{\bar y=0}$ and $\partial^\A\partial^\db J_{\A\db}|_{\bar y=0}$.

Explicitly, $J^Z$ is expressed in terms of ${\cal B}'$ as
\ie\label{jzaa}
J^Z_{\A\db} &= {1\over 2z} \left[\partial_{z^\A} (\sigma^z)^\C{}_\db \partial_\C
+\partial_{\bar z^\db} (\sigma^z)_\A{}^\dc\partial_\dc \right]{\cal B}|_{z=\bar z=0}
-{1\over 4z}\sigma^z_{\A\db} (\partial_z \partial_y + \partial_{\bar z}\partial_{\bar y}){\cal B}|_{z=\bar z=0}
\\
&~~~ + {1\over 2z}(\partial_{z^\A}\bar y_\db+\partial_{\bar z^\db} y_\A - \partial_{z^\A}\partial_{\bar z^\db}){\cal B}|_{z=\bar z=0}
\fe
In components, (\ref{bseqn}) can be written as
\ie\label{bcomp}
\partial_{z^\A} {\cal B} &= -S_\A*B+B*\bar\pi(S_\A) \\
&=\int_0^1 dt\,t \left[ \left(z_\A B(-tz,\bar y)K(t)\right)*B-B*\left(
z_\A B(-tz,-\bar y)K(t)\right) \right] ,\\
\partial_{\bar z^\db} {\cal B} &= -S_\db*B+B*\pi(S_\db) \\
&=\int_0^1 dt\,t \left[ \left(\bar z_\db B(y,-t\bar z)\bar K(t)\right)*B-B*\left(
\bar z_\db B(-y,-t\bar z)\bar K(t)\right) \right].
\fe
where we have used the linearized relation between $S$ and $B$, (\ref{sbrel}), and we have
suppressed the spacetime dependence of the fields in writing the above equations.
Note that $\partial_{z^\A}\partial_{\bar z^\db}{\cal B}=0$ at the second order.
Also observe that $\partial_y\partial_z{\cal B}|_{z=0}=\partial_{\bar y}\partial_{\bar z}{\cal B}|_{\bar z=0}=0$,
where the indices are contract, i.e. $\partial_y\partial_z=\epsilon^{\B\A}{\partial\over\partial y^\A}{\partial\over\partial z^\B}$ etc.
It follows that
\ie\label{ppj}
\partial^\A\partial^\db J^Z_{\A\db}|_{z=\bar z=0}=0,
\fe
in fact, without the need to set $\bar y$ to zero.
If we further set $\bar y=0$, it is not hard to see that
\ie
\partial_{z^\A} {\cal B}|_{z=\bar z=\bar y=0}=0.
\fe
This is because $z,\bar z$ and $\bar y$ are completely contracted under $*$-product in the first equation
of (\ref{bcomp}); while the $y^\A$ in $K(t)$ are not entirely contracted with $B$, we may replace either $(z^\A y_\A)*(\cdots)|_{z=0}$ or $(\cdots)*(z^\A y_\A)|_{z=0}$ by $y_\A\partial_y^\A(\cdots)|_{z=0}$.
One then observes that the two terms in the integrand in the first equation of (\ref{bcomp}) in fact
cancel each other, when $z,\bar z,\bar y$ are all set to zero at the end.
Note however that $\partial_{\bar z^\db} {\cal B}|_{z=\bar z=\bar y=0}$ does not vanish,
according to the second equation of (\ref{bcomp}).
Collecting these properties of ${\cal B}'$, we can simplify (\ref{jzaa}) when $\bar y$ is set to zero,
\ie
&J^Z_{\A\db}|_{\bar y=0} = {1\over 2z}\left(y+\sigma^z\partial_{\bar y} \right)_\A
\partial_{\bar z^\db}{\cal B}|_{z=\bar z=\bar y=0},
\fe
and therefore
\ie\label{yyj}
y^\A J^Z_{\A\db}|_{\bar y=0} &= {1\over 2z}(y\sigma^z\partial_{\bar y})
\partial_{\bar z^\db}{\cal B}|_{z=\bar z=\bar y=0}\\
&=\left.{1\over 2z}\int_0^1 dt\,t (y\sigma^z\partial_{\bar y})\left[ \left(\bar z_\db B(y,-t\bar z)\bar K(t)\right)*B-B*\left(
\bar z_\db B(-y,-t\bar z)\bar K(t)\right) \right]\right|_{\bar z=\bar y=0}\\
&=\left.-{1\over 2z}\int_0^1 dt\,t^2 (y\sigma^z)^\dc\left[ \left(\bar z_\db\bar z_\dc B(y,-t\bar z)\bar K(t)\right)*B-B*\left(
\bar z_\db\bar z_\dc B(-y,-t\bar z)\bar K(t)\right) \right]\right|_{\bar z=\bar y=0}\\
&~~~\left.+{1\over 2z}\int_0^1 dt\,t (y\sigma^z)^\dc\left[ \left(\bar z_\db B(y,-t\bar z)\bar K(t)\right)*
\partial_{\dc}B-\partial_{\dc}B*\left(
\bar z_\db B(-y,-t\bar z)\bar K(t)\right) \right]\right|_{\bar z=\bar y=0} \\
&=\left.-{1\over 2z}\int_0^1 dt\,t (y\sigma^z)^\dc\left\{ t\left[ \bar y_\db\bar y_\dc B(y,-t\bar y)
,B \right]_* -  \left\{ \bar y_\db B(y,-t\bar y),
\partial_{\dc}B \right\}_*\right\}\right|_{\bar y=0} \\
&=\left.{1\over 2z}\int_0^1 dt\,t (1-t) (y\sigma^z)^\dc\left[ \bar y_\db\bar y_\dc B(y,-t\bar y)
,B \right]_* \right|_{\bar y=0}.
\fe
In the above manipulation, we have frequently replaced the star product with $\bar y$ or $\bar z$ by derivatives on $\bar y$ and $\bar z$, or vice versa, as these variables are set to zero in the end.
(\ref{yyj}) and (\ref{ppj}) are all we need for the $J^Z$ contribution to $J(y)$ in (\ref{eqncc}).

Now let us turn to $J^Y$. It can be split into to terms, $J^Y=J^\Omega + J'^Y$, where
\ie\label{jomy}
& J^\Omega = -\Omega*B+B*\pi(\Omega),\\
& J'^Y = \left.-W'*B+B*\pi(W')\right|_{z=\bar z=0}.
\fe
We will also write $J'=J^Z+J^Y$, and $J_\mu = J^\Omega_\mu + J'_\mu$.

Let us examine the structure of $J'^Y$.
At the linearized order, recall from (\ref{wpr})
\ie
W' &= {\hat z^\A dx_\A{}^\db\over 2z} \int_0^1 dt\,(1-t)(\partial_\db - t(\sigma^z)^\C{}_\db \hat z_\C) B(-t \hat z,\bar y) K(t) + c.c.
\fe
We have
\ie
J'^Y_{\A\db}|_{\bar y=0} dx^{\A\db} &= -\left.W'*B+B*\pi(W')\right|_{z=\bar z=\bar y=0} \\
&=-{1\over 2z}\int_0^1 dt(1-t) \left.\left[\bar y d{\bf x} (\partial_y-t\sigma^z\bar y)B(y,-t\bar y),B \right]_*\right|_{\bar y=0}
\fe
It immediately follows that $\partial^\A\partial^\db J'^Y_{\A\db}|_{\bar y=0}=0$, as in the case of $J^Z$, because
it involves expression of the form $$\left.\partial_{y^\A} \left[(z^\A f(z,\bar y,zy))*\tilde B\right]\right|_{z=\bar z=0}=0$$ and the analogous complex conjugate expressions.
On the other hand, when restricting $J'^Y$ itself to $\bar y=0$, we have
\ie
y^\A J'^Y_{\A\db}|_{\bar y=0} = {1\over 2z}\int_0^1 dt(1-t) \left.y^\A\left[\bar y_\db (\partial_y-t\sigma^z\bar y)_\A B(y,-t\bar y),B \right]_*\right|_{\bar y=0}.
\fe
Combining this with (\ref{yyj}),
\ie
y^\A J^Z_{\A\db}|_{\bar y=0}
&=\left.{1\over 2z}\int_0^1 dt\,t (1-t) y^\A\left[ \bar y_\db(\sigma^z\bar y)_\A B(y,-t\bar y)
,B \right]_* \right|_{\bar y=0},
\fe
we obtain the contributions from $J'_\mu$,
\ie\label{yjp}
&y^\A J'_{\A\db}|_{\bar y=0} = {1\over 2z}\int_0^1 dt(1-t) \left.y^\A\left[\bar y_\db \partial_\A B(y,-t\bar y),B \right]_*\right|_{\bar y=0},\\
&\partial^\A\partial^\db J'_{\A\db}|_{\bar y=0}=0.
\fe
Suppose we have two sources on the boundary, at points $\vec x=0$ and $\vec x=\vec x_1$, of spin and polarization $(s,\varepsilon)$ and $(\tilde s,{\tilde\epsilon})$ respectively. We will denote $\tilde x = x-\vec x_1$,
$\vec{\tilde x}=\vec x-\vec x_1$, and similarly use the ``$\sim$" notation for all the variables associated with the spin $\tilde s$ current. Recall the expressions for
the spin-$s$ boundary-to-bulk propagator for the master field $B$,
\ie
&B(x|y,\bar y) = {1\over 2}Ke^{-y\Sigma \bar y} T(x|y,\varepsilon)^s + c.c.,\\
&\Sigma = \sigma^z-{2z\over x^2}{\bf x},\\
&T(x|y,\varepsilon) = {K^2\over z} y{\bf x}{\slash\!\!\!\varepsilon}\sigma^z {\bf x} y
={1\over 8z}(y(1-\Sigma\sigma^z)\lambda)^2.
\fe
In the last step above, we traded the null polarization vector $\varepsilon$ for a spinor $\lambda$, defined by
$2({\slash\!\!\!\varepsilon}\sigma_z)_{\A\B} = \lambda_\A\lambda_\B$, $2({\slash\!\!\!\varepsilon}\sigma_z)_{\da\db}
=\bar\lambda_\da \bar\lambda_\db$, with $\bar\lambda=\sigma^z\lambda$ (the factor of $2$ here is just our choice of convention).
Similarly, we must include the boundary-to-bulk propagators for the source of spin $\tilde s$ at $\vec x_1$.
Plugging these into (\ref{yjp}), we arrive at the expression
\ie\label{wywy}
y^\A J'_{\A\db}|_{\bar y=0} &={z\over 8x^2\tilde x^2}\int_0^1 dt\,(1-t) y^\A\left[\bar y_\db \partial_\A\left\{ e^{t\bar y\Sigma y} \left(T(y)^s+\bar T(-t\bar y)^s\right)\right\},\right.\\
&~~~\left.e^{-y\tilde\Sigma\bar y}\left( \tilde T(y)^{\tilde s}+\bar{\tilde T}(\bar y)^{\tilde s} \right) \right]_{*,\bar y=0} +(x\leftrightarrow \tilde x, ~s\leftrightarrow\tilde s).
\fe
We will defer the explicit computation of $J^\Omega_\mu$ to later. For now, let us point out that $\partial^\A\partial^\db J^\Omega_{\A\db}|_{\bar y=0}$ in general does
not vanish, unlike for $J'_\mu$.

Now we would like to compute $C(s,\tilde s;s')$, by extracting the $(2s',0)$ term in the
$(y,\bar y)$ expansion of $J(y)$. By counting powers of $y$ while contracting all $\bar y$'s, it is not
hard to see that
$J'_\mu$ contributes to the spin $s'$ field if $s'\geq |s-\tilde s|$, while $J^\Omega_\mu$ contributes if
$s'<s+\tilde s$. We may encounter three different cases:

(1) $s'\geq s+\tilde s$. Only $J'$ contributes.

(2) $s'<|s-\tilde s|$. Only $J^\Omega$ contributes.

(3) $|s-\tilde s|\leq s' < s+\tilde s$. Both $J'$ and $J^\Omega$ contribute.

There is also a special exceptional case:

(4) $s=\tilde s$, $s'=0$. In this case, the contributions from both $J'$ and $J^\Omega$ vanish, so that
naively we would conclude $C(s,s;0)=0$ for all $s$. We will see later that this is in fact
not the case, by ``analytically continuing" from $C(s_1,s_2;0)$ for $s_1\not=s_2$. This is
presumably due to a singular behavior related to the nonlocality of Vasiliev theory, which we do not fully understand.

There is a particularly simple case, when the triangular inequality among the three spins is strictly {\sl not} obeyed: if $s'> s+\tilde s$, then $\tilde s<s'-s$, and $s<s'-\tilde s$. So $C(s,\tilde s;s')$ receives contribution only from $J'$, while $C(s,s';\tilde s)$ and $C(\tilde s,s';s)$ receive contribution only from $J^\Omega$. We expect
\ie
C_{s_1 s_2 s_3} = {a_{s_3}\over a_{s_1}a_{s_2}}C(s_1,s_2;s_3)
\fe
to be the coefficient of the normalized three-point function, which should be symmetric in $(s_1, s_2, s_3)$.

In this section, we will compute explicitly $C(s,\tilde s;0)$, which receives contribution from $J^\Omega$ only.
They determine the three-point function coefficients $C_{0s\tilde s}$ up to a normalization factor of the form
$a(s)a(\tilde s)$. We will find agreement with the conjecture that the dual CFT is the free $O(N)$ vector theory, or the critical $O(N)$ model when the boundary condition for the scalar field is such that the dual operator has dimension $\Delta=2$ instead of $\Delta=1$.

Later, in section 6, we will consider the case when the outcoming field is of nonzero spin.
In particular, we will compute $C(0,s;s')$ in the case $s>s'$, which receives contribution from $J^\Omega$ only.
The result will allow us to determine the ratio among the normalization factor $a(s)$'s, when combined with
our result for $C(s,s';0)$. We will find that the two results are consistent with the
structure of the three-point function constrained by higher spin symmetry, and further, strikingly, in complete agreement
with the free $O(N)$ vector theory.

At the end of section 6, we will also consider the computation of $C(0,0;s)$ for both $\Delta=1$ and $\Delta=2$ boundary conditions on the bulk scalar field.
This coefficient receives contribution from $J'$ alone.
Our result for $C^{\Delta=1}(0,0;s)$ is however inconsistent with the other three-point function computations,
and our result for $C^{\Delta=2}(0,0;s)$ simply vanishes. We believe that this is an artifact due to the singularly
nonlocal behavior of Vasiliev theory, which requires a subtle regularization
which we do not fully understand. Presumably, the correct answer will be obtained if we take the two source spins
to be different, then analytically continue in the spins, and take the limit when the two spins coincide (both being zero in this case).

\subsection{Solving for $B$ at second order}

In this section, we will solve for $B(\vec x,z|y,\bar y)$ near the boundary $z\to 0$, from
(\ref{eqncc}). Let us write the LHS of (\ref{eqncc}) explicitly in Poincar\'e coordinates. First,
using our formula for $\omega_0^L$,
\ie
y^\C \bar y^\dd \nabla_{\C\dd} B^{(2s,0)} & = y^\A \bar y^\db \left[ \partial_{\A\db} + {1\over 2z}(\sigma^z)^\C{}_\db y_\A\partial_\C
-{s\over 2z}\sigma^z_{\A\db}  \right] B^{(2s,0)} \\
&= \left( y^\A \bar y^\db \partial_{\A\db} +{s\over 2z} y\sigma^z \bar y \right) B^{(2s,0)},
\fe
and then
\ie
&\partial^\A \partial^\db \nabla_{\A\db} z y^\C \bar y^\dd \nabla_{\C\dd} B^{(2s,0)}  =
\partial^\A \partial^\db \nabla_{\A\db} z \left( y^\C \bar y^\dd \partial_{\C\dd} +{s\over 2z} y\sigma^z \bar y \right) B^{(2s,0)} \\
&=\partial^\A \partial^\db \left[ \partial_{\A\db} + {1\over 2z}(\sigma^z)^\tau{}_\db y_\A\partial_\tau
+{1\over 2z}(\sigma^z)_\A{}^{\dot\tau} \bar y_\db \partial_{\dot\tau}
-{s+1\over 2z}\sigma^z_{\A\db}  \right] z \left( y^\C \bar y^\dd \partial_{\C\dd} +{s\over 2z} y\sigma^z \bar y \right) B^{(2s,0)} \\
&=\left[ \partial^\A \partial^\db \partial_{\A\db} - {s+3\over 2z}(\partial_y \sigma^z\partial_{\bar y}) \right] z \left( y^\C \bar y^\dd \partial_{\C\dd} +{s\over 2z} y\sigma^z \bar y \right) B^{(2s,0)} \\
&=\left[ -\partial^\A y^\C \partial_{\A}{}^\db z \partial_{\C\db} - {s+3\over 4}\partial_y \sigma^z {\slash\!\!\!\partial}
\,y + {s\over 4}\partial_y {\slash\!\!\!\partial} \sigma^z y+ {s(s+1)(s+3)\over 2z} \right] B^{(2s,0)} \\
&=\left[ -z\partial^\A y^\C \partial_{\A}{}^\db \partial_{\C\db} + {1\over 2}\partial^\A y^\C (\sigma^z)_{\A}{}^\db \partial_{\C\db} - {s+3\over 4}\partial_y \sigma^z {\slash\!\!\!\partial}
\,y + {s\over 4}\partial_y {\slash\!\!\!\partial} \sigma^z y+ {s(s+1)(s+3)\over 2z} \right] B^{(2s,0)} \\
&=\left[ -{s+1\over 2}z\partial^{\mu} \partial_{\mu}  - {s+2\over 4}\partial_y \sigma^z {\slash\!\!\!\partial}
\,y + {s\over 4}\partial_y {\slash\!\!\!\partial} \sigma^z y+ {s(s+1)(s+3)\over 2z} \right] B^{(2s,0)} \\
&=(s+1)\left[ -{1\over 2}z\partial^{\mu} \partial_{\mu}  + \partial_z -{1\over 2} y \sigma^z \hat{\slash\!\!\!\partial}
\,\partial_y + {s(s+3)\over 2z} \right] B^{(2s,0)}
\fe
Note that our convention for ${\slash\!\!\!\partial}$ is $\partial_{\A\db} = -{1\over 2}\sigma^\mu_{\A\db}\partial_\mu=-{1\over 2}{\slash\!\!\!\partial}_{\A\db}$. The equation (\ref{eqncc}) is now
\ie
\left[ z^2\partial^{\mu} \partial_{\mu}  -2z \partial_z + z y \sigma^z \hat{\slash\!\!\!\partial}
\,\partial_y  - {(s-2)(s+1)} \right] B^{(2s,0)} =-{2z\over s+1} J
\fe
The solution takes the form
\ie
B^{(2s,0)}(\vec x,z|y) = \int {d^3\vec x_0 dz_0\over z_0^4} G(\vec x,z;\vec x_0,z_0|y,\partial_{y_0}) \left[ -{2z_0\over s+1} J(x_0,z_0|y_0)
\right]
\fe
where $G(\vec x,z;\vec x_0,z_0|y,\partial_{y_0})$ is the Green's function for $B^{(2s,0)}$.
Define
\ie
&{\bf L}=z^2\partial^{\mu} \partial_{\mu}  -2z \partial_z + z y \sigma^z \hat{\slash\!\!\!\partial}
\,\partial_y  - {(s-2)(s+1)}.
\fe
The Green's function obeys
\ie\label{lg}
{\bf L}\cdot G(x,z;x',z'|y,\partial_{y'}) = (z')^4\delta(z-z')\delta^3(\vec x-\vec x') e^{y\partial_{y'}}|_{y'=0}.
\fe
In momentum space, we can write (\ref{lg}) as
\ie
&\left[ z^2 \partial_z^2-2z\partial_z-(s-2)(s+1)-z^2 p^2+izy\sigma^z {\slash\!\!\! p}\partial_y\right]\tilde G(z,z';p|y,\partial_{y'})\\
&= (z')^4\delta(z-z') {(y\partial_{y'})^{2s}\over (2s)!},
\fe
The small $z,z'$ limit is equivalent to the $p\to 0$ limit, where the equation reduces to
\ie
&\left[ z^2 \partial_z^2-2z\partial_z-(s-2)(s+1) \right]\tilde G(z,z';0|y,\partial_{y'}) = (z')^4\delta(z-z') {(y\partial_{y'})^{2s}\over (2s)!},
\fe
The solution is
\ie
& G(z,z';0|y,\partial_{y'}) = {z^{s+1}(z')^{2-s}\over 2s-1}{(y\partial_{y'})^{2s}\over (2s)!},~~~~~z<z';\\
& G(z,z';0|y,\partial_{y'}) = {z^{2-s}(z')^{s+1}\over 2s-1}{(y\partial_{y'})^{2s}\over (2s)!},~~~~~z>z'.
\fe
Fourier transforming back to position space, it follows that in the limit $z,z'\to 0$, $z>z'$,
\ie
G(\vec x,z;\vec x',z'|y,\partial_{y'}) \to (z')^{s+1} {z^{2-s}\over 2s-1}\delta^3(\vec x-\vec x'){(y\partial_{y'})^{2s}\over (2s)!}.
\fe

We will not need the
explicit form of $G(\vec x,z;\vec x_0,z_0|y,\partial_{y_0})$, since we are only interested in the behavior of $B^{(2s,0)}$ near a point $\vec x$ on
the boundary.
In the $z'\to 0$ limit, the Green's function reduces to,
\ie
G(\vec x,z;\vec x',z'|y,\partial_{y'}) \to (z')^{s+1} {\cal K}(\vec x-\vec x',z|y,\partial_{y'}),
\fe
where ${\cal K}$ is understood to act on a function homogenous in $y'_\A$ of degree $2s$. It satisfies
the equation and boundary condition
\ie\label{ckeqn}
& {\bf L}\cdot {\cal K}(\vec x,z|y,\partial_{y'}) =0,\\
&{\cal K}(\vec x,z|y,\partial_{y'})\to {z^{2-s}\over 2s-1}\delta^3(\vec x){(y\partial_{y'})^{2s}\over (2s)!},
~~~~z\to 0.
\fe
Importantly, we note that while in the case $s=0$,
${\cal K}$ is the boundary-to-bulk propagator for the scalar field,
for $s>0$ ${\cal K}$ is {\sl not} the same as the boundary-to-bulk propagator of $B^{(2s,0)}(\vec x,z|y)$ we derived earlier.
While the latter is also annihilated by ${\bf L}$, it does not obey the boundary condition of
(\ref{ckeqn}), and in particular its integral over $\vec x$ vanishes (unlike $\Phi_s(\vec x,z|y,\bar y)$).

Working in momentum space, we have
\ie\label{lpsp}
& \left[ \partial_z^2-2s\partial_z{1\over z} - p^2- {i\over z} y{\slash\!\!\! p}\sigma^z\partial_y \right] z^{s-1}\tilde{\cal K}(\vec p,z|y,\lambda) =0,\\
&\tilde {\cal K}(\vec p,z|y,\lambda)\to {z^{2-s}\over 2s-1} {(y\lambda)^{2s}\over (2s)!},
~~~~z\to 0.
\fe
We may write ${i\over 2}y{\slash\!\!\! p}\sigma^z\partial_y=\vec p\cdot\vec \ell$, where
$\vec\ell$ acts on $\tilde{\cal K}$ as an angular momentum operator of total spin $s$. The ``states" of angular momentum ${\hat p}\cdot\vec\ell=m$ along $\hat p={\vec p\over p}$ direction is given by
\ie
(y\lambda + iy\hat{\slash\!\!\!p}\sigma^z\lambda)^{s+m}(y\lambda - iy\hat{\slash\!\!\!p}\sigma^z\lambda)^{s-m}
\fe
On each ${\hat p}\cdot\vec\ell=m$ state, the equation (\ref{lpsp}) is solved by
confluent hypergeometric functions of the second kind. The ${\hat p}\cdot\vec\ell=m$ component of
$\tilde{\cal K}$ takes the form
\ie\label{psim}
&\tilde\psi_m(\vec p,z|y,\lambda) = {2^{-2s} z^{2-s} e^{-pz}\over 2s-1}
{U(m+1-s,2-2s|2pz)\over U(m+1-s,2-2s|0)}
{(y\lambda + iy\hat{\slash\!\!\!p}\sigma^z\lambda)^{s+m}\over (s+m)!}{(y\lambda - iy\hat{\slash\!\!\!p}\sigma^z\lambda)^{s-m}\over (s-m)!}
\\
&={2^{-2s} z^{2-s} \over 2s-1}\int_0^\infty dt\, e^{-pz(1+2t)}
{t^{m-s}(1+t)^{-m-s} \over B(m+1-s,2s-1)}
{(y\lambda + iy\hat{\slash\!\!\!p}\sigma^z\lambda)^{s+m}\over (s+m)!}{(y\lambda - iy\hat{\slash\!\!\!p}\sigma^z\lambda)^{s-m}\over (s-m)!},
\fe
for $m=-s+1,\cdots, s$. When $m\not= s$,
the integral representation in the second line should be understood as defined by analytic continuation in $s$.
The $m=-s$ case is special. In momentum space, there seems to be no solution with the desired boundary condition (\ref{lpsp}) at $z=0$.
Rather, there is a solution that dies off at $z\to \infty$ and behaves like $z^{s+1}$ near $z=0$,
\ie
\tilde\psi_{-s}'(\vec p,z|y,\lambda) &= z^{s+1} e^{-pz} f(p){(y\lambda - iy\hat{\slash\!\!\!p}\sigma^z\lambda)^{2s}\over (2s)!}.
\fe
where $f(p)$ is an arbitrary function of the momentum. For $f(p)= p^{2s-1}$, the Fourier transform of $\tilde\psi_{-s}$ gives
\ie
\psi_{-s}'(\vec x,z|y,\lambda) &= {1\over 2\pi^2} z^{s+1}{(y{\slash\!\!\!\partial}\sigma^z \lambda)^{2s}\over (2s)!}{1\over x^2}
= {2^{2s+1}\over \pi^2 } (y{\bf x}\sigma^z \lambda)^{2s}{z^{s+1}\over (x^2)^{2s+1}}.
\fe
This is nothing but the boundary-to-bulk propagator for $B^{(2s,0)}(\vec x,z|y)$ we derived previously.

Let us return to (\ref{psim}), and Fourier transform back to position space,
\ie
&\psi_m(\vec x,z|y,\lambda) = {2^{-2s} z^{2-s} \over 2s-1}\int_0^\infty dt
{t^{m-s}(1+t)^{-m-s} \over B(m+1-s,2s-1)}\\
&~~~\times{(-{1\over 2t+1}y\partial_z\lambda + y \hat {\slash\!\!\!\partial}\sigma^z\lambda)^{s+m}\over (s+m)!}{(-{1\over 2t+1}y\partial_z\lambda - y \hat{\slash\!\!\!\partial}\sigma^z\lambda)^{s-m}\over (s-m)!}\int {d^3\vec p\over (2\pi)^3} {e^{-pz(1+2t)+i\vec p\cdot\vec x}\over p^{2s}}
\\
&= {2^{-2s} z^{2-s} \over 2s-1}\int_0^\infty dt
{t^{m-s}(1+t)^{-m-s} \over B(m+1-s,2s-1)} \left.\left[{(y \sigma^z{\slash\!\!\!\partial}\lambda)^{s+m}\over (s+m)!}{( y {\slash\!\!\!\partial}\sigma^z\lambda)^{s-m}\over (s-m)!}\int {d^3\vec p\over (2\pi)^3} {e^{-pz+i\vec p\cdot\vec x}\over p^{2s}}\right]\right|_{z\to (2t+1)z}
\\
&= {2^{-2s} z^{2-s} \over 2s-1}\int_0^\infty dt {t^{m-s}(1+t)^{-m-s} \over B(m+1-s,2s-1)} \\
&~~~\times \left.\left\{ {(y \sigma^z{\slash\!\!\!\partial}\lambda)^{s+m}\over (s+m)!}{( y {\slash\!\!\!\partial}\sigma^z\lambda)^{s-m}\over (s-m)!} \left[-{\Gamma(2-2s)(x^2)^{s-1}\over 2\pi^2|\vec x|}{
\sin\left(2(s-1)\arctan{|\vec x|\over z} \right) } \right] \right\}\right|_{z\to (2t+1)z}
\fe
Note that although the factor $\Gamma(2-2s)$ seems to diverge for positive integer $s$,
the above expression should be understood as defined via analytic continuation in $s$;
upon taking $2s$ derivatives $(y \sigma^z{\slash\!\!\!\partial}\lambda)^{s+m}( y {\slash\!\!\!\partial}\sigma^z\lambda)^{s-m}$, the divergent term vanishes, leaving a finite result at integer values of
$s$.


The behavior of the outcoming spin-$s$ field near the boundary is now given by
\ie
&B^{(2s,0)}(\vec x,z\to 0)  \to  z^{s+1} \int {dz_0 d^3\vec x_0\over z_0^4} {\cal K}(\vec x-\vec x_0, z_0|y,\partial_{y_0}) \left[ -{2z_0\over s+1} J(\vec x_0,z_0;\vec x_1,\vec x_2|y_0) \right] .
\fe
where $\vec x_1, \vec x_2$ represent the positions of the two boundary sources. Here
we have suppressed the polarization of the source currents. We will see that the three point function coefficient of
the spin-$s$ current, $\langle
J_s\cdots\rangle$, receives contribution from only the {\sl helicity}-$s$
part of $B^{(2s,0)}(\vec x,z\to 0)$. This is extracted from the $\hat p\cdot\ell=s$ part of ${\cal K}$ above. We will denote
the helicity-$s$ part of the propagator ${\cal K}$ by ${\cal K}_{(s)}$, which is given explicitly by
\ie
&{\cal K}_{(s)}(\vec x,z|y,\lambda) =
{2^{-2s} z^{2-s} }\int_0^\infty dt
{(1+t)^{-2s}  } \\
&~~~\times \left.\left\{ {(y \sigma^z{\slash\!\!\!\partial}\lambda)^{2s}\over (2s)!} \left[-{\Gamma(2-2s)(x^2)^{s-1}\over 2\pi^2|\vec x|}{
\sin\left(2(s-1)\arctan{|\vec x|\over z} \right) } \right] \right\}\right|_{z\to (2t+1)z}
\fe
Away from $\vec x=0$, ${\cal K}_{(s)}$ has an expansion around $z=0$ of the form
\ie\label{kzser}
{\cal K}_{(s)}(\vec x,z|y,\lambda) = z^{2-s}\sum_{n=0}^\infty a_n^{(s)}(\vec x|y,\lambda) z^n + z^{s+1}\log(z)
\sum_{n=0}^\infty b_n^{(s)}(\vec x|y,\lambda).
\fe
Importantly, $b_0^{(s)}(\vec x|y,\lambda)$ is given by
\ie
b_0^{(s)}(\vec x|y,\lambda) = {\cal N}_s{(y{\bf \hat x}\sigma^z \lambda)^{2s}\over (2s)!(x^2)^{2s+1}},
~~~~~{\cal N}_s = {2^{2s-1}s\over \pi^2}.
\fe
where ${\bf \hat x}\equiv \vec x\cdot\vec \sigma$. The other helicity components ${\cal K}_{(m)}$, for $m<s$,
when expanded near $z=0$, will only have the first branch of (\ref{kzser}) and not the second branch with the $\log(z)$ factor.

The scalar field is a special case. For $s=0$, the dual operator can have dimension
$\Delta=1$ or $\Delta=2$. ${\cal K}_{(0)}$ is simply given by
\ie\label{ksimple}
&{\cal K}_{(0)}^{\Delta=1}(\vec x,z) = {1\over 2\pi^2}{z\over \vec x^2+z^2},\\
&{\cal K}_{(0)}^{\Delta=2}(\vec x,z) = {1\over \pi^2}{z^2\over (\vec x^2+z^2)^2},\\
\fe

The three-point function coefficient $C(s_1,s_2;s)$ will be computed from
\ie
&\lim_{z\to 0} z^{-s-1} B_{h=s}^{(2s,0)}(\vec x,z) = \int {dz_0 d^3\vec x_0\over z_0^4} {\cal K}_{(s)}(\vec x-\vec x_0, z_0|y,\partial_{y_0}) \left[ -{2z_0\over s+1} J(\vec x_0,z_0;\vec x_1,\vec x_2|y_0) \right] .
\fe

A particularly interesting limit is
when the two sources collide, $\vec\delta = \vec x_2-\vec x_1\to 0$. We can extract the coefficient
of the three-point function from this limit alone. Let us rescale the coordinates by defining
$\hat\delta = \vec\delta/\delta$, $z_0=\delta z'$, $\vec x_0 = \delta\vec x'$. Then
\ie
\lim_{z\to 0} z^{-s-1} B_{h=s}^{(2s,0)}(\vec x,z)
&= -{2\delta\over s+1}\int {dz' d^3\vec x'\over (z')^3} {\cal K}_{(s)}(\vec x-\delta\vec x', \delta z'|y,\partial_{y'}) J(\delta\vec x',\delta z'; \vec x_1, \vec x_2|y')
\\
&\to -{2\delta^{-s_1-s_2-2}\over s+1}\int {dz' d^3\vec x'\over (z')^3} {\cal K}_{(s)}(\vec x, \delta z'|y,\partial_{y'}) J(\vec x', z'; 0, \hat\delta|y')
\fe
in the $\delta\to 0$ limit. In the second step we used the scaling property of $J$ which follows
from our boundary-to-bulk propagators, and we used translational invariance to set $\vec{x}_1=0$.

For the scalar, this is given by
\ie\label{scalarccc}
&\lim_{z\to 0} z^{-1} B^{(0,0)}_{\Delta=1}(\vec x,z) \to -{\delta^{-s_1-s_2-1}\over\pi^2|\vec x|^2}\int {dz' d^3\vec x'} (z')^{-2} J(\vec x', z'; 0, \hat\delta),\\
&\lim_{z\to 0} z^{-1} B^{(0,0)}_{\Delta=2}(\vec x,z) \to -{2\delta^{-s_1-s_2}\over\pi^2|\vec x|^4}\int {dz' d^3\vec x'} (z')^{-1} J(\vec x', z'; 0, \hat\delta),~~~~~\delta\to 0.
\fe
On the other hand, in the $s>0$ case, it will turn out that in the $\delta\to 0$ limit, the only term in
(\ref{kzser}) that contributes is the term of order $z^{s+1}\log(z)$,
\ie\label{spinsccc}
&\lim_{z\to 0} z^{-s-1} B_{h=s}^{(2s,0)}(\vec x,z|y) \\
&~~\to -{\cal N}_s{2\delta^{s-s_1-s_2-1}\over(s+1)(\vec x^2)^{2s+1}}\int {dz' d^3\vec x'} (z')^{s-2}\log(z') {(y{\bf\hat x}\sigma^z\partial_{y'})^{2s}\over (2s)!}
J(\vec x', z'; 0, \hat\delta|y')\\
&~~=-{\cal N}_s{2\delta^{s-s_1-s_2-1}\over(s+1)(\vec x^2)^{2s+1}}\int {dz' d^3\vec x'} (z')^{s-2}\log(z')
J(\vec x', z'; 0, \hat\delta|{\bf\hat x}\sigma^z y),~~~~~\delta\to 0.
\fe
We will see that the terms of order $(z')^{-s-1},\cdots,(z')^{s-2}$, multiplied by $J$, integrate to zero.
In particular, ${\cal K}_{(m)}$ for $m<s$ will not contribute to (\ref{spinsccc}) in the limit $\delta/|\vec x|\to 0$,
and it is sufficient to consider the $m=s$ component alone.
Note that the scaling in $\delta$ and $|\vec x|$ of (\ref{scalarccc}) and (\ref{spinsccc})
are the ones expected of the three-point function of spin $s_1, s_2$ currents are $\vec x_1=0$, $\vec x_2=\hat\delta$
with a spin-$s$ current at $\vec x$ in the boundary CFT. To extract the coefficients $C(s_1,s_2;s)$, it remains to compute the integrals
\ie\label{scint}
&I^{\Delta=1}(\hat\delta)=\int d^3\vec x' dz' (z')^{-2} J(\vec x',z';
0,\hat\delta),\\
&I^{\Delta=2}(\hat\delta)=\int d^3\vec x' dz' (z')^{-1} J(\vec x',z';
0,\hat\delta),
\fe
in the $s=0$ case, and
\ie\label{ssint}
I_s(\hat\delta,y)=\int d^3\vec x' dz' (z')^{s-2}\log(z') J(\vec x',z';
0,\hat\delta|y)
\fe
in the $s>0$ case. In the next subsections, we will carry out the computation of three-point functions in detail.
The formal consistency of the AdS/CFT dictionary for Vasiliev theory is discussed in appendix A.

\subsection{The computation of $C(0,s;0)$}

In this subsection, we will carry out the explicit computation of the three-point function of
two scalars and one spin-$s$ current. More precisely, we will take the spin-$s$ field to be the outcoming field,
i.e. we will compute $C(0,s;0)$. As discussed before, this coefficient only receives contribution
from $J^\Omega$, and is simpler than the computation of $C(0,0;s)$, which we will defer to later sections.

We will take the spin-$s$ source to be at $\vec x=0$ on the boundary, and the scalar source at
$\vec x=\vec\delta$. When we write the first fields $B,\Omega$ etc. we mean the boundary-to-bulk propagators
sourced by the spin-$s$ operator. On the other hand, we write $\tilde x= x-\vec\delta$, and denote by $\tilde B(\tilde x|y,\bar y)$ etc. the fields sourced by the scalar operator. We shall first compute
$J_\mu dx^\mu = -\hat W*\widetilde B+\widetilde B*\pi(\hat W)-\hat {\widetilde W}* B+B*\pi(\hat {\widetilde W})$,
and then $J(y)|_{y=0}\equiv J^{(0)}$ which is the source for the outcoming scalar master field $B$ at the second order.
As we have seen, there is no contribution from the terms involving $W'$ and ${\widetilde W}'$ to
the scalar components of $J^{(0)}$, and the only contribution
comes from $J^\Omega_\mu dx^\mu=-\Omega^{(s)}*\tilde B^{(0)}+\tilde B^{(0)}*\pi(\Omega^{(s)})$, where
the superscripts indicate the corresponding spins.

Recall
our normalization convention $\Omega^{(s-1,s-1)}_{\A\db}={n_s\over z}\partial_\A\partial_\db\Phi_s$,
where $n_s = {(s!)^2\over 2\tilde N_s (2s)!}{1\over s} = {\pi^{3\over 2}\over 4s}{\Gamma(s-{1\over 2})\over s!}$.
$J^{(0)}_{\A\db}=J^{(0)}_\mu\sigma^\mu_{\A\db}$ is given by
\ie\label{jztemp}
J_{\A\db}^{(0)} &= -{n_s\over z}\left(\partial_\A\partial_\db \Phi_s * \tilde B-\tilde B*\pi(\partial_\A\partial_\db \Phi_s) \right)
- (\Omega^1_{\A\db}+\Omega^{-1}_{\A\db})*\tilde B +\tilde B*\pi(\Omega^1_{\A\db}+\Omega^{-1}_{\A\db}) + \cdots \\
&=-{n_s\over z}\{\partial_\A\partial_\db \Phi_s, \tilde B\}_* - [\Omega^1_{\A\db}+\Omega^{-1}_{\A\db},B]_* + \cdots
\fe
where $\cdots$ stands for terms that appear only at degree $(2,2)$ and higher in $(y,\bar y)$, and will not contribute
in our computation of $J(y)$ below. In the second line, we used the fact that $\Phi_s$ is of degree $(s,s)$ in $(y,\bar y)$, hence $\partial_\A\partial_\db\Phi_s$
is odd under $\pi$, for even spin $s$; $\Omega^1_{\A\db}=\Omega^{(s,s-2)}_{\A\db}$ and $\Omega^{-1}_{\A\db}$, on the other hand,
are even under $\pi$. Note that the term $[\Omega^1_{\A\db}+\Omega^{-1}_{\A\db},B]_*$ only contributes at degree $(1,1)$ in $(y,\bar y)$,
via the term $\partial^\A\partial^\db [\Omega^1_{\A\db}+\Omega^{-1}_{\A\db},B]_*|_{y=\bar y=0}\sim
[\Omega^1_{++}+\Omega^{-1}_{++},B]_*|_{y=\bar y=0}$. By counting powers in $y$ and $\bar y$, one sees that the latter vanishes identically.
So in fact only the first term in the second line of (\ref{jztemp}) will contribute to $J(y)$.

Now from the definition of $J(y)$ (\ref{eqncc}), we have
\ie\label{jze}
J^{(0)} &= \partial^\A \partial^\db \nabla_{\A\db} z J_{\C\dd}^{(0,0)} y^\A \bar y^\db - {1\over 2}\partial^\A\partial^\db J_{\A\db}^{(1,1)} \\
&= \left.-{z\over 2}\partial_y ({\slash\!\!\!\partial}-{2\over z}\sigma^z) {\slash\!\!\! J}y
- {1\over 2}\partial^\A\partial^\db J_{\A\db}\right|_{y=\bar y=0} \\
&= \left.-n_s\left[-{1\over 2} ({\slash\!\!\!\partial}-{3\over z}\sigma^z)^{\A\db} \{\partial_\A\partial_\db \Phi_s, \tilde B\}_*
- {1\over 2z}\{\partial_\A\partial_\db \Phi_s, \partial^\A\partial^\db \tilde B\}_*\right]\right|_{y=\bar y=0} \\
&= \left.-n_s\left[-{1\over 2} ({\slash\!\!\!\partial}-{3\over z}\sigma^z)^{\A\db} \{\partial_\A\partial_\db \Phi_s, \tilde B\}_*
- {s^2\over 2z}\{ \Phi_s, \tilde B\}_*\right]\right|_{y=\bar y=0} \\
&=-n_s\int d^4u d^4v (e^{uv+\bar u\bar v}+e^{-uv-\bar u\bar v}) \left[ {1\over 2} \partial_u({\slash\!\!\!\partial}-{3\over z}\sigma^z)\partial_{\bar u} - {s^2\over 2z} \right] \Phi_s(x|u,\bar u) \tilde B(\tilde x|v,\bar v)\\
&=-n_s\int d^4u d^4v (e^{uv+\bar u\bar v}+e^{-uv-\bar u\bar v}) \left[ {1\over 2} v({\slash\!\!\!\partial}-{3\over z}\sigma^z)\bar v - {s^2\over 2z} \right] \Phi_s(x|u,\bar u) \tilde B(\tilde x|v,\bar v)
\fe
In the second line we used the notation ${\slash\!\!\! J}=J_\mu\sigma^\mu$, and the formula (\ref{jysimplify}).
Note that unlike in $J'$, here $\partial^\A\partial^\db J^\Omega_{\A\db}$ does not vanish. In the last two lines above,
we used the integral representation of the star product, and traded $\partial_u$ for $v$ via integration by part.
Recall the boundary-to-bulk propagators for $\Phi_s(x|y,\bar y)$ and $\tilde B(\tilde x|y,\bar y)$,
\ie
& \Phi_s(x|y,\bar y) = {\tilde N_s\over (s!)^2}{z^{s+1}\over (x^-)^s}\partial_+^{2s}{(y{\bf x}\bar y)^s\over x^2},\\
& \tilde B(\tilde x|y,\bar y) = \tilde K e^{-y\tilde\Sigma \bar y}.
\fe
Using these, (\ref{jze}) becomes
\ie
J^{(0)} &= -{1\over 2s(2s)!}\int d^4u d^4v (e^{uv+\bar u\bar v}+e^{-uv-\bar u\bar v}) \left[ {1\over 2} v({\slash\!\!\!\partial}-{3\over z}\sigma^z)\bar v - {s^2\over 2z} \right] {z^{s+2}\over (x^-)^s\tilde x^2}e^{-v\tilde\Sigma \bar v}\partial_+^{2s}{(u{\bf x}\bar u)^s\over x^2}
\fe
In order to extract the three point function, we only need to calculate the integral (\ref{scint}),
\ie\label{ioa}
&I^{\Delta=1}(\hat\delta) = \int d^3\vec x dz\, z^{-2}J^{(0)}(\vec x,z;0,\hat\delta) \\
&= {1\over 2s(2s)!}\int {d^3\vec x dz\over 2z^3}\int d^4u d^4v (e^{uv+\bar u\bar v}+e^{-uv-\bar u\bar v}) ( v\sigma^z\bar v + s^2 ) {z^{s+2}\over (x^-)^s\tilde x^2}e^{-v\tilde\Sigma \bar v}\partial_+^{2s}{(u{\bf x}\bar u)^s\over x^2} \\
&= {1\over 2s(2s)!}\int {d^3\vec x dz\over 2z^3}\int d^4u d^4v (e^{uv+\bar u\bar v}+e^{-uv-\bar u\bar v})e^{-v\sigma^z \bar v} ( v\sigma^z\bar v + s^2 ) {z^{s+2}\over (x^-)^s\tilde x^2}e^{{2z\over \tilde x^2}v{\bf\tilde x} \bar v}\partial_+^{2s}{(u{\bf x}\bar u)^s\over x^2},
\fe
where in the first step we have integrated by part on $z$. To proceed,
we need a generating function
\ie
I_\lambda &\equiv \int d^4u d^4v (e^{uv+\bar u\bar v}+e^{-uv-\bar u\bar v}) ( v\sigma^z\bar v + s^2 ) e^{\lambda u{\bf x}\bar u-v\tilde\Sigma \bar v} \\
&= \left.( \partial_j\sigma^z\partial_{\bar j} + s^2 )\right|_{j=\bar j=0}\int d^4u d^4v (e^{uv+\bar u\bar v}+e^{-uv-\bar u\bar v})  e^{\lambda u{\bf x}\bar u-v\tilde\Sigma \bar v+jv+\bar j\bar v}\\
&= {2\over\det(1-\lambda\tilde\Sigma{\bf x})} \left.( \partial_j\sigma^z\partial_{\bar j} + s^2 )\right|_{j=\bar j=0} e^{\lambda j{\bf x}(1-\lambda\tilde\Sigma{\bf x})^{-1}\bar j} \\
&= {2\over\det(1-\lambda\tilde\Sigma{\bf x})}\left(-\lambda {\rm Tr}\left[\sigma^z{\bf x}(1-\lambda\tilde\Sigma{\bf x})^{-1} \right]+s^2 \right) \\
&= {2\over\det(1-\lambda\tilde\Sigma{\bf x})}\left[-2\lambda {z-\lambda x^2(1-{2z^2\over\tilde x^2})\over \det(1-\lambda\tilde\Sigma{\bf x})}+s^2 \right].
\fe
In the fourth line, ${\rm Tr}$ is the trace over chiral spinors. Later on when there is ambiguity, we will denote by ${\rm Tr}_+$
the trace over chiral indices and ${\rm Tr}_-$ for the trace over anti-chiral indices. Similarly, $\det$ here is understood as the determinant of
$2\times 2$ matrix. Further define
\ie
&\Xi=\det(1-\lambda\tilde\Sigma {\bf x}) = 1-2\lambda z + 4\lambda z{x\cdot\tilde x\over \tilde x^2}+\lambda^2 x^2,\\
&\xi = 1-2\lambda z + 4\lambda z{x\cdot\tilde x\over \tilde x^2}.
\fe
We can write the integral (\ref{ioa}) as
\ie
I^{\Delta=1}(\hat\delta) &= {s!\over 2s(2s)!}\int {d^3\vec x dz\over 2z^3}
{z^{s+2}\over (x^-)^s\tilde x^2} \partial_+^{2s} \left( {1\over x^2} \left.I_\lambda\right|_{\lambda^s} \right)
\\
&= {s!\over 2s(2s)!}\int {d^3\vec x dz }{z^{s-1}\over (x^-)^s\tilde x^2} \left.\partial_{x^+}^{2s} {1\over x^2\Xi}\left[-2\lambda {z-\lambda x^2(1-{2z^2\over\tilde x^2})\over \Xi}+s^2 \right]\right|_{\lambda^s}\\
&= {(s-1)!\over 2(2s)!} \int {d^3\vec x dz }{z^{s-1}\over (x^-)^s\tilde x^2} \left.\partial_{x^+}^{2s} {1\over x^2}\left(-2{\lambda z\over \xi^2}+{s^2\over\xi} \right)\right|_{\lambda^s},
\fe
where $|_{\lambda^s}$ means to take the coefficient of $\lambda^s$, when expanded in powers of $\lambda$. In the second and
the third line, note that $\partial_{x^+}$, as opposed to $\partial_+$, by definition acts on $x^+$ only and not on $\tilde x^+$. In the last step
we made use of the simple fact that no polynomials of degree $\geq 2s$ in $x$ appear on the RHS of $\partial_{x^+}^{2s}$.
So, we have
\ie
&I^{\Delta=1}(\hat\delta) = {(s-1)!\over 2(2s)!}\int {d^3\vec x dz }{z^{2s-1}\over (x^-)^s\tilde x^2} \partial_{x^+}^{2s} {1\over x^2}\left[-s2^{s}(1-2{x\cdot\tilde x\over \tilde x^2})^{s-1} + s^2 2^s(1-2{x\cdot\tilde x\over \tilde x^2})^s \right]
\\
&= {(s-1)!\over 2(2s)!} \int {d^3\vec x dz }{z^{2s-1}\over (x^-)^s\tilde x^2}
\sum_{n=0}^{2s}{2s\choose n}\left(\partial_{x^+}^{2s-n} {1\over x^2}\right)\partial_{x^+}^n\left[-s2^{s}(1-2{x\cdot\tilde x\over \tilde x^2})^{s-1} + s^2 2^s(1-2{x\cdot\tilde x\over \tilde x^2})^s \right] \\
&= {(s-1)!\over 2}\int {d^3\vec x dz }\,z^{2s-1}
\sum_{n=0}^{s}{s\choose n}{(x^-)^{s-n}(\tilde x^-)^n\over (x^2)^{2s-n+1}(\tilde x^2)^{n+1}}\\
&~~~~\times\left[-(s-n)2^{s}(1-2{x\cdot\tilde x\over \tilde x^2})^{s-n-1} + s^2 2^{s}(1-2{x\cdot\tilde x\over \tilde x^2})^{s-n} \right] \\
&= s 2^{2s-1} s!\int {d^3\vec x dz }{z^{2s-1}\over(x^2)^{s+1}\tilde x^2}
\left[({1\over 2}-{1\over 2s}){x^-\over x^2}-{x^-(x\cdot\tilde x)\over x^2\tilde x^2}+{\tilde x^-\over 2\tilde x^2}\right] \left[{x^-\over 2x^2}-{x^-(x\cdot\tilde x)\over x^2\tilde x^2}+{\tilde x^-\over 2\tilde x^2}\right]^{s-1}
\\
&= s 2^{s-1} s!\int {d^3\vec x dz }{z^{2s-1}\over(x^2)^{s+1}(\tilde x^2)^{s+1}}
\left[{x^-\over x^2}\left(1-{\tilde x^2\over s}\right)-\delta^-\right] \left({{x^-\over x^2}-\delta^-}\right)^{s-1}
\fe
Using the integral formula derived in Appendix B,
\ie
\int {d^3\vec x dz }{z^{2s-1}(x^-)^k\over(x^2)^{s+1+k}(\tilde x^2)^{s+1+n}} = J(2s-1,s+1+n,s+1+k,k,0)(\delta^-)^k,
\fe
with $J(\cdots)$ given by (\ref{jfive}),
we arrive at
\ie
I^{\Delta=1}(\hat\delta) &= {2^{-s-1}\pi^{5\over 2}\Gamma(s+{1\over 2})}(\delta^-)^s = {\pi^{5\over 2}\over 2}\Gamma(s+{1\over 2}) (\hat\delta\cdot\vec\varepsilon)^s.
\fe
In the last step we restored the null polarization vector $\vec\varepsilon$ of the spin-$s$ current.
Now using (\ref{scalarccc}), we find the behavior of the outcoming scalar field near the boundary $z\to 0$,
\ie\label{bzsz}
\lim_{z\to 0}z^{-1} B^{(0,0)}(\vec x,z) &\to -{I^{\Delta=1}(\hat\delta)\over \pi^2}{\delta^{-s-1}\over |\vec x|^2}\\
&\equiv C(0,s;0){(\vec\delta\cdot\vec\varepsilon)^s \over |\vec x|^2\delta^{2s+1}}
\fe
in the limit $\delta/|\vec x|\to 0$. The coefficient $C(0,s;0)$ is given by
\ie
C(0,s;0) &= -{\pi^{1\over 2}\over 2}\Gamma(s+{1\over 2}).
\fe

Let us compare (\ref{bzsz}) with the boundary behavior of the boundary-to-bulk propagator for $B^{(0,0)}(\vec x,z)$,
\ie\label{bzz}
\lim_{z\to 0}z^{-1} B_{prop}^{(0,0)}(\vec x,z) \to {1\over |\vec x|^2}.
\fe
The relative coefficient between (\ref{bzsz}) and $(\ref{bzz})$ determines the coefficient of the three-point function up to
certain factors that depends only on our normalization convention of the boundary-to-bulk propagators.
More precisely, in the limit where the two sources collide, the corresponding three-point function in the boundary CFT has the form
\ie\label{czsz}
\left\langle J_0(0) J_s(\vec \delta;\vec\varepsilon) J_0(\vec x) \right\rangle
&\to {g \over a_s} C(0,s;0){(\vec\delta\cdot\vec\varepsilon)^s \over |\vec x|^2\delta^{2s+1}},~~~~{\delta\over |\vec x|}\to 0.
\fe
where the position dependence on the RHS is fixed by conformal symmetry. More generally, before taking the limit ${\delta\over |\vec x|}\to 0$,
the structure of the above three point function, up to the overall coefficient, is fixed by conformal symmetry, which
we will derive explicitly using free field theory. The factor $a_s$ in (\ref{czsz}) is a normalization factor associated to
the boundary-to-bulk propagator for the spin-$s$ current. This is a priori not determined, since we do not know the normalization of the
two-point function of the operator dual to the spin-$s$ gauge field. $g$ is the overall coupling constant of Vasiliev theory,
which must be put in by hand, since we have been using only the bulk equation of motion, and not the action.

While (\ref{czsz}) does not by itself give the three point function $\left\langle J_0(0) J_s(\vec \delta;\vec\varepsilon) J_0(\vec x) \right\rangle$,
due to the ambiguity in $a_s$, we note that $a_s$ is a normalization factor that has to do with only
boundary-to-bulk propagator. We will be able to fix the relative normalization of $a_s$'s by computing, say, $C(0,0;s)$, which is related
to $C(0,s;0)$ by the symmetry properties of the three-point function of higher spin currents.


\subsection{$C(s,\tilde s;0)$}

We will now compute the three-point function coefficient of one scalar with two higher spin currents, in particular
$C(s,\tilde s;0)$. Alternatively, we could also consider $C(0,s;\tilde s)$, whose computation is more involved and will be deferred
to later sections. Note that as in the previous subsection, even though we have not yet fixed the normalization factor
$a_s$, knowing $C(s,\tilde s;0)$ we will be able to determine the normalized three-point function coefficient up to a factor
of the form $f(s)f(\tilde s)$, i.e. factorized normalization factors. The comparison of the non-factorized part of $C(s,\tilde s;0)$
to that of the free $O(N)$ theory would provide a highly nontrivial check of Klebanov-Polyakov conjecture.

We have seen that only $J^\Omega$ contributes to the computation of $C(s,\tilde s;0)$. Without loss of generality, let us assume $s>\tilde s$.
We are interested in the outcoming scalar field near the boundary. For this purpose we only need to consider the $(0,0)$ and $(1,1)$ components of $J_{\A\db}^{(0)}$ (superscript 0 indicating the scalar component) in its $(y,\bar y)$ expansion.
\ie
J_{\A\db}^{(0)} &= -\Omega_{\A\db}^{\tilde s}*\tilde B^{(s-1+\tilde s,s-1-\tilde s)}+\tilde B^{(s-1+\tilde s,s-1-\tilde s)}*\pi(\Omega_{\A\db}^{\tilde s})\\
&~~~
-\Omega_{\A\db}^{-\tilde s}*\tilde B^{(s-1-\tilde s,s-1+\tilde s)}+\tilde B^{(s-1-\tilde s,s-1+\tilde s)}*\pi(\Omega_{\A\db}^{-\tilde s})
+\cdots
\\
&=-\left\{\Omega_{\A\db}^{\tilde s}+\Omega_{\A\db}^{-\tilde s},\tilde B \right\}_*+\cdots
\fe
where $\cdots$ are terms involving other components of $\Omega_{\A\db}$, which do not contribute to $J(y)$ for the same reason as discussed
in the previous subsection.
In particular, the analogous terms with the spin $s$ and $\tilde s$ fields exchanged do not contribute.
By our gauge choice, $\Omega^{\tilde s}_{\A-}=\Omega^{-\tilde s}_{-\db}=0$, and so
\ie
J^{(0)} &= \left.-{z\over 2}\partial_y({\slash\!\!\!\partial} - {2\over z}\sigma^z){\slash\!\!\! J} y
-{1\over 2}\partial^\A\partial^\db J_{\A\db}\right|_{y=\bar y=0}
\\
&=\left.{z\over 2} ({\slash\!\!\!\partial} - {2\over z}\sigma^z)^{\A\db}
\left\{\Omega_{\A\db}^{\tilde s}+\Omega_{\A\db}^{-\tilde s},\tilde B \right\}_*
+{1\over 2}\partial^\A\partial^\db \left\{\Omega_{\A\db}^{\tilde s}+\Omega_{\A\db}^{-\tilde s},\tilde B \right\}_*
\right|_{y=\bar y=0}
\\
&=\left.{z\over 2} ({\slash\!\!\!\partial} - {2\over z}\sigma^z)^{\A\db}
\left\{\Omega_{\A\db}^{\tilde s}+\Omega_{\A\db}^{-\tilde s},\tilde B \right\}_*
+{1\over 2}\partial^\A\left\{\Omega_{\A\db}^{\tilde s},\partial^\db \tilde B \right\}_*
+{1\over 2}\partial^\db \left\{\Omega_{\A\db}^{-\tilde s},\partial^\A\tilde B \right\}_*
\right|_{y=\bar y=0}
\\
&=\left.{z\over 2} ({\slash\!\!\!\partial} - {2\over z}\sigma^z)^{\A\db}
\left\{\Omega_{\A\db}^{\tilde s}+\Omega_{\A\db}^{-\tilde s},\tilde B \right\}_*
+{1\over 2}\partial^\A\left[\Omega_{\A+}^{\tilde s}, \tilde B \right]_*
+{1\over 2}\partial^\db \left[\Omega_{+\db}^{-\tilde s},\tilde B \right]_*
\right|_{y=\bar y=0}
\\
&=\left.{z\over 2(s^2-\tilde s^2)} ({\slash\!\!\!\partial} - {2\over z}\sigma^z)^{\A\db}
\left\{\partial_\A\partial_\db\Omega_{++}^{\tilde s}-y_\A\partial_\db \Omega_{-+}^{\tilde s}
+\partial_\A\partial_\db\Omega_{++}^{-\tilde s}-\bar y_\db\partial_\A \Omega_{+-}^{-\tilde s},\tilde B \right\}_*
\right.\\
&~~~\left.+{1\over 2}\left[\Omega_{-+}^{\tilde s}+\Omega_{+-}^{-\tilde s}, \tilde B \right]_*
+{1\over 2}\left\{\Omega_{++}^{\tilde s}+\Omega_{++}^{-\tilde s}, \tilde B \right\}_*
\right|_{y=\bar y=0}
\fe
where we have repeatedly traded $\partial_y, \partial_{\bar y}$ with $y, \bar y$ under $*$ product, as $y,\bar y$ are set to zero in the end.
Let us split $J$ into two parts, $J^+$ and $J^-$, with
\ie
J^+ &= \left.{z\over 2(s^2-\tilde s^2)} ({\slash\!\!\!\partial} - {2\over z}\sigma^z)^{\A\db}
\left\{\partial_\A\partial_\db\Omega_{++}^{\tilde s}-y_\A\partial_\db \Omega_{-+}^{\tilde s},\tilde B \right\}_*
+{1\over 2}\left[\Omega_{-+}^{\tilde s}, \tilde B \right]_*
+{1\over 2}\left\{\Omega_{++}^{\tilde s}, \tilde B \right\}_*
\right|_{y=\bar y=0}
\fe
and $J^-$ the analogous expression with $\tilde s\to -\tilde s$, $y$ and $\bar y$ exchanged.
Note that $\left.\left[\Omega_{-+}^{\tilde s}, \tilde B \right]_*\right|_{y=\bar y=0} = 0$.
In fact, as we have seen earlier, the boundary-to-bulk propagators for $\Omega_{-+}^n$ are zero
in our gauge choice, and so
\ie
J^+ &= \left.{z\over 2(s^2-\tilde s^2)} ({\slash\!\!\!\partial} - {2\over z}\sigma^z)^{\A\db}
\left\{\partial_\A\partial_\db\Omega_{++}^{\tilde s},\tilde B \right\}_*
+{1\over 2}\left\{\Omega_{++}^{\tilde s}, \tilde B \right\}_*
\right|_{y=\bar y=0}
\fe

Now we shall make use of the formula
\ie
&\Omega^{\tilde s}_{++} = {2^{-\tilde s-2}\over (2s-1)!}{z^s\over (x^-)^{s+\tilde s}}
(y{\bf x}\sigma^{-z}{\bf x}y)^{\tilde s} \partial_+^{2s} {(y{\bf x}\bar y)^{s-\tilde s} \over x^2}, \\
&\tilde B = {1\over 2} \tilde K e^{-y\tilde\Sigma \bar y} (T(y)^{\tilde s} + \bar T(\bar y)^{\tilde s}),
\fe
for the boundary-to-bulk propagator of the spin-$s$ field at $\vec x=0$ (be aware that in our notation,
$\Omega^{\tilde s}\equiv \Omega^{(s-1+\tilde s,s-1-\tilde s)}$ is a spin-$s$ component of the $W$ master field, with
grading $\tilde s$),
and for the $B$ master field of the spin-$\tilde s$ field at $\vec x=\vec\delta$.
Using the integral representation of the star product, $J^+$ can be written as
\ie
J^+ &= -{1\over 2}\int d^4 u d^4 v \cosh(uv+\bar u\bar v)
\left[ \partial_u{z\over s^2-\tilde s^2}({\slash\!\!\!\partial}-{2\over z}\sigma^z)\partial_{\bar u}
-1\right]\Omega_{++}^{\tilde s}(x|u,\bar u) \tilde K e^{-v\tilde\Sigma \bar v} T(v)^{\tilde s}
\\
&=- {2^{-\tilde s-3}\over (2s-1)!}
\int d^4 u d^4 v \cosh(uv+\bar u\bar v)
\left[ v{z\over s^2-\tilde s^2}({\slash\!\!\!\partial}-{2\over z}\sigma^z) {\bar v}
-1\right] \\
&~~~\times {z^{s+\tilde s+1}\over (x^-)^{s+\tilde s}(\tilde x^2)^{2\tilde s+1}}
e^{-v\tilde\Sigma \bar v} (v{\bf\tilde x}{\slash\!\!\!\tilde\varepsilon}\sigma^{z}{\bf\tilde x} v)^{\tilde s} (u{\bf x}\sigma^{-z}{\bf x}u)^{\tilde s} \partial_+^{2s} {(u{\bf x}\bar u)^{s-\tilde s} \over x^2}
\fe
For simplicity, we will now assume that the polarization vector $\vec\varepsilon$ of the spin-$s$ current coincides
with the polarization vector $\vec{\tilde \varepsilon}$, namely $\tilde\varepsilon=\varepsilon$, and therefore in the light cone
coordinates, $\slash\!\!\!\tilde\varepsilon=
{1\over 2}\sigma^-$. This is all we need in order to extract the coefficient of the corresponding three-point function.
Now we have
\ie
J^+
&= -{2^{-2\tilde s-3}\over (2s-1)!}
\int d^4 u d^4 v \cosh(uv+\bar u\bar v)
\left[ v{z\over s^2-\tilde s^2}({\slash\!\!\!\partial}-{2\over z}\sigma^z) {\bar v}
-1\right] \\
&~~~\times {z^{s+\tilde s+1}\over (x^-)^{s+\tilde s}(\tilde x^2)^{2\tilde s+1}}
e^{-v\tilde\Sigma \bar v} (v{\bf\tilde x}\sigma^{-z}{\bf\tilde x} v)^{\tilde s} (u{\bf x}\sigma^{-z}{\bf x}u)^{\tilde s} \partial_+^{2s} {(u{\bf x}\bar u)^{s-\tilde s} \over x^2}
\\
&= -s\cdot{2^{-s-\tilde s-2}}
\int d^4 u d^4 v \cosh(uv+\bar u\bar v)
\left[ v{z\over s^2-\tilde s^2}({\slash\!\!\!\partial}-{2\over z}\sigma^z) {\bar v}
-1\right] \\
&~~~\times {z^{s+\tilde s+1}\over (x^2)^{2s+1}(\tilde x^2)^{2\tilde s+1}}
e^{-v\tilde\Sigma \bar v} (v{\bf\tilde x}\sigma^{-z}{\bf\tilde x} v)^{\tilde s} (u{\bf x}\sigma^{-z}{\bf x}u)^{\tilde s} (u{\bf x}\sigma^-{\bf x}\bar u)^{s-\tilde s}.
\fe
Then, the integral (\ref{scint}) is given by
\ie
I_+^{\Delta=1}(\hat\delta) &= \int d^3 x dz\,z^{-2}J^+\\
&=   s\cdot{2^{-s-\tilde s-2}} \int d^3 x dz {z^{s+\tilde s-1}\over (x^2)^{2s+1}(\tilde x^2)^{2\tilde s+1}}
\int d^4 u d^4 v \cosh(uv+\bar u\bar v)
\left( {v\sigma^z\bar v \over s^2-\tilde s^2}+1\right) \\
&~~~\times
e^{-v\tilde\Sigma \bar v} (v{\bf\tilde x}\sigma^{-z}{\bf\tilde x} v)^{\tilde s} (u{\bf x}\sigma^{-z}{\bf x}u)^{\tilde s} (u{\bf x}\sigma^-{\bf x}\bar u)^{s-\tilde s}
\\
&=   s{2^{-s-\tilde s-2}} (s-\tilde s)!((2\tilde s)!)^2 \int d^3 x dz {z^{s+\tilde s-1}\over (x^2)^{2s+1}(\tilde x^2)^{2\tilde s+1}}
\\
&~~~\left.\left.\times\left( {\partial_j\sigma^z\partial_{\bar j} \over s^2-\tilde s^2}+1\right)\right|_{j=\bar j=0}\int d^4 u d^4 v \cosh(uv+\bar u\bar v)
e^{-v\tilde\Sigma \bar v+jv+\bar j\bar v} e^{\zeta(v{\bf\tilde x}+u{\bf x})\bar\lambda} e^{\eta (u{\bf x}\sigma^-{\bf x}\bar u)}\right|_{\zeta^{4\tilde s}\eta^{s-\tilde s}}.
\fe
In the first step above, we integrated by part in $z$. In the second step, we turned the integral into a generating function
that involves a Gaussian integral only, and extract the coefficient of $\zeta^{4\tilde s}\eta^{s-\tilde s}$ in the end.
We have also introduced a ``polarization spinor" $\lambda$, or $\bar\lambda=\sigma^z\lambda$, which are related to the polarization vector by
$\sigma^{-z}_{\A\B}=\bar\lambda_\A\bar\lambda_\B$, or equivalently $\sigma^{-z}_{\da\db}=\bar\lambda_\da\bar\lambda_\db$. For instance,
we can then write $(u{\bf x}\sigma^{-z}{\bf x}u)=(u{\bf x}\bar\lambda)^2$.

After performing the Gaussian integral, we have
\ie
&I_+^{\Delta=1}(\hat\delta) = s{2^{-s-\tilde s-2}} (s-\tilde s)!((2\tilde s)!)^2 \int d^3 x dz {z^{s+\tilde s-1}\over (x^2)^{2s+1}(\tilde x^2)^{2\tilde s+1}} {1\over \det(1-\eta {\bf x}\sigma^-{\bf x}\tilde\Sigma)}
\\
&~~~\left.\left.\times\left( {\partial_j\sigma^z\partial_{\bar j} \over s^2-\tilde s^2}+1\right)\right|_{j=\bar j=0}e^{(j+\zeta\bar\lambda {\bf\tilde x})(1-\eta {\bf x}\sigma^-{\bf x}\tilde\Sigma)^{-1} \eta{\bf x}
\sigma^-{\bf x}\bar j} \cosh\left[ (j+\zeta\bar\lambda {\bf\tilde x})(1-\eta {\bf x}\sigma^-{\bf x}\tilde\Sigma)^{-1}
\zeta {\bf x}\bar\lambda\right] \right|_{\zeta^{4\tilde s}\eta^{s-\tilde s}}
\\
&= s{2^{-s-\tilde s-2}} (s-\tilde s)!((2\tilde s)!)^2 \int d^3 x dz {z^{s+\tilde s-1}\over (x^2)^{2s+1}(\tilde x^2)^{2\tilde s+1}} {1\over \det(1-\eta {\bf x}\sigma^-{\bf x}\tilde\Sigma)}
\\
&~~~\times \exp\left[ \zeta^2 \bar\lambda {\bf\tilde x}(1-\eta {\bf x}\sigma^-{\bf x}\tilde\Sigma)^{-1}
{\bf x}\bar\lambda\right] \left\{ 1 - {\eta \over s^2-\tilde s^2}{\rm Tr}_- \left[\sigma^z(1-\eta {\bf x}\sigma^-{\bf x}\tilde\Sigma)^{-1} {\bf x}
\sigma^-{\bf x}\right]
\right.\\
&~~~~~~\left.\left.
-{\zeta^2\eta \over s^2-\tilde s^2}
\bar\lambda{\bf\tilde x}(1-\eta {\bf x}\sigma^-{\bf x}\tilde\Sigma)^{-1} {\bf x}\sigma^-{\bf x}
\sigma^z(1-\eta {\bf x}\sigma^-{\bf x}\tilde\Sigma)^{-1} {\bf x}\bar\lambda
\right\}\right|_{\zeta^{4\tilde s}\eta^{s-\tilde s}}
\\
&= s{2^{-s-\tilde s-2}} (s-\tilde s)!((2\tilde s)!)^2 \int d^3 x dz {z^{s+\tilde s-1}\over (x^2)^{2s+1}(\tilde x^2)^{2\tilde s+1}} {1\over \det(1-\eta {\bf x}\sigma^-{\bf x}\tilde\Sigma)}
\\
&~~~\times \exp\left[ \zeta^2 {\rm Tr}_+\left( (1-\eta {\bf x}\sigma^-{\bf x}\tilde\Sigma)^{-1}
{\bf x}\sigma^{-z}{\bf\tilde x}\right)\right] \left\{ 1 - {\eta \over s^2-\tilde s^2}{\rm Tr}_- \left[(1-\eta {\bf x}\sigma^-{\bf x}\tilde\Sigma)^{-1} {\bf x}
\sigma^-{\bf x}\sigma^z\right]
\right.\\
&~~~~~~\left.\left.
-{\zeta^2\eta \over s^2-\tilde s^2}
{\rm Tr}_+\left[(1-\eta {\bf x}\sigma^-{\bf x}\tilde\Sigma)^{-1} {\bf x}\sigma^-{\bf x}
\sigma^z(1-\eta {\bf x}\sigma^-{\bf x}\tilde\Sigma)^{-1} {\bf x}\sigma^{-z}{\bf \tilde x}\right]
\right\}\right|_{\zeta^{4\tilde s}\eta^{s-\tilde s}}
\fe
where ${\rm Tr}_+$ and ${\rm Tr}_-$ stand for the trace
over the chiral and the anti-chiral sector, respectively.
To proceed, let us collect the following useful formulae,
\ie
& \Xi\equiv \det(1-\eta {\bf x}\sigma^-{\bf x}\tilde\Sigma) = 1-\eta\{\tilde\Sigma,{\bf x}\sigma^-{\bf x}\}\\
&~~~= 1-\eta\{\sigma^z - {2z{\bf\tilde x}\over \tilde x^2}, 2x^-{\bf x}-x^2\sigma^-\}\\
&~~~=1-4z\eta\left[x^- - {2x^-(x\cdot\tilde x)\over \tilde x^2} + {x^2\tilde x^-\over\tilde x^2} \right]\\
&~~~=1+2z\eta \left[\left(1 - {2x\cdot\tilde x \over \tilde x^2} + {x^2 \over\tilde x^2} \right)\bar\lambda{\bf x}\sigma^z\bar\lambda - {x^2\over\tilde x^2} \bar\lambda{\slash\!\!\!\delta}\sigma^z\bar\lambda \right]\\
&~~~=1+\eta{2z\over\tilde x^2} \left( \bar\lambda{\bf x}\sigma^z\bar\lambda - x^2 \bar\lambda{\slash\!\!\!\delta}\sigma^z\bar\lambda \right),
\\
& {\rm Tr}_+\left[ (1-\eta {\bf x}\sigma^-{\bf x}\tilde\Sigma)^T {\bf x}\sigma^{-z}{\bf\tilde x} \right]
= {\rm Tr}_+\left[ {\bf x}\sigma^{-z}{\bf\tilde x} - \eta\tilde\Sigma{\bf x}\sigma^-{\bf x}{\bf x}\sigma^{-z}{\bf\tilde x} \right]
=\bar\lambda{\bf\tilde x}{\bf x}\bar\lambda,\\
& {\rm Tr}\left[ (1-\eta {\bf x}\sigma^-{\bf x}\tilde\Sigma)^T {\bf x}\sigma^-{\bf x}\sigma^z \right]
=2z {\rm Tr} \left[ (1-\eta {\bf x}\sigma^-{\bf x}\tilde\Sigma)^T {\bf x}\sigma^- \right]
=4zx^-,\\
& {\rm Tr}_+\left[(1-\eta {\bf x}\sigma^-{\bf x}\tilde\Sigma)^T {\bf x}\sigma^-{\bf x}
\sigma^z(1-\eta {\bf x}\sigma^-{\bf x}\tilde\Sigma)^T {\bf x}\sigma^{-z}{\bf\tilde x}\right]
={\rm Tr}_+\left( {\bf x}\sigma^-{\bf x}
\sigma^z {\bf x}\sigma^{-z}{\bf\tilde x}\right) = 4zx^-(\bar\lambda{\bf\tilde x}{\bf x}\bar\lambda).
\fe
Using them, we can simplify the expression for $I_+^{\Delta=1}(\hat\delta)$ drastically,
\ie
&I_+^{\Delta=1}(\hat\delta) =  s{2^{-s-\tilde s-2}} (s-\tilde s)!((2\tilde s)!)^2 \int d^3 x dz {z^{s+\tilde s-1}\over (x^2)^{2s+1}(\tilde x^2)^{2\tilde s+1}} {1\over \det(1-\eta {\bf x}\sigma^-{\bf x}\tilde\Sigma)}
\\
&~~~\left.\times \exp\left({\zeta^2\bar\lambda{\bf\tilde x}{\bf x}\bar\lambda\over
\det(1-\eta {\bf x}\sigma^-{\bf x}\tilde\Sigma)}\right) \left[ 1 - {\eta \over s^2-\tilde s^2}{4zx^-\over \det(1-\eta {\bf x}\sigma^-{\bf x}\tilde\Sigma)}\left(1+{\zeta^2\bar\lambda{\bf\tilde x}{\bf x}\bar\lambda\over
\det(1-\eta {\bf x}\sigma^-{\bf x}\tilde\Sigma)} \right)
\right]\right|_{\zeta^{4\tilde s}\eta^{s-\tilde s}}
\\
&=  s{2^{-s-\tilde s-2}} (s-\tilde s)!(2\tilde s)! \int d^3 x dz {z^{s+\tilde s-1}\over (x^2)^{2s+1}(\tilde x^2)^{2\tilde s+1}}
\left.  {(\bar\lambda{\bf x}{\slash\!\!\!\delta}\bar\lambda)^{2\tilde s}\over
\Xi^{2\tilde s+1}}  \left[ 1 + {\eta \over s^2-\tilde s^2}{2z(\bar \lambda{\bf x}\sigma^z\bar \lambda)\over \Xi}\left(1+2\tilde s \right)
\right]\right|_{\eta^{s-\tilde s}}
\\
&=  s{2^{-2\tilde s-2}} (s+\tilde s)! \int d^3 x dz {z^{2s-1}(\bar\lambda{\bf x}{\slash\!\!\!\delta}\bar\lambda)^{2\tilde s}
\over (x^2)^{2s+1}(\tilde x^2)^{s+\tilde s+1}}  \left[
\left( \bar\lambda{\bf x}\sigma^z\bar\lambda - x^2 \bar\lambda{\slash\!\!\!\delta}\sigma^z\bar\lambda \right)^{s-\tilde s} \right.
\\
&~~~~~\left. - {\tilde x^2(\bar \lambda{\bf x}\sigma^z\bar \lambda)\over s+\tilde s}
\left( \bar\lambda{\bf x}\sigma^z\bar\lambda - x^2 \bar\lambda{\slash\!\!\!\delta}\sigma^z\bar\lambda \right)^{s-\tilde s-1}
\right]
\\
&=2^{-2s-2\tilde s-2}\pi^{5\over 2}\Gamma(s+\tilde s+{1\over 2}) (\bar\lambda{\slash\!\!\!\delta}\sigma^z\bar\lambda)^{s+\tilde s}.
\fe
In the last step, we have again used the integration formulae in Appendix B.
Similarly, we have an identical contribution from $I_-^{\Delta=1}(\hat\delta) = 2^{-2s-2\tilde s-2}\pi^{5\over 2}\Gamma(s+\tilde s+{1\over 2})(\bar\lambda{\slash\!\!\!\delta}\sigma^z\bar\lambda)^{s+\tilde s}$. Putting them together, we find
\ie
I^{\Delta=1}(\hat\delta)&=2^{-2s-2\tilde s-1}\pi^{5\over 2}\Gamma(s+\tilde s+{1\over 2})(\bar\lambda{\slash\!\!\!\delta}\sigma^z\bar\lambda)^{s+\tilde s}
\\
&={\pi^{5\over 2}\over 2}\Gamma(s+\tilde s+{1\over 2})(\hat\delta\cdot \vec\varepsilon)^{s+\tilde s}
\fe
where in the last step we replaced $\bar\lambda\bar\lambda$ by $\sigma^{-z}=2 {\slash\!\!\!\varepsilon}\sigma^z$.
Now we have the boundary behavior of the outcoming scalar field $B^{(0,0)}(\vec x,z)$,
\ie
\lim_{z\to 0} z^{-1} B^{(0,0)}(\vec x|z) &\to C(s,\tilde s;0){(\vec\delta\cdot \vec\varepsilon)^{s+\tilde s}\over |\vec x|^2\delta^{2s+2\tilde s+1}}\\
\fe
in the $\delta/|\vec x|\to 0$ limit, where the coefficient $C(s,\tilde s;0)$ is given by
\ie\label{csszres}
C(s,\tilde s;0) &= -{\pi^{1\over 2}\over 2}\Gamma(s+\tilde s+{1\over 2})
\fe
Note that even though we have assumed $s>\tilde s$ in the computation, the result for $C(s,\tilde s;0)$ is symmetric in $s$ and $\tilde s$
by a naive ``analytic continuation" in the spins. A particularly intriguing case is when $\tilde s=s$. Naively,
the three point function $C(s,s;0)$ vanishes identically, as discussed earlier. In fact, there are no such cubic couplings in the bulk Lagrangian!
But a formal extrapolation from $C(s,\tilde s;0)$ for $s\not=\tilde s$ suggests that in fact
$C(s,s;0)=-{\pi^{1\over 2}\over 2}\Gamma(2s+{1\over 2})$. We believe that this is a singular feature of Vasiliev theory. For instance,
if we assume that there is a non-derivative cubic coupling involving three scalar fields, with boundary condition such that they have dual dimension $\Delta=1$, then the corresponding tree level three-point function would diverge, from the integration of the
product of three $\Delta=1$ boundary-to-bulk propagators over $AdS_4$. While it is a priori unclear how to regularize such a computation,
we have seen that by a formal analytic continuation we can compute such three point functions in Vasiliev theory.
Similarly, we suspect that there are ``vanishing" derivative couplings involving a scalar and a pair of spin-$s$ fields, together with a divergent
bulk integral gives the nonzero coefficients $C(s,s;0)$. Potentially, if one can extend Vasiliev theory to $AdS_d$ for $d=4-\epsilon$,
such three-point functions could be computed using dimensional regularization.

We expect that corresponding three-point function in the dual CFT to behave as
\ie
\left\langle J_s(0;\vec\varepsilon) J_{\tilde s}(\vec \delta;\vec\varepsilon) J_0(\vec x) \right\rangle
&\to g {a_0\over a_sa_{\tilde s}} C(s,\tilde s;0){(\vec\delta\cdot\vec\varepsilon)^{s+\tilde s} \over |\vec x|^2\delta^{2s+2\tilde s+1}},~~~~{\delta\over |\vec x|}\to 0.
\fe
In the next section, we will compare our result (\ref{csszres}) to that of the {\sl free} $O(N)$ vector theory in three dimensions.

\subsection{Comparison to the free $O(N)$ vector theory}

In this section, we consider the free CFT of $N$ massless scalar fields in three dimensions, in the $O(N)$ singlet
sector. We may alternatively think of the theory as defined by gauging the $O(N)$ symmetry and then taking the gauge coupling to zero.
We will first examine the spectrum of operators, which consists of higher spin currents, and compute their correlation functions.

Let us denote the $N$ massless scalar fields by $\phi^i$, $i=1,\cdots,N$. A class of primary operators are spin-$s$ currents of the form
\begin{equation}
J_{\mu_1\cdots\mu_s} = \phi^i \partial_{(\mu_1}\cdots \partial_{\mu_s)} \phi^i+\cdots
\end{equation}
where $\cdots$ stands for similar terms with the derivatives distributed in various ways on the two $\phi^i$'s. We demand that $J_s$ are conserved currents, with the indices $\mu_1,\cdots,\mu_s$ symmetric and traceless.
These conditions fix $J_{\mu_1\cdots\mu_s}$ up to an overall normalization. It is convenient to introduce a
polarization vector $\vec\varepsilon$ and write $J_s(\vec\varepsilon)=J_{\mu_1\cdots\mu_s}\varepsilon^\mu_1\cdots\varepsilon^\mu_n$, or
$J_s$ for short. The explicit form of $J_s$ will be determined shortly. Note that in the $O(N)$ theory only the even spin conserved currents exists;
the $\phi^i$ bilinear operators with an odd number of derivatives are always descendants of the even spin currents.

The currents can be packaged in a generating function
\begin{equation}
{\cal O}_f(\vec x;{\varepsilon})=\phi^i(\vec x) f({\varepsilon}_\mu,\overrightarrow\partial_\mu,\overleftarrow\partial_\mu) \phi^i(\vec x) = \sum_{s=0}^\infty J_{\mu_1\cdots\mu_s}(\vec x) {\varepsilon}^{\mu_1} \cdots {\varepsilon}^{\mu_s}.
\end{equation}
The conservation and traceless condition on the currents can be implement on the function $f(\vec {\varepsilon},\vec u,\vec v)$ as
\begin{equation}\label{feqn}
(\vec u+\vec v)\cdot \vec\partial_{\varepsilon} f=\vec\partial_{\varepsilon}^2 f = 0.
\end{equation}
Further, by the massless equations of motion, we may assume
$u^2=v^2=0$ in $f(\vec {\varepsilon},\vec u,\vec v)$. The equations (\ref{feqn}) can be solved in three dimensions by
\begin{equation}
e^{\vec\alpha_\pm\cdot \vec{\varepsilon}},~~~~~\vec \alpha_\pm = \vec u-\vec v\pm \sqrt{-2\over u\cdot v} \vec u \times \vec v.
\end{equation}
In particular, we may take the function $f$ to be
\begin{equation}
f(\vec {\varepsilon},\vec u,\vec v) = {e^{\vec\alpha_+\cdot {\vec\varepsilon}}+e^{\vec\alpha_-\cdot {\vec\varepsilon}}\over 2} =
e^{(u-v)\cdot {\varepsilon}} \cosh\left[\sqrt{2(u\cdot v) {\varepsilon}^2-4 (u\cdot {\varepsilon}) (v\cdot {\varepsilon})}\right].
\end{equation}
Correspondingly, the generating operator ${\cal O}(\vec x;\vec{\varepsilon})$ is given by
\begin{equation}
\begin{aligned}
{\cal O}(\vec x;{\vec\varepsilon}) &= \phi^i(x-{\varepsilon}) \sum_{n=0}^\infty {\left(2{\varepsilon}^2\overleftarrow\partial_x\cdot \overrightarrow\partial_x -4 ({\varepsilon}\cdot \overleftarrow\partial_x)({\varepsilon}\cdot \overrightarrow\partial_x)\right)^n\over (2n)!} \phi^i(x+{\varepsilon})
\\
& = \phi^i\phi^i(x)+\phi^i(x)\overleftrightarrow\partial_\mu \phi^i(x) {\varepsilon}^\mu\\
&+{1\over 2}\left[\phi^i(x) \overleftrightarrow\partial_\mu
\overleftrightarrow\partial_\nu \phi^i(x) -2 \partial_{(\mu}\phi^i(x)
\partial_{\nu)} \phi^i(x)+2\delta_{\mu\nu} \partial^\rho\phi^i(x) \partial_\rho\phi^i(x)\right] {\varepsilon}^\mu {\varepsilon}^\nu+
\cdots
\end{aligned}
\end{equation}
In the second line, we exhibited the spin 0, spin 1 (which vanishes identically in the $O(N)$ theory) and spin 2 (the stress-energy tensor) currents explicitly.
The connected $n$-point functions of the currents can be easily computed via
\ie
\left\langle \prod_{i=1}^n {\cal O}(\vec x_i;\vec{\varepsilon}_i) \right\rangle &= {2^{n-1}N\over n}\sum_{\sigma\in S_n}
P_\sigma \overrightarrow{\prod_{i=1}^n} \left[\cosh(\sqrt{2{\varepsilon}_i^2 \overleftarrow\partial_i\cdot
\overrightarrow\partial_i-4({\varepsilon}_i\cdot\overleftarrow\partial_i)
({\varepsilon}_i\cdot\overrightarrow\partial_i)})\right.\\
&~~~~~\left.\times {1\over |x_{i}
-x_{i+1}+{\varepsilon}_{i} + {\varepsilon}_{i+1}|}\right]
\fe
where $P_\sigma$ stands for the permutation on $(\vec x_i;{\vec \varepsilon}_i)$ by $\sigma$, and the product is understood to be of cyclic order;
$\overleftarrow\partial$ and $\overrightarrow\partial$ act on their neighboring propagators only.
In particular, the two-point function can be written as
\begin{equation}
\begin{aligned}
&\left\langle {\cal O}(\vec x_1;{\vec \varepsilon}_1) {\cal O}(\vec x_2;{\vec \varepsilon}_2) \right\rangle = 2N{1\over |x_{12}+{\varepsilon}_1+{\varepsilon}_2| }\cosh(\sqrt{2{\varepsilon}_1^2 \overleftarrow\partial_1\cdot
\overrightarrow\partial_1-4({\varepsilon}_1\cdot\overleftarrow\partial_1)({\varepsilon}_1\cdot\overrightarrow\partial_1)})
\\
&~~~\times\cosh(\sqrt{2{\varepsilon}_2^2 \overleftarrow\partial_2\cdot
\overrightarrow\partial_2-4({\varepsilon}_2\cdot\overleftarrow\partial_2)({\varepsilon}_2\cdot\overrightarrow\partial_2)})
{1\over |x_{12}-{\varepsilon}_1-{\varepsilon}_2|}.
\end{aligned}
\end{equation}
It is not immediately obvious that by expanding this expression in powers of $\varepsilon_1$ and $\varepsilon_2$, we will find
an orthogonal basis of currents. Without loss of generality, we can assume that ${\varepsilon}$ is a null polarization vector. We will sometimes
work with the light cone
coordinates, such that $\varepsilon^+=1,\varepsilon^-=\varepsilon^\perp=0$. Then the current $J_s$ can be written explicitly as
\ie
J_s(\vec x,\vec \varepsilon) =  \sum_{n=0}^{s/2} {(-4)^n\over (2n)!}\sum_{k=0}^{s-2n}{(-)^k\over k!(s-2n-k)!}\partial_+^{n+k}\phi^i \partial_+^{s-n-k} \phi^i.
\fe
We will assume $s$ is even from now on.
The two-point function is then evaluated as
\begin{equation}\nonumber\begin{aligned}
&\left\langle J_{s_1}(\vec x,\vec\varepsilon_1) J_{s_2}(0, \vec\varepsilon_2) \right\rangle \\
&= 2N\left.{1\over |x| }e^{(\varepsilon_1+ \varepsilon_2)\cdot\overleftarrow\partial}\cos\left(2\sqrt{\varepsilon_1\cdot\overleftarrow\partial \varepsilon_1\cdot \overrightarrow\partial}\right)\cos\left(2\sqrt{\varepsilon_2\cdot\overleftarrow\partial \varepsilon_2\cdot \overrightarrow\partial}\right)e^{-(\varepsilon_1+\varepsilon_2)\cdot\overrightarrow\partial} {1\over |x|}\right|_{\varepsilon_1^{s_1}
\varepsilon_2^{s_2}}
\\
&= 2N\sum_{k=0}^{s_1}\sum_{\ell=0}^{s_2}{4^{k+\ell}\over (2k)!(2\ell)!}
\sum_{n=0}^{s_1-2k}\sum_{m=0}^{s_2-2\ell} {(-)^{m+n+k+\ell}\over n!m!(s_1-2k-n)!(s_2-2\ell-m)!}\\
&~~~~\times \left[(\varepsilon_1\cdot\partial)^{k+n}
(\varepsilon_2\cdot\partial)^{\ell+m}{1\over |x| }\right] \cdot \left[ (\varepsilon_1\cdot\partial)^{s_1-k-n}
(\varepsilon_2\cdot\partial)^{s_2-\ell-m} {1\over |x|}\right]
\end{aligned}\end{equation}
\ie
&= 2N\sum_{k=0}^{s_1}\sum_{\ell=0}^{s_2}{4^{k+\ell}\over (2k)!(2\ell)!}
\sum_{n=0}^{s_1-2k}\sum_{m=0}^{s_2-2\ell} {(-)^{m+n+k+\ell+a+b}\over n!m!(s_1-2k-n)!(s_2-2\ell-m)!} \sum_{a=0}^{k+n}\sum_{b=0}^{s_1-k-n}   \\
&~~~~\times {k+n\choose a}{s_1-k-n\choose b}{2^{s_1+s_2-a-b}\over \pi}\Gamma(n+m+k+\ell-a+{1\over 2}) \\
&~~~~\times \Gamma(s_1+s_2-n-m-k-\ell-b+{1\over 2}){(\ell+m)!\over (\ell+m-a)!}{(s_2-\ell-m)!\over (s_2-\ell-m-b)!}
\\
&~~~~\times
{(\varepsilon_1\cdot\varepsilon_2)^{a+b}(\varepsilon_1\cdot x)^{s_1-a-b}(\varepsilon_2\cdot x)^{s_2-a-b}\over (x^2)^{s_1+s_2-a-b+1}},
\fe
The summation can be performed, giving the result
\ie
&\left\langle J_{s}(\vec x,\vec\varepsilon_1) J_{\tilde s}(0, \vec\varepsilon_2) \right\rangle = N\delta_{s\tilde s} c_s
{2^{3s}\pi^{-{1\over 2}}\Gamma(s+{1\over 2})\over s!} {\left(\varepsilon_1\cdot\varepsilon_2\, x^2
-2 \varepsilon_1\cdot x \,\varepsilon_2\cdot x \right)^s\over (x^2)^{2s+1}}.
\label{JJe1e2}
\fe
Here $c_s=1$ for $s\ge 2$ and $c_0=2$.
It is often easier to work under the assumption $\varepsilon_1=\varepsilon_2=\varepsilon$, in which case we simply have
\ie\label{tpfree}
\left\langle J_{s}(\vec x,\vec\varepsilon) J_s(0, \vec\varepsilon) \right\rangle 
&=N c_s {2^{2s}\pi^{-{1\over 2}}\Gamma(s+{1\over 2})\over s!}{(x^-)^{2s}\over (x^2)^{2s+1}},
\fe
where we used the light cone variable $x^-=2\varepsilon\cdot x$.

Now we will calculate the three-point functions of these conserved currents. As a warm up, let us first consider
the correlation function of two scalar operators and one spin-$s$ current,
\ie
&\left\langle J_0(\vec x_1) J_0(\vec x_2) J_s(0;\vec \varepsilon) \right\rangle
= 8N\left.{1\over |x_{12}| }{1\over |x_2|}e^{\lambda\overleftarrow\partial_+}\cos\left(2\lambda\sqrt{\overleftarrow\partial_+ \overrightarrow\partial_+}\right)
e^{-\lambda\overrightarrow\partial_+}{1\over |x_1|}\right|_{\lambda^s} \\
&~~~= 8N{2^{2s}\pi^{-{1\over 2}}\Gamma(s+{1\over 2})\over |x_{12}| } \sum_{n=0}^s {(-)^{n}\over (2n)!(2s-2n)!} \left(
\partial_+^{s-n}{1\over |x_2|} \right) \left(\partial_+^n
{1\over |x_1|}\right) \\
&~~~= 8N{\pi^{-{1\over 2}}\Gamma(s+{1\over 2})\over |x_{12}| } \sum_{n=0}^s {(-)^{s-n}\over n!(s-n)!} {(x_1^-)^n(x_2^-)^{s-n}\over |x_1|^{2n+1} |x_2|^{2s-2n+1}} \\
&~~~= 8N{\pi^{-{1\over 2}}\Gamma(s+{1\over 2})\over s!|x_{12}| |x_1||x_2|}\left({x_1^-\over x_1^2}-{x_2^-\over x_2^2}\right)^s \\
&~~~= 8N{2^s\pi^{-{1\over 2}}\Gamma(s+{1\over 2})\over s!|x_{12}| |x_1||x_2|}(\vec\varepsilon\cdot\vec\Delta)^s,
\fe
where in the first line, $\partial_+$ is defined to act on both $\vec x_1$ and $\vec x_2$. In the last line, $\vec\Delta = {\vec x_1\over x_1^2}-{\vec x_2\over x_2^2}$. In the limit where the two scalar operators collide, $\vec x_{12}=\vec\delta\to 0$, we have
\ie\label{czzsstr}
\langle J_0(\vec x) J_0(\vec x-\vec \delta) J_s(0;\vec\varepsilon) \rangle \to 8N{2^s \pi^{-{1\over 2}}\Gamma(s+{1\over 2})\over s!|\delta| (x^2)^{s+1}}\left[\vec\varepsilon\cdot(\vec\delta - {2\delta\cdot x\over x^2}\vec x)\right]^s
\fe
This will be compared to $C(0,0;s)$ in Vasiliev theory.
On the other hand, in the limit where one scalar collide with the spin-$s$ current, say $\vec x_2=\vec\delta\to 0$, we have
\ie\label{czszstr}
\langle J_0(\vec x) J_0(\vec \delta) J_s(0;\vec\varepsilon) \rangle \to 8N{2^s\pi^{-{1\over 2}}\Gamma(s+{1\over 2})\over s!|\delta| (x^2)^{s+1}}(\vec\varepsilon\cdot\vec\delta)^s
\fe
This coefficient should be compared to $C(0,s;0)$. Note the different polarization dependence in the two limits (\ref{czzsstr}) and
(\ref{czszstr}). These indeed agree with the structure of the propagator ${\cal K}(\vec x,z|y,\partial_{y'})$ we used to
compute the boundary expectation value of the $B$ master field. We can normalize the two point function
(\ref{tpfree}) to $c_s(\vec\varepsilon\cdot \vec x)^{2s} (x^2)^{-2s-1}$, by defining a normalized current
\ie
J^{norm}_s(\vec x;\vec\varepsilon) = N^{-{1\over 2}}2^{-2s}\sqrt{\pi^{1\over 2} s!\over \Gamma(s+{1\over 2})} J_s(\vec x;\vec\varepsilon).
\fe
The normalized three point function coefficients $C_{00s}$ for the free $O(N)$ theory are given by
\ie
C_{00s}^{free} = N^{-{1\over 2}}2^{3-s}\pi^{-{1\over 4}} \sqrt{\Gamma(s+{1\over 2})\over s!}.
\fe

Next, let us examine the three-point function of one scalar operator and two higher spin currents.
\ie
&\langle J_{s_1}(\vec x_1,\vec \varepsilon_1) J_{s_2}(\vec x_2,\vec \varepsilon_2) J_0(0) \rangle \\
&= 8N \left.{1\over |x_1|}e^{-\varepsilon_1\cdot\overleftarrow{\partial_1}} \cos(2\sqrt{\varepsilon_1\cdot \overleftarrow{\partial_1}\varepsilon_1\cdot\overrightarrow{\partial_1}}) e^{\varepsilon_1\cdot\overrightarrow{\partial_1}}{1\over |x_{12}|}
e^{-\varepsilon_2\cdot\overleftarrow{\partial_2}}
\cos(2\sqrt{\varepsilon_2\cdot \overleftarrow{\partial_2}\varepsilon_2\cdot\overrightarrow{\partial_2}})
e^{\varepsilon_2\cdot\overrightarrow{\partial_2}} {1\over |x_2|}\right|_{\varepsilon_1^{s_1}\varepsilon_2^{s_2}}
 \\
&=8N 2^{2s_1+2s_2}\pi^{-1}\Gamma(s_1+{1\over 2})\Gamma(s_2+{1\over 2})\sum_{n=0}^{s_1}\sum_{m=0}^{s_2} {(-)^{n+m}\over (2n)!(2m)!(2s_1-2n)!(2s_2-2m)!}\\
&~~~\times\left[(\varepsilon_1\cdot\partial_1)^{s_1-n}{1\over |x_1|} \right] \left[(\varepsilon_2\cdot\partial_2)^{s_2-m}{1\over |x_2|} \right] \left[(\varepsilon_1\cdot\partial_1)^{n}(\varepsilon_2\cdot\partial_2)^{m}{1\over |x_{12}|} \right]  \\
&=8N\pi^{-1}\Gamma(s_1+{1\over 2})\Gamma(s_2+{1\over 2})\sum_{n=0}^{s_1}\sum_{m=0}^{s_2} { 2^{s_1+s_2+n}\over (2n)!m!(s_1-n)!(s_2-m)!}\\
&~~~\times{(\varepsilon_1\cdot x_1)^{s_1-n}(\varepsilon_2\cdot x_2)^{s_2-m}\over |x_1|^{2s_1-2n+1}|x_2|^{2s_2-2m+1}} \left[(\varepsilon_1\cdot\partial_1)^{n}{(\varepsilon_2\cdot x_{12})^m\over |x_{12}|^{2m+1}} \right] \\
&=8N\pi^{-1}\Gamma(s_1+{1\over 2})\Gamma(s_2+{1\over 2})\sum_{n=0}^{s_1}\sum_{m=0}^{s_2} { 2^{s_1+s_2+n}\over (2n)!m!(s_1-n)!(s_2-m)!} {(\varepsilon_1\cdot x_1)^{s_1-n}(\varepsilon_2\cdot x_2)^{s_2-m}\over |x_1|^{2s_1-2n+1}|x_2|^{2s_2-2m+1}} \\
&~~~\times\sum_{k=0}^n{(-)^k\over 2^{k}} {n\choose k} {(m!)^2(2m+2k)!\over (2m)!(m+k)!(m-n+k)!}
{(\varepsilon_1\cdot\varepsilon_2)^{n-k}(\varepsilon_1\cdot x_{12})^k(\varepsilon_2\cdot x_{12})^{m-n+k}\over |x_{12}|^{2m+2k+1}}
\fe
Let us focus on the special case $\varepsilon_1=\varepsilon_2=\varepsilon$,
\ie\label{sdd}
&\langle J_{s_1}(\vec x_1,\vec \varepsilon) J_{s_2}(\vec x_2,\vec \varepsilon) J_0(0) \rangle
\\
&=8N 2^{s_1+s_2}\pi^{-1}\Gamma(s_1+{1\over 2})\Gamma(s_2+{1\over 2})\sum_{n=0}^{s_1}\sum_{m=0}^{s_2} { (-)^n (2m+2n)! \over (2n)!(2m)!(s_1-n)!(s_2-m)!(n+m)!} \\
&~~~\times
{(\varepsilon\cdot x_1)^{s_1-n}(\varepsilon\cdot x_2)^{s_2-m}(\varepsilon\cdot x_{12})^{n+m} \over |x_1|^{2s_1-2n+1}|x_2|^{2s_2-2m+1}|x_{12}|^{2m+2n+1}}
\fe
In the limit where the two higher spin currents collide, $\vec x_{12}=\vec\delta\to 0$, corresponding to
the coefficient $C(s_1,s_2;0)$, we have
\ie\label{csszfree}
& \langle J_{s_1}(\vec x,\vec\varepsilon) J_{s_2}(\vec x-\vec\delta,\vec\varepsilon) J_0(0) \rangle \to
8N 2^{s_1+s_2}\pi^{-{1\over 2}}{\Gamma(s_1+s_2+{1\over 2})\over s_1!s_2!}
{(\vec\varepsilon\cdot \vec\delta)^{s_1+s_2} \over x^2|\delta|^{2s_1+2s_2+1}}
\fe
Observe that, modulo the factorized normalization factor associated with each current, this three-point function has precisely the same
dependence on $s_1$ and $s_2$ (namely, $\Gamma(s_1+s_2+{1\over 2})$) as the tree-level three-point function coefficient $C(s_1,s_2;0)$ of Vasiliev theory! We would like to emphasize that the computation of $C(s_1,s_2;0)$ in the previous section was highly nontrivial: a priori, it wasn't even obvious that $C(s_1,s_2;0)$ would be an analytic function in $s_1$ and $s_2$. Also, recall $C(s,s;0)$ is naively zero in Vasiliev theory, and we argued that
its appropriately regularized answer should be given by the analytic continuation from $C(s_1,s_2;0)$ for $s_1\not=s_2$.
As expected, the coupling constant $g$ of Vasiliev theory scales like $N^{-{1\over 2}}$ of the free $O(N)$ vector theory.
Let us emphasize that the relative normalization on the spin-$s$ currents in Vasiliev theory can be fixed
independently, through the computation of $C(0,s_1;s_2)$, as we will perform in section 6.1. We will then see a complete agreement with
(\ref{csszfree}), and will determine the precise relation between $g$ and $N$.

On the other hand, in the limit $\vec x_2=\vec\delta\to 0$, corresponding to $C(0,s_1;s_2)$,
we expect $\langle J_{s_1}(\vec x,\varepsilon) J_{s_2}(\vec \delta,\varepsilon) J_0(0) \rangle$ to scale like $|x|^{-2s_1-2}|\delta|^{s_1-s_2-1}$.
To simplify the expression let us further restrict to the case $\varepsilon\cdot x=0$, so that (\ref{sdd}) becomes
\ie
& \langle J_{s_1}(\vec x,\vec\varepsilon) J_{s_2}(\vec \delta,\vec\varepsilon) J_0(0) \rangle_{\vec\varepsilon\cdot \vec x=0}
=
8N2^{s_1+s_2}\pi^{-1}\Gamma(s_1+{1\over 2})\Gamma(s_2+{1\over 2}) \\
&~~~~\times\sum_{m=0}^{s_2} {(-)^m(2m+2s_1)! \over (2s_1)!(2m)!(s_2-m)!(s_1+m)!}
{(\varepsilon\cdot \delta)^{s_1+s_2} \over |x||\delta|^{2s_2-2m+1}|x-\delta|^{2s_1+2m+1}}
\\
&\to 8N2^{s_1+s_2}\pi^{-1}{\Gamma(s_1+{1\over 2})\Gamma(s_2+{1\over 2}) \over s_1!s_2!}
{(\varepsilon\cdot \delta)^{s_1+s_2} \over |x|^{2s_1+2}|\delta|^{2s_2+1}}.
\fe
We will consider the corresponding computation in Vasiliev theory in later sections.

Finally, let us consider the case of three general spins $(s_1,s_2,s_3)$, but with the simplification that all the polarization
vectors are the same.
\ie
&\langle J_{s_1}(\vec x_1,\vec \varepsilon_1) J_{s_2}(\vec x_2,\vec \varepsilon) J_{s_3}(\vec x_3,\vec \varepsilon) \rangle \\
&= 8N \left.{1\over |x_1|}e^{-\varepsilon_1\cdot\overleftarrow{\partial_1}} \cos(2\sqrt{\varepsilon_1\cdot \overleftarrow{\partial_1}\varepsilon_1\cdot\overrightarrow{\partial_1}}) e^{\varepsilon_1\cdot\overrightarrow{\partial_1}}{1\over |x_{12}|}
e^{-\varepsilon_2\cdot\overleftarrow{\partial_2}}
\cos(2\sqrt{\varepsilon_2\cdot \overleftarrow{\partial_2}\varepsilon_2\cdot\overrightarrow{\partial_2}})
e^{\varepsilon_2\cdot\overrightarrow{\partial_2}} {1\over |x_2|}\right|_{\varepsilon_1^{s_1}\varepsilon_2^{s_2}}
\\
&=8N\pi^{-{3\over 2}}\sum_{n_i=0}^{s_i} \prod_{i=1}^3 2^{2s_i}\Gamma(s_i+{1\over 2})
{(-)^{n_i}\over (2n_i)!(2s_i-2n_i)!}\\
&~~~\times \left[(\varepsilon\cdot\partial_1)^{n_1}(\varepsilon\cdot\partial_2)^{s_2-n_2}{1\over |x_{12}|} \right] \
\left[(\varepsilon\cdot\partial_2)^{n_2}(\varepsilon\cdot\partial_3)^{s_3-n_3}{1\over |x_{23}|} \right]
\left[(\varepsilon\cdot\partial_3)^{n_3}(\varepsilon\cdot\partial_1)^{s_1-n_1}{1\over |x_{31}|} \right]
\\
&=8N\pi^{-{3\over 2}}\sum_{n_i=0}^{s_i} \prod_{i=1}^3 2^{2s_i}\Gamma(s_i+{1\over 2})
{1\over (2n_i)!(2s_i-2n_i)!}\\
&~~~\times \left[(\varepsilon\cdot\partial_1)^{n_1-n_2+s_2}{1\over |x_{12}|} \right] \
\left[(\varepsilon\cdot\partial_2)^{n_2-n_3+s_3}{1\over |x_{23}|} \right]
\left[(\varepsilon\cdot\partial_3)^{n_3-n_1+s_1}{1\over |x_{31}|} \right]. \\
\fe
Further, writing $\vec x_1=\vec x$, $\vec x_2=\vec x-\vec \delta$, $\vec x_3=0$, and assuming $\vec \varepsilon\cdot \vec x=0$, we
find the answer
\ie\label{freegeneral}
&\langle J_{s_1}(\vec x,\vec \varepsilon_1) J_{s_2}(\vec x-\vec \delta,\vec \varepsilon) J_{s_3}(0,\vec \varepsilon) \rangle_{\vec \varepsilon\cdot \vec x=0} \\
&=8N\pi^{-{3\over 2}}\sum_{n_i=0}^{s_i} \prod_{i=1}^3 2^{2s_i}\Gamma(s_i+{1\over 2})
{1\over (2n_i)!(2s_i-2n_i)!}\\
&~~~\times \left[(\varepsilon\cdot\partial_\delta)^{n_1-n_2+s_2}{1\over |\delta|} \right] \
\left[(\varepsilon\cdot\partial_x)^{n_2-n_3+s_3}{1\over |x-\delta|} \right]
\left[(-\varepsilon\cdot\partial_x)^{n_3-n_1+s_1}{1\over |x|} \right] \\
&=8N \pi^{-{3\over 2}} \prod_{i=1}^3 2^{2s_i}\Gamma(s_i+{1\over 2})
\sum_{n_2=0}^{s_2}{1\over (2n_2)!(2s_2-2n_2)!(2s_1)!(2s_3)!}{1\over |x|} \\
&~~~\times \left[(\varepsilon\cdot\partial_\delta)^{s_1-n_2+s_2}{1\over |\delta|} \right] \
\left[(\varepsilon\cdot\partial_x)^{n_2+s_3}{1\over |x-\delta|} \right]
\\
&\to 8N\pi^{-{3\over 2}} \prod_{i=1}^3 {2^{2s_i}\Gamma(s_i+{1\over 2})\over (2s_i)!} {1\over |x|}
\left[(\varepsilon\cdot\partial_\delta)^{s_1+s_2}{1\over |\delta|} \right] \
\left[(\varepsilon\cdot\partial_x)^{s_3}{1\over |x-\delta|} \right] \\
&\to 8N\pi^{-{3\over 2}} \prod_{i=1}^3 {2^{3s_i}\Gamma(s_i+{1\over 2})\over (2s_i)!} {1\over |x|}
{\Gamma(s_1+s_2+{1\over 2})\over \Gamma({1\over 2})}{(\varepsilon\cdot\delta)^{s_1+s_2}\over |\delta|^{2s_1+2s_2+1}}
{\Gamma(s_3+{1\over 2})\over \Gamma({1\over 2})}{(\varepsilon\cdot\delta)^{s_3}\over |x|^{2s_3+1}}
\\
&= 8N2^{s_1+s_2+s_3}{\Gamma(s_1+s_2+{1\over 2})\Gamma(s_3+{1\over 2})\over \pi (s_1)!(s_2)!(s_3)!} {(\varepsilon\cdot\delta)^{s_1+s_2+s_3}\over |x|^{2s_3+2}|\delta|^{2s_1+2s_2+1}}.
\fe

\section{The $\Delta=2$ Scalar and the Critical $O(N)$ Model}

In this section we consider the alternative boundary condition for the bulk scalar field, such that its dual operator has
dimension $\Delta=2$. The boundary-to-bulk propagator for the scalar field has the form $C(\vec x,z)=K^2= z^2/(\vec x^2+z^2)^2$.
Let us now solve for the boundary-to-bulk propagator for the scalar component of the master field $B$, analogously to section 3.
\ie
B^{\Delta=2}(x|y,\bar y) &= \sum_{n=0}^\infty {1\over (n!)^2}{1\over z^n} (-z^2 y{\slash\!\!\!\partial}\bar y)^n
C(x)
\\
&= \sum_{n=0}^\infty {n+1\over n!}{1\over z^n} (-z y\Sigma\bar y)^n
K^2
\\
&= K^2 (1-y\Sigma \bar y) e^{-y\Sigma\bar y}.
\fe
When the outcoming field is a $\Delta=2$ scalar, we must also use the propagator ${\cal K}_{(0)}^{\Delta=2}$
as in (\ref{ksimple}) and (\ref{scalarccc}). In the next two subsections, we will repeat our previous computation of
$C(0,s;0)$ and $C(s,\tilde s;0)$  with the $\Delta=2$ scalars (note that for the $\Delta=2$ scalar, $C(0, s;0)$ is not a special case of $C(s,\tilde s;0)$), and then compare with the leading $1/\sqrt{N}$
three-point functions of the critical $O(N)$ model.

\subsection{$C(0,s;0)$}

Now let us compute the three-point function coefficient $C(0,s;0)$, where spin-$0$ refers to the $\Delta=2$ scalar, and
$s>0$. Note that unlike the $\Delta=1$ case, where the scalar is treated on equal footing as the higher spin fields, the $\Delta=2$ scalar is distinguished from the higher spin fields.

Analogously to the $\Delta=1$ case, we need to compute the integral (\ref{scalarccc}), or (\ref{scint}).
$I^{\Delta=2}(\hat\delta)$ is given by
\ie
I^{\Delta=2}(\hat \delta) & = \int d^3\vec x' dz'\,(z')^{-1} J(\vec x',z';0,\hat\delta)
\\
&=  {1\over 2s(2s)!}\int {d^3\vec x dz\over 2z^2} \int d^4u d^4v \left(e^{uv+\bar u\bar v}+e^{-uv-\bar u\bar v}\right)
(2v\sigma^z\bar v+s^2)\\
&~~~\times {z^{s+3}\over (x^-)^s (\tilde x^2)^2} (1-v\tilde\Sigma\bar v)e^{-v\tilde\Sigma \bar v} \partial_+^{2s}
{(u{\bf x}\bar u)^s\over x^2}
\\
&=  {s!\over 2s(2s)!}\int {d^3\vec x dz\over 2z^2}
{z^{s+3}\over (x^-)^s (\tilde x^2)^2} \partial_{x^+}^{2s}
\left[{1\over x^2}I_\lambda^{\Delta=2}|_{\lambda^s}\right].
\fe
Note the additional factor of 2
in front of $v\sigma^z\bar v$ in the second line, coming from integration by part in $z$.
Be aware that in the last line, $\partial_{x^+}$ acts only on $x$ and not $\tilde x$.
The generating function $I^{\Delta=2}_\lambda$ is given by
\ie
I_\lambda^{\Delta=2}  &=
\int d^4u d^4v \left(e^{uv+\bar u\bar v}+e^{-uv-\bar u\bar v}\right)
(2v\sigma^z\bar v+s^2)(1-v\tilde\Sigma\bar v)e^{-v\tilde\Sigma \bar v+\lambda (u{\bf x}\bar u)}
\\
&=\left. {2 \over \det(1-\lambda\tilde\Sigma{\bf x})}(1-\partial_j \tilde\Sigma \partial_{\bar j})
(2\partial_j \sigma^z \partial_{\bar j}+s^2) e^{\lambda j {\bf x}(1-\lambda\tilde\Sigma{\bf x})^{-1}\bar j}\right|_{j=\bar j=0} \\
&= {2 \over \det(1-\lambda\tilde\Sigma{\bf x})}\left\{ \left[1+\lambda {\rm Tr}(\tilde \Sigma {\bf x}(1-\lambda\tilde\Sigma{\bf x})^{-1})\right]
\left[s^2-2\lambda {\rm Tr}(\sigma^z{\bf x}(1-\lambda\tilde\Sigma{\bf x})^{-1}) \right] \right. \\
&~~~\left. -2\lambda^2 {\rm Tr}\left[\tilde\Sigma {\bf x}(1-\lambda\tilde\Sigma{\bf x})^{-1}
\sigma^z {\bf x} (1-\lambda\tilde\Sigma{\bf x})^{-1} \right] \right\} \\
&= {2 \over \Xi}\left\{ \left[1+ 2\lambda{z(1-{2x\cdot\tilde x\over \tilde x^2})-\lambda x^2\over \Xi} \right]
\left[s^2-4\lambda {z-\lambda x^2 (1-{2z^2\over\tilde x^2})\over\Xi} \right] \right. \\
&~~~\left. -{2\lambda^2\over \Xi^2} \left[
4z^2(1-{2x\cdot\tilde x\over\tilde x^2})-4\lambda z x^2+2(\lambda^2 x^2-1)x^2(1-{2z^2\over\tilde x^2})\right] \right\}
\fe
where we define, as before,
\ie
&\Xi = 1-2\lambda z+4\lambda z {x\cdot\tilde x\over \tilde x^2} + \lambda^2 x^2,\\
&\xi = 1-2\lambda z+4\lambda z {x\cdot\tilde x\over \tilde x^2}.
\fe
When acting on with $\partial_{x^+}^{2s}$, we can equivalently replace $\Xi$ by $\xi$ in $I_\lambda^{\Delta=2}$,
and write
\ie
\left.\partial_{x^+}^{2s}\left({1\over x^2} I_\lambda^{\Delta=2}\right)\right|_{\lambda^s} &=
2\left.\partial_{x^+}^{2s} {1\over x^2 }\left( {s^2+4\lambda z\over \xi^2}
-{8\lambda z\over\xi^3} \right)\right|_{\lambda^s} \\
&=2\,\partial_{x^+}^{2s} {2^sz^{s}\over x^2}\left[ {s^2}\cdot (s+1)(1-{2x\cdot\tilde x\over\tilde x^2})^s -
2s^2(1-{2x\cdot\tilde x\over\tilde x^2})^{s-1} \right].
\fe
Now continuing on the integral $I^{\Delta=2}(\hat\delta)$,
\begin{equation}\nonumber\begin{aligned}
&I^{\Delta=2}(\hat\delta) = {s!\over 2s (2s)!}\int d^3\vec x dz {2^s z^{2s+1}\over (x^-)^s (\tilde x^2)^2}
\partial_{x^+}^{2s} {1\over x^2}\left[ {s^2}\cdot (s+1)(1-{2x\cdot\tilde x\over\tilde x^2})^s -2s^2(1-{2x\cdot\tilde x\over\tilde x^2})^{s-1} \right] \\
&= {s!\over 2s (2s)!}\int d^3\vec x dz {2^s z^{2s+1}\over (x^-)^s (\tilde x^2)^2}\sum_{n=0}^{2s}{2s\choose n}
\left(\partial_{x^+}^{2s-n} {1\over x^2}\right)\\
&~~~~\times\partial_{x^+}^n\left[ {s^2}\cdot (s+1)(1-{2x\cdot\tilde x\over\tilde x^2})^s -2s^2(1-{2x\cdot\tilde x\over\tilde x^2})^{s-1} \right]\\
&= 2^{s-1}{s!\over s}\int d^3\vec x dz\, z^{2s+1}\sum_{n=0}^{s}{s\choose n}
{(x^-)^{s-n}(\tilde x^-)^n\over (x^2)^{2s-n+1}(\tilde x^2)^{n+2}}\\
&~~~~\times\left[ {s^2}\cdot (s+1)(1-{2x\cdot\tilde x\over\tilde x^2})^{s-n} -2{s-n\over s}s^2(1-{2x\cdot\tilde x\over\tilde x^2})^{s-n-1} \right]
\end{aligned}\end{equation}
\ie
&= 2^{s-1}{s!\over s}\int d^3\vec x dz {z^{2s+1}\over (x^2)^{s+1}(\tilde x^2)^{s+2}}\left[ {s^2}\cdot (s+1)\left({x^-\over x^2}-{\delta^-}\right)^{s} - 2s^2 {x^-\tilde x^2\over x^2}\left(
{x^-\over x^2}-{\delta^-} \right)^{s-1} \right] \\
&= 2^{-s-2}\pi^{5\over 2} s\Gamma(s+{1\over 2}) (\delta^-)^s \\
&= {\pi^{5\over 2}\over 4} s\Gamma(s+{1\over 2}) (\hat\delta\cdot\vec\varepsilon)^s.
\fe
Using (\ref{scalarccc}), we find the boundary expectation value of the outgoing $\Delta=2$ scalar,
\ie
\lim_{z\to 0} z^{-2}B_{\Delta=2}^{(0,0)}(\vec x,z)
\to -{2\delta^{-s-1}\over \pi^2|\vec x|^4} I^{\Delta=2}(\hat\delta)
= C^{\Delta=2}(0,s;0) { (\vec\delta\cdot\vec\varepsilon)^s\over |\vec x|^4 \delta^{2s+1}}
\fe
in the $\delta\to 0$ limit. The coefficient $C^{\Delta=2}(0,s;0)$ is given by
\ie
C^{\Delta=2}(0,s;0) &= -{\pi^{1\over 2}\over 2} s\Gamma(s+{1\over 2}).
\fe
Taking into account the still undetermined normalization factors on the boundary-to-bulk propagators,
the corresponding normalized three-point function in Vasiliev theory is related by
\ie
C^{\Delta=2}_{00s} = g{a_0' a_s\over a_0'} C^{\Delta=2}(0,s;0) =g a_s C^{\Delta=2}(0,s;0)
\fe
where $a_0'$ is the normalization factor associated with the $\Delta=2$ scalar operator.
Here $g$ and $a_s$ are the same coupling constant and normalization factors as in the $\Delta=1$ case.

\subsection{$C(s,\tilde s;0)$}

Next, let us turn to the computation of $C(s,\tilde s;0)$, where $s$ and $\tilde s$ are nonzero spins,
and the outgoing spin-0 field is subject to $\Delta=2$ boundary condition.
Without loss of generality, we will assume $s>\tilde s>0$. The expression of the source $J_{(0)}$ of
for the spin-0 component of $B(\vec x, z|y,\bar y)$ at second order is identical to the $\Delta=1$ case.
The only difference occurs when we integrate $J_{(0)}$ with the propagator ${\cal K}_{\Delta=2}$ to obtain the boundary
expectation value. We perform the computation in the case where the polarization vectors $\vec\varepsilon$ of the the spin $s$
and spin $\tilde s$ currents are identical, with ${\slash\!\!\!\varepsilon}=\vec\varepsilon\cdot\vec\sigma = {1\over 2}\sigma^-$,
$\sigma^{-z}_{\da\db}=\bar\lambda_\da\bar\lambda_\db$.
Now we simply need to replace the integral $I_+^{\Delta=1}(\hat\delta)$ in
section 4.4 by
\begin{equation}\nonumber\begin{aligned}
&I_+^{\Delta=2}(\hat\delta) = \int d^3\vec x dz\, z^{-1}J^{+}_{(0)}(\vec x,z;0,\hat\delta)\\
&= s\cdot{2^{-s-\tilde s-2}} \int d^3 x dz {z^{s+\tilde s}\over (x^2)^{2s+1}(\tilde x^2)^{2\tilde s+1}}
\int d^4 u d^4 v \cosh(uv+\bar u\bar v)
\left( {2v\sigma^z\bar v \over s^2-\tilde s^2}+1\right) \\
&~~~\times
e^{-v\tilde\Sigma \bar v} (v{\bf\tilde x}\sigma^{-z}{\bf\tilde x} v)^{\tilde s} (u{\bf x}\sigma^{-z}{\bf x}u)^{\tilde s} (u{\bf x}\sigma^-{\bf x}\bar u)^{s-\tilde s}
\end{aligned}\end{equation}
\ie
&= s{2^{-s-\tilde s-2}} (s-\tilde s)!(2\tilde s)! \int d^3 x dz {z^{s+\tilde s}\over (x^2)^{2s+1}(\tilde x^2)^{2\tilde s+1}}
\left.  {(\bar\lambda{\bf x}{\slash\!\!\!\delta}\bar\lambda)^{2\tilde s}\over
\Xi^{2\tilde s+1}}  \left[ 1 + {2\eta \over s^2-\tilde s^2}{2z(\bar \lambda{\bf x}\sigma^z\bar \lambda)\over \Xi}\left(1+2\tilde s \right)
\right]\right|_{\eta^{s-\tilde s}}
\\
&= s{2^{-2\tilde s-2}} (s+\tilde s)! \int d^3 x dz {z^{2s}(\bar\lambda{\bf x}{\slash\!\!\!\delta}\bar\lambda)^{2\tilde s}
\over (x^2)^{2s+1}(\tilde x^2)^{s+\tilde s+1}}  \left[
\left( \bar\lambda{\bf x}\sigma^z\bar\lambda - x^2 \bar\lambda{\slash\!\!\!\delta}\sigma^z\bar\lambda \right)^{s-\tilde s} \right.
\\
&~~~~~\left. - {2\tilde x^2(\bar \lambda{\bf x}\sigma^z\bar \lambda)\over s+\tilde s}
\left( \bar\lambda{\bf x}\sigma^z\bar\lambda - x^2 \bar\lambda{\slash\!\!\!\delta}\sigma^z\bar\lambda \right)^{s-\tilde s-1}
\right]
\\
&=2^{-2s-2\tilde s-2}\pi^2 s\Gamma(s+\tilde s) (\bar\lambda{\slash\!\!\!\delta}\sigma^z\bar\lambda)^{s+\tilde s}
\\
&={\pi^2\over 4} s\Gamma(s+\tilde s) (\hat\delta\cdot\vec\varepsilon)^{s+\tilde s}.
\fe
In the second line, we again note the factor of 2 in front of $v\sigma^z\bar v$, from integration by
part in $z$, which is different from the $\Delta=1$ case. Similarly, there is an identical contribution
from $I_-^{\Delta=2}(\hat\delta)$.
It then follows from (\ref{scalarccc}) that
\ie
\lim_{z\to 0} z^{-2} B_{\Delta=2}^{(0,0)}(\vec x,z) \to -{2\delta^{-s-\tilde s}\over \pi^2|\vec x|^4} I^{\Delta=2}(\hat\delta)
= C^{\Delta=2}(s,\tilde s;0) { (\vec\delta\cdot\vec\varepsilon)^{s+\tilde s}\over |\vec x|^4 \delta^{2s+2\tilde s}}
\fe
where the coefficient $C^{\Delta=2}(s,\tilde s;0)$ is given by
\ie
C^{\Delta=2}(s,\tilde s;0) = -s\Gamma(s+\tilde s),~~~~~s>\tilde s.
\fe
$C^{\Delta=2}(s,\tilde s;0)$ is by definition symmetric in $s$ and $\tilde s$, and so is given by
${1\over 2}\tilde s\Gamma(s+\tilde s)$ for $s<\tilde s$. Note that unlike the $\Delta=1$ case,
$C^{\Delta=2}(s,\tilde s;0)$ is not formally an analytic function in $s$ and $\tilde s$,
as one takes $s$ across $\tilde s$. However, our result does suggest a ``regularized answer"
for the naively singular three-point function coefficient $C^{\Delta=2}(s,s;0)$, as the two spins coincide.

As we will see in the next subsection, in the critical $O(N)$ vector model, it is more convenient to compute the three point function $\langle J_s (\vec x_1) J_{\tilde s}(\vec x_2) \alpha(\vec x_3)\rangle$ with $\vec x_3$ integrated over
the three-dimensional spacetime; here $\alpha$ is the scalar operator of classical scaling dimension $\Delta=2$.
To make the comparison, we would like to consider the same computation in Vasiliev theory, namely
\ie
\int d^3\vec x \left\langle J_s(\vec \delta,\vec\varepsilon) J_{\tilde s}(0,\vec\varepsilon)
{\cal O}_{\Delta=2}(\vec x)\right\rangle.
\fe
For this purpose we can no longer take the $\delta/|\vec x|\to 0$ limit, but instead must use the full expression
of ${\cal K}_{\Delta=2}(\vec x,z)$. We have
\ie
\int d^3\vec x\lim_{z\to 0} z^{-2}B^{(0,0)}_{\Delta=2}(\vec x,z)
&= -{2\delta^{-s-\tilde s-1}\over \pi^2}\int d^3\vec x \int {d^3\vec x' dz'\over (z')^3}\left[ {z'\over (\vec x-\vec x')^2+(z')^2} \right]^2
J_{(0)}(\vec x',z';0,\hat\delta)
\\
&= -{2\delta^{-s-\tilde s-1}}\int {d^3\vec x' dz'\over (z')^2}
J_{(0)}(\vec x',z';0,\hat\delta)
\fe
The integral in the last line is in fact identical to that in the computation of $C(s,\tilde s;0)$ in the
$\Delta=1$ theory. We then obtain the result
\ie\label{ftres}
\int d^3\vec x\lim_{z\to 0} z^{-2}B^{(0,0)}_{\Delta=2}(\vec x,z)
&= -\pi^{5\over 2}\Gamma(s+\tilde s+{1\over 2}) {(\vec\delta\cdot\vec\varepsilon)^{s+\tilde s}\over
\delta^{2s+2\tilde s+1}}.
\fe
Combined with the normalization factor $g {a_sa_{\tilde s}\over a_0'}$, this gives the integrated three-point
function
$\int d^3\vec x\langle J_s (\vec \delta,\vec\varepsilon) J_{\tilde s}(0,\vec\varepsilon) {\cal O}_{\Delta=2}(\vec x)\rangle$
in Vasiliev theory.

\subsection{Comparison with the critical $O(N)$ vector model}

In this subsection, we will consider the critical $O(N)$ vector model in dimension $2<d<4$,\footnote{We will only need the results in $d=3$ to compare with Vasiliev theory in this paper. It is nevertheless useful to have the formulae in general $d$.} and calculate the three-point function of the scalar operator and higher spin currents to leading order in the $1/N$ expansion, namely $N^{-{1\over 2}}$. While the computation described in this subsection
have already appeared in \cite{Lang:1990ni,Lang:1990re,Lang:1991jm,Lang:1991kp,Lang:1992zw}, the explicit results are not immediately available. The goal of this subsection
is to extract the three-point function coefficients and compare to the conjectured dual Vasiliev theory,
following the approach of Lang and R\"uhl \cite{Lang:1990ni}.

The $O(N)$ vector model in $d$-dimensions can be defined by the
path integral
\ie
&\int D\vec S D\alpha \exp\left\{ - {N } \int d^d x \left[ {1\over 2}(\partial_\mu \vec S)^2
+{1\over 2}\alpha\left(\vec S^2-{1\over g}\right) \right] \right\} \\
&= \int D\alpha \exp\left\{ -{N\over 2} \left[ {\rm Tr}\ln (-\Delta+\alpha) - {1\over g}\int d^dx \alpha \right] \right\},
\fe
where $\vec S = (S_1,\cdots, S_N)$ are $N$ scalar fields, and $\alpha(x)$ is a Lagrangian multiplier field.
$g$ is a coupling constant that will be taken to infinity at the critical point.
In the second step we integrated out $\vec S$ to obtain an effective action in $\alpha$.
At the leading order in $1/N$ expansion, the expectation value of $\alpha$, which plays the role of mass
square of the scalar fields $\vec S$, is given by the
critical point of the $\alpha$-effective action,
\be
\alpha = m^2,~~~~\int {d^dp\over (2\pi)^d}{1\over p^2+m^2} = {1\over g},
\ee
where $m^2$ is solved to be
\be
m^2 = \left[ (4\pi)^{-{d\over 2}}\Gamma(1-{d\over 2}) g \right]^{-{2\over d-2}}
\ee
by analytic continuation in dimension $d$. In $d>2$, the critical point $m=0$ is achieved by
sending $g\to\infty$. The authors of \cite{Lang:1990ni} considered a field $\tilde\alpha$ related to $\alpha$ by $\alpha=i\tilde \alpha$, which should be thought of as a real field in the path integral description.
From the CFT perspective, it will be more convenient to work with $\alpha$, which has positive
two-point function in position space.
The effective propagator for $\alpha$, after integrating out $\vec S$,
is\footnote{Note that our $\tilde\gamma$ differs from the notation $\gamma$ in \cite{Lang:1990ni} by a sign,
since we are working with $\alpha$ instead of $\tilde\alpha$.}
\ie
& \tilde G(p) = {\tilde\gamma\over N} \,(p^2)^{2-{d\over 2}},~~~~\tilde\gamma = 2(4\pi)^{d\over 2} {\sin {\pi d\over 2}\over \pi} {\Gamma(d-2)\over \Gamma({d\over 2}-1)},
\fe
or in position space,
\ie\label{gxa}
& G(x) = \int {d^dp\over (2\pi)^d}\tilde G(p)e^{ip\cdot x} = {1\over N}{2^{d+2}\sin{\pi d\over 2}\Gamma({d-1\over 2})\over \pi^{{3\over 2}}\Gamma({d\over 2}-2)} {1\over (x^2)^2}
\equiv {\gamma_\alpha\over N}{1\over (x^2)^2}.
\fe
Note that for $d=3$, $\tilde\gamma=-16$, $\gamma_\alpha={16\over\pi^2}>0$.
The propagator for $\vec S$ is the standard one,
\be
{\delta_{ab}\over N}{1\over p^2},~~~{\rm or}~~~{\delta_{ab}\over N} {\Gamma({d\over 2}-1)\over 4\pi^{d\over 2}}{1\over (x^2)^{{d\over2}-1}}\equiv {\gamma_S\over N}{\delta_{ab}\over (x^2)^{{d\over2}-1}},
\ee
and the $\alpha S_a S_b$ vertex comes with coefficient $-N\delta_{ab}$.

We want to compute the three-point functions involving $\alpha(x)$ and higher spin currents $J_s(x,\varepsilon)$,
to the first nontrivial in the $1/N$ expansion. While in general we don't know a priori the expression of $J_s$
in terms of the fundamental fields, we will extract them from the OPE of a pair of $S_a(x)$ fields.

\begin{figure}
\begin{center}
\includegraphics[width=50mm]{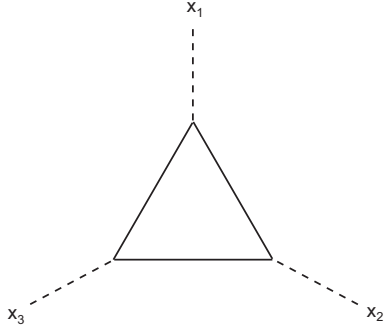}
\parbox{13cm}{
\caption{
The 1-loop contribution to $\langle\alpha\alpha\alpha\rangle$. The dotted lines represent the effective propagators
$G(x)$ of $\alpha$, while the solid lines are propagators of $\vec S$.
\label{triangle-aaa}}}
\end{center}
\end{figure}

Let us start with the 3-point function $\langle \alpha(x_1)\alpha(x_2)\alpha(x_3) \rangle$. Our convention is such that the two-point function of $\alpha(x)$ scales like $1/N$. The leading contribution to $\langle\alpha\alpha\alpha\rangle$ is of order $1/N^2$, and so if we normalize the two-point function of $\alpha$,
the corresponding 3-point function will scale like $1/\sqrt{N}$.
This comes from a 1-loop diagram as shown in figure 2.
It is evaluated as
\ie\label{aaafn}
&\left\langle \alpha(x_1)\alpha(x_2)\alpha(x_3) \right\rangle\\
&= -{1\over N^2} \int d^dy_1 d^dy_2 d^dy_3 {\gamma_\alpha^3\gamma_S^3\over (x_1-y_1)^4(x_2-y_2)^4(x_3-y_3)^4
(y_{12}^2)^{{d\over2}-1}(y_{23}^2)^{{d\over2}-1}(y_{31}^2)^{{d\over2}-1}} \\
&= -{\gamma_\alpha^3\gamma_S^3\over N^2} v(2,{d\over2}-1,{d\over2}-1)^2v(2,1,d-3){1\over x_{12}^2x_{23}^2x_{31}^2}\\
&= -{\gamma_\alpha^3\gamma_S^3\over N^2} \pi^{{3d\over2}} {\Gamma({d\over2}-2)^2\Gamma({d\over2}-1)\Gamma(3-{d\over2})\over \Gamma({d\over2}-1)^4\Gamma(d-3)} {1\over x_{12}^2x_{23}^2x_{31}^2}
\fe
where we have used the graphical rules of \cite{Kazakov:1983ns} in the second line. We have used the definitions
in \cite{Kazakov:1983ns} for the coefficients
\ie
& a(t) = {\Gamma({d\over 2}-t)\over \Gamma(t)},\\
& v(t_1,t_2,t_3) = \pi^{d\over 2} a(t_1) a(t_2) a(t_3).
\fe
Note that (\ref{aaafn}) vanishes at $d=3$. This has been observed in \cite{Sezgin:2003pt} to match with
the fact that there is no scalar cubic coupling in the bulk Vasiliev theory.

\begin{figure}\label{GRAPH1}
\begin{center}
\includegraphics[width=40mm]{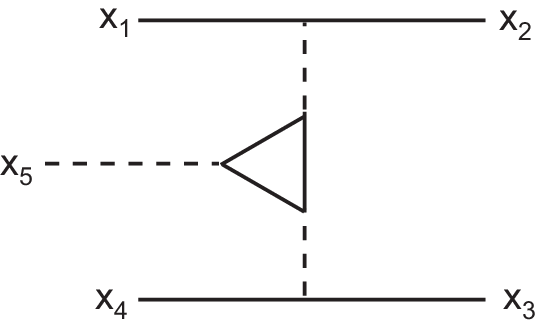}
~~~~~~\includegraphics[width=40mm]{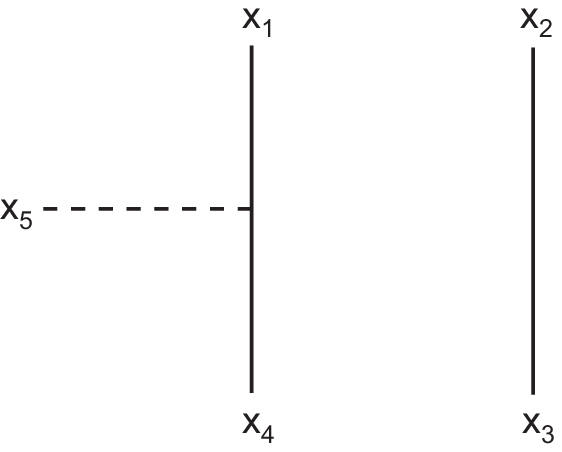}
~~~~~~\includegraphics[width=40mm]{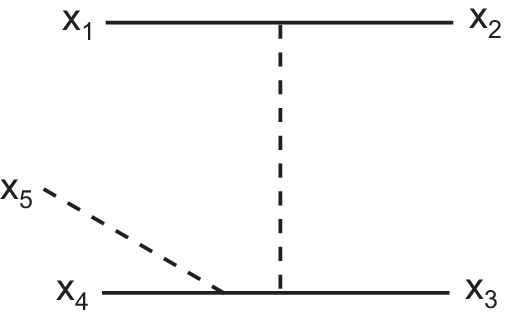}
\\
(a)~~~~~~~~~~~~~~~~~~~~~~~~~~~~~~~(b)~~~~~~~~~~~~~~~~~~~~~~~~~~~~~~(c)
\parbox{13cm}{
\caption{
The leading $1/N$ contributions to $\langle SSSS\alpha\rangle$, from which we will extract $\langle
J_s J_{s'} \alpha\rangle$.
\label{SSSSa}}}
\end{center}
\end{figure}

Next, we will investigate the three point function of $\alpha$ with two higher spin currents, $J_s$ and $J_{\tilde s}$.
The idea is to consider the five point function $\langle S_a(x_1)S_b(x_2)S_c(x_3)S_d(x_4)\alpha(x_5) \rangle$,
expanded around the limit $x_{12},x_{34}\to 0$. The three point function $\langle
J_s(x_1,\varepsilon_1) J_{\tilde s}(x_3,\tilde\varepsilon) \alpha(x_5)\rangle$ will be extracted
from the channel $\delta_{ab}\delta_{cd} (\varepsilon\cdot x_{12})^s (\tilde\varepsilon\cdot x_{34})^{\tilde s}$.\footnote{Note that the higher spin primary currents in the critical $O(N)$ model are not expressed
in terms of $\vec S$ bilinear in the same way as in the free $O(N)$ theory, even at leading order in $1/N$,
due to the exchange of $\alpha$. As pointed out in \cite{Lang:1990re}, even in the large $N$ limit, the currents of the critical $O(N)$ model cannot be embedded
in the Hilbert space of the free $O(N)$ CFT.}
The scaling in $N$ in the $SS$ OPE is of the form $SS\sim N^{-{3\over 2}}\sum J_s$. The leading nontrivial contribution in the $1/N$ expansion of $\langle SSSS\alpha\rangle$ is of order $N^{-4}$, corresponding to the normalized three-point function $\langle J J (\sqrt{N}\alpha)\rangle$ at order $N^{-{1\over 2}}$.
The relevant diagrams are a 1-loop triangle diagram (figure 3(a)), a disconnected tree diagram (figure 3(b)),
and a connected tree diagram (figure 3(c)).

The one-loop diagram in figure 3(a) is evaluated as
\ie\label{ssssaint}
& -{\delta_{ab}\delta_{cd} \over N^4} \gamma_\alpha^3 \gamma_S^7 \int d^dy_1 d^dy_2 d^dy_3d^dz_1 d^dz_2\\
&~\times{1\over (x_5-y_3)^4(y_1-z_1)^4(y_2-z_2)^4(x_1-z_1)^{d-2}(x_2-z_1)^{d-2}(x_3-z_2)^{d-2}(x_4-z_2)^{d-2} y_{12}^{d-2} y_{23}^{d-2} y_{31}^{d-2} }\\
&= -{\delta_{ab}\delta_{cd} \over N^4} \gamma_\alpha^3 \gamma_S^7 v(2,d/2-1,d/2-1)^2 v(2,1,d-3) \\
&~~\times \int d^dz_1 d^dz_2
{1\over (x_5-z_1)^2(x_5-z_2)^2z_{12}^2 (x_1-z_1)^{d-2}(x_2-z_1)^{d-2}(x_3-z_2)^{d-2}(x_4-z_2)^{d-2}}
\fe
where we write $x^n$ for $(x^2)^{n/2}$ for short.
We are interested in comparing the overall coefficient of $\langle J_sJ_{\tilde s}\alpha\rangle$
to that of Vasiliev theory. For this purpose, it is sufficient to consider the three-point function
with the position of $\alpha(x)$ integrated out. Integrating over $x_5$ drastically simplifies (\ref{ssssaint});
it reduces to
\ie\label{ssssaloopred}
& -{\delta_{ab}\delta_{cd} \over N^4} \gamma_\alpha^3 \gamma_S^7 v(2,d/2-1,d/2-1)^2 v(2,1,d-3) v(1,1,d-2) \\
&~~\times \int d^dz_1 d^dz_2
{1\over z_{12}^{6-d} (x_1-z_1)^{d-2}(x_2-z_1)^{d-2}(x_3-z_2)^{d-2}(x_4-z_2)^{d-2}} \\
&= -{\delta_{ab}\delta_{cd} \over N^4} \gamma_\alpha^3 \gamma_S^3 v(2,d/2-1,d/2-1)^2 v(2,1,d-3) v(1,1,d-2)
{(4\pi)^{d/2} 2^{d-6} \over a(d-3)}\\
&~~\times \int {d^dP\over (2\pi)^d} I(x_{21},P)I(x_{34},P) (P^2)^{3-d} e^{iP\cdot x_{13}}
\fe
where we have Fourier transformed $z_{12}$ into the momentum variable $P$, and defined
\ie
&I(x,P) \equiv \int {d^dk\over (2\pi)^d} {e^{ik\cdot x}\over k^2(P-k)^2} \\
&=(4\pi)^{-d/2}{\pi\over \sin{\pi d\over 2}}(P^2)^{d/2-2}\int_0^1 d\xi \left[\xi(1-\xi) \right]^{d/2-2} e^{iP\cdot x(1-\xi)} \\
&~~\times\sum_{k=0}^\infty {1\over k!} \left[ {1\over \Gamma(k+3-d/2)} \left( {\xi(1-\xi)x^2P^2\over 4} \right)^{k+2-d/2}
-{1\over \Gamma(k+d/2-1)} \left( {\xi(1-\xi)x^2P^2\over 4} \right)^k \right]
\fe
where in the second line, we have integrated out $k$ using Feynman parameterization, and expanded the resulting Bessel function in powers of $P^2$. As pointed out in \cite{Lang:1990ni}, the higher spin currents $J_s$ arise from only the second branch, involving integer powers of $x$ for general non-integer values of dimension $d$. In fact,
the correlation function involving the primary fields $J_s$ come from the $k=0$ terms in the second branch only, whereas the $k>0$ terms are contributions from the descendants of $J_s$. The spin-$s$ component is therefore
extracted from the $k=0$, ${\cal O}((x\cdot P)^s)$ term in $I(x,P)$, which we denote by
\ie
I_{(s)}(x,P) &= -{(4\pi)^{-d/2}\over \Gamma({d\over 2}-1)}{\pi\over \sin{\pi d\over 2}}(P^2)^{{d\over 2}-2}\int_0^1 d\xi \left[\xi(1-\xi) \right]^{{d\over 2}-2} {(iP\cdot x(1-\xi))^s\over s!} \\
&= - {2^{-d}\pi^{1-{d\over 2}} \Gamma({d\over 2}+s-1)\over \sin({\pi d\over 2})s!\Gamma(d+s-2)}
(P^2)^{{d\over 2}-2} (ix\cdot P)^s
\fe
The contribution to the integrated three-point function $\int d^d\vec x \langle J_s(\vec \delta, \varepsilon)
J_{\tilde s}(0,\tilde\varepsilon) \alpha(\vec x) \rangle$ is extracted from (\ref{ssssaloopred})
to be
\ie
& -{\delta_{ab}\delta_{cd} \over N^4} \gamma_\alpha^3 \gamma_S^3 v(2,d/2-1,d/2-1)^2 v(2,1,d-3) v(1,1,d-2)
{(4\pi)^{d/2} 2^{d-6} \over a(d-3)}\\
&~~ \times \int {d^d\vec P\over (2\pi)^d} I_{(s)}(\varepsilon,P)I_{(\tilde s)}(\tilde\varepsilon,P) (P^2)^{3-d} e^{i\vec P\cdot \vec\delta}\\
&= -{\delta_{ab}\delta_{cd} \over N^4} \gamma_\alpha^3 \gamma_S^3 v(2,d/2-1,d/2-1)^2 v(2,1,d-3) v(1,1,d-2)
{(4\pi)^{d/2} 2^{d-6} \over a(d-3)}\cdot {2^{-2d}\pi^{2-d}\over \sin^2({\pi d\over 2})}\\
&~~\times {\Gamma({d\over 2}+s-1)\over s!\Gamma(d+s-2)}{\Gamma({d\over 2}+\tilde s-1)\over \tilde s!\Gamma(d+\tilde s-2)} (\varepsilon\cdot\partial)^s(\tilde\varepsilon\cdot\partial)^{\tilde s} \left[
{\Gamma({d\over 2}-1)\over 4\pi^{d\over 2}} {1\over (\delta^2)^{{d\over 2}-1}}\right].
\fe

Next, we consider the contribution from the disconnected tree diagram in figure 3(b),
of the form $\langle S_a(x_1)S_c(x_3)\rangle \langle S_b(x_2)S_d(x_4)\alpha(x_5)\rangle$.
There are 4 such diagrams, related by
exchanging $S_a(x_1)$ with $S_b(x_2)$, and $S_c(x_3)$ with $S_d(x_4)$.
After integration over the position of $\alpha(x_5)$, they give
\ie\label{disctree}
& -{1\over N^3} \gamma_\alpha \gamma_S^3 v(2,d/2-1,d/2-1)v(1,1,d-2)\\
&\times \left[
\delta_{ac}\delta_{bd}\left({1\over (x_{13}^2)^{d/2-1}}+{1\over (x_{24}^2)^{d/2-1}}\right)
+\delta_{ad}\delta_{bc} \left({1\over (x_{14}^2)^{d/2-1}}+{1\over (x_{23}^2)^{d/2-1}}\right) \right]
\fe
Since we are only interested in $O(N)$-singlets, we will restrict to the $\delta^{ab}\delta^{cd}$ channel
of (\ref{disctree}), which is obtained as $N^{-2}\delta_{ab}\delta_{cd}
\sum_{e,f}\langle S_e(x_1)S_f(x_3)\rangle \langle S_e(x_2)S_f(x_4)\alpha(x_5)\rangle+3~{\rm more}$,
namely
\ie
&  -{\delta_{ab}\delta_{cd}\over N^4} \gamma_\alpha \gamma_S^3 v(2,d/2-1,d/2-1)v(1,1,d-2)\\
&\times \left[
{1\over (x_{13}^2)^{d/2-1}}+{1\over ((x_{13}+x_{21}+x_{34})^2)^{d/2-1}}
+{1\over ((x_{13}+x_{34})^2)^{d/2-1}}+{1\over ((x_{13}+x_{21})^2)^{d/2-1}} \right]
\fe
The contribution to $\int d^d x\langle J_s(\delta,\varepsilon)J_{\tilde s}(0,\tilde\varepsilon)\alpha(x)\rangle$ is extracted as
\ie
&  -{\delta_{ab}\delta_{cd}\over N^4} \gamma_\alpha \gamma_S^3 v(2,d/2-1,d/2-1)v(1,1,d-2){1\over s!\tilde s!}(\varepsilon\cdot \partial)^s(\tilde\varepsilon\cdot\partial)^{\tilde s}{1\over (\delta^2)^{d/2-1}}
\fe
Finally, let us consider the contribution from the 4 connected tree diagrams in figure 3(c). After integrating out the position of $\alpha(x)$, applying repeatedly the
graphical rules of \cite{Kazakov:1983ns}, we obtain
\ie
& -{\delta_{ab}\delta_{cd}\over N^4} \gamma_\alpha^2 \gamma_S^3 v(2,d/2-1,d/2-1)^2 v(1,1,d-2)
{(4\pi)^{d/2}\over 4 a(d/2-1)}\\
&~~\times \int {d^dP\over (2\pi)^d} (P^2)^{1-d/2} \left[ I(x_{34},P)(e^{iP\cdot x_{13}}+e^{iP\cdot x_{23}})
+I(x_{21},P)(e^{iP\cdot x_{13}}+e^{iP\cdot x_{14}}) \right]
\fe
Extracting the contribution to $\int d^d x\langle J_s(\delta,\varepsilon)J_{\tilde s}(0,\tilde\varepsilon)\alpha(x)\rangle$, we have
\ie
& {\delta_{ab}\delta_{cd}\over N^4} \gamma_\alpha^2 \gamma_S^3 v(2,d/2-1,d/2-1)^2 v(1,1,d-2)
{(4\pi)^{d/2}\over 4 a(d/2-1)}\\
&~~\times {2^{-d}\pi^{1-{d\over 2}}\over\sin({\pi d\over 2})s!\tilde s!} \int {d^dP\over (2\pi)^d} \left[{ \Gamma({d\over 2}+s-1)\over \Gamma(d+s-2)}{(iP\cdot x_{34})^s(iP\cdot x_{12})^{\tilde s} }
+(s\leftrightarrow\tilde s,x_{34}\leftrightarrow x_{12})\right] {e^{iP\cdot x_{13}}\over P^2}\\
&\to {\delta_{ab}\delta_{cd}\over N^4} \gamma_\alpha^2 \gamma_S^3 v(2,d/2-1,d/2-1)^2 v(1,1,d-2)
{2^{d-4}\Gamma(d/2-1)\over a(d/2-1)}\\
&~~\times {2^{-d}\pi^{1-{d\over 2}}\over\sin({\pi d\over 2})s!\tilde s!} \left[{ \Gamma({d\over 2}+s-1)\over \Gamma(d+s-2)}
+ { \Gamma({d\over 2}+\tilde s-1)\over \Gamma(d+\tilde s-2)}\right] (\varepsilon\cdot\partial)^s(\tilde\varepsilon\cdot\partial)^{\tilde s}\left[
{\Gamma({d\over 2}-1)\over 4\pi^{d\over 2}} {1\over (\delta^2)^{{d\over 2}-1}}\right].
\fe
Putting these together, the total contribution from the diagrams in figure 3 is given by
\ie
& \int d^d x \left\langle \overline J_s(\delta,\varepsilon)\overline J_{\tilde s}(0,\tilde\varepsilon)\alpha(x)\right\rangle\\
& = {C_d\over N} f(s)f(\tilde s) (\varepsilon\cdot\partial)^s(\tilde\varepsilon\cdot\partial)^{\tilde s}
 {1\over (\delta^2)^{{d\over 2}-1} } ,
\fe
where
\ie
& C_d = {\pi^{-{d\over 2}}\over 2}\Gamma({d\over 2}-1),\\
& f(s) = {1\over s!}\left[ 1 - 2^{d-2}\pi^{-{1\over 2}}{\Gamma({d-1\over 2})\Gamma({d\over 2}+s-1)\over
\Gamma(d+s-2)} \right].
\fe
Here the notation $\overline J_s(x)$ standards for the primary currents $J_s$ together with a linear combination
of descendants of lower spin currents, which appears on the RHS of the OPE
\ie
S_{a}(x) S_b(0)\sim {\delta_{ab}\over N^{3\over 2}} x^{\mu_1}\cdots x^{\mu_s} \overline J_{\mu_1\cdots\mu_s}(0)
+\cdots
\fe
If we further set $\tilde\varepsilon=\varepsilon$ to be null, and $d=3$, we obtain
\ie
\int d^3\vec x \left\langle \overline J_s(\vec\delta,\varepsilon)\overline J_{\tilde s}(0, \varepsilon)\alpha(\vec x)\right\rangle
&= {1\over 2\pi N} f(s)f(\tilde s) (\varepsilon\cdot\partial)^{s+\tilde s}
 {1\over |\vec\delta| } \\
&=
{1\over 2\pi^{3\over 2} N} f(s)f(\tilde s) 2^{s+\tilde s}\Gamma(s+\tilde s+{1\over 2})
{(\vec\varepsilon\cdot \vec\delta)^{s+\tilde s}\over \delta^{2s+2\tilde s+1}},
\fe
where $f(s)$ is now given by
\ie\label{fsfac}
f(s) = {1\over s!}\left[ 1 - 2{\Gamma(s+{1\over 2})\over
\pi^{1\over 2}s!} \right].
\fe
The mixed currents $\overline J_s$ are related to the primaries in a general form
$\overline J_s(x,\varepsilon) = J_s(x,\varepsilon) + \sum_{r=0}^{s-2} c(s,r) (\varepsilon\cdot \partial)^{s-r} J_r(x,\varepsilon)$.
In order to determine the operator mixing, we consider the two-point function of $\overline J_s$, by extracting from the four-point function
$\langle S_a(x_1)S_b(x_2)S_c(x_3)S_d(x_4)\rangle$, expanding it in the channel
$\delta_{ab}\delta_{cd} {\cal O}(x_{12}^s){\cal O}(x_{34}^{\tilde s})$. At leading nontrivial
order in $1/N$, there are two disconnected tree diagrams related by exchanging $x_3$ and $x_4$,
and a connected tree diagram with an $\alpha$ propagator. The total contribution in the $\delta_{ab}\delta_{cd}$ channel is
\ie
& {\gamma_S^2\over N^3} \delta_{ab}\delta_{cd} \left[ {1\over (x_{13}^2)^{{d\over 2}-1}(x_{24}^2)^{{d\over 2}-1}}
+{1\over (x_{14}^2)^{{d\over 2}-1}(x_{23}^2)^{{d\over 2}-1}} \right] \\
& + {\gamma_\alpha \gamma_S^4\over N^3}\delta_{ab}\delta_{cd} \int d^d z_1 d^d z_2 {1\over z_{12}^4 (x_1-z_1)^{d-2} (x_2-z_1)^{d-2}
(x_3-z_2)^{d-2}(x_4-z_2)^{d-2}}.
\fe
As before, we can turn the integration over $z_1,z_2$ in the second line into a momentum integral of the form
$\int {d^dP\over (2\pi)^d} I(x_{21},P)I(x_{34},P) (P^2)^{2-{d\over 2}} e^{iP\cdot x_{13}}$. Expanding this in $x_{12}, x_{34}$, we can extract the two-point function
\ie
&\langle \overline J_s(x,\varepsilon) \overline J_{\tilde s}(0,\tilde\varepsilon) \rangle
 = {\gamma_S^2\over s!\tilde s!} \left\{ {1\over x^{d-2}}(\varepsilon\cdot\partial)^s(\tilde\varepsilon\cdot\partial)^{\tilde s}
{1\over x^{d-2}} + \left[(\varepsilon\cdot\partial)^s{1\over x^{d-2}}\right]\left[(\tilde\varepsilon\cdot\partial)^{\tilde s}
{1\over x^{d-2}}\right] \right\}\\
&~~~
+ \gamma_\alpha {2^{2d-8}\Gamma(d-2)\Gamma({d\over 2}-2)\over \Gamma(2-{d\over 2})}
{2^{-2d}\pi^{2-d}\over \sin^2({\pi d\over 2})} {\Gamma(s+{d\over 2}-1)\over s!\Gamma(s+d-2)}
{\Gamma(\tilde s+{d\over 2}-1)\over \tilde s!\Gamma(\tilde s+d-2)}
(\varepsilon\cdot\partial)^s(\tilde\varepsilon\cdot\partial)^{\tilde s} {1\over x^{2d-4}}.
\fe
Restricting to $d=3$, and setting $\tilde\varepsilon=\varepsilon$, we have
\ie
&\langle \overline J_s(x,\varepsilon) \overline J_{\tilde s}(0,\varepsilon) \rangle
= {2^{s+\tilde s-4}\Gamma(s+{1\over 2})\Gamma(\tilde s+{1\over 2})\over \pi^3 s!\tilde s!}
\left[ 1 - 2{(s+\tilde s)!\over s!\tilde s!}+ {\sqrt{\pi} \Gamma(s+\tilde s+{1\over 2})\over \Gamma(s+{1\over 2})\Gamma(\tilde s+{1\over 2})} \right]
{(\varepsilon\cdot x)^{s+\tilde s}\over (x^2)^{s+\tilde s+1}}.
\fe
This allows us to determine the operator mixing,
\ie
\overline J_s(x,\varepsilon) = {1\over \sqrt{N}}{\Gamma(s+{1\over 2})\over 2\pi^{3\over 2}}\sum_{r=1}^{s/2} {(2r)!\over (s-2r)!(s+2r)!} (\varepsilon\cdot\partial)^{s-2r}J_{2r}(x,\varepsilon)
\fe
where $J_s(x,\varepsilon)$ are normalized such that
\ie
&\langle J_s(x,\varepsilon) J_{\tilde s}(0,\varepsilon) \rangle = \delta_{s\tilde s} N 2^{4s}\pi^{-{1\over 2}} {\Gamma(s+{1\over 2})\over s!}{(\varepsilon\cdot x)^{2s}\over (x^2)^{2s+1}},
\fe
i.e. the same normalization convention as the current $J_s$ in the free $O(N)$ theory in section 4.5.
The normalization factor $f(s)$ in (\ref{fsfac}) can be written as
\ie
f(s) = {4\Gamma(s+{1\over 2})\over \sqrt{\pi}} \sum_{r=1}^{s/2} {1\over (s+2r)!(s-2r)!}.
\fe
From this, we determine the integrated three point function of the primary currents,
\ie\label{delttmp}
\int d^3\vec x \left\langle J_s(\vec\delta,\varepsilon) J_{\tilde s}(0, \varepsilon)\alpha(\vec x)\right\rangle
&= {32\pi \over s!\tilde s!}  (\varepsilon\cdot\partial)^{s+\tilde s}
 {1\over |\vec\delta| } \\
&= {32\pi^{1\over 2} }  {2^{s+\tilde s}\Gamma(s+\tilde s+{1\over 2})\over s!\tilde s!}
{(\vec\varepsilon\cdot \vec\delta)^{s+\tilde s}\over \delta^{2s+2\tilde s+1}}.
\fe
To make comparison with our result (\ref{ftres}) in Vasiliev theory, let us define $J_0'(\vec x) = N\gamma_\alpha^{-{1\over 2}}
\alpha(\vec x)$, so that the two-point function of $J_0'(\vec x)$ is normalized in the same way as $J_s$'s.
Now we expect
\ie\label{compttmp}
{\int d^3\vec x \left\langle J_s(\vec\delta,\varepsilon) J_{\tilde s}(0, \varepsilon)J_0'(\vec x)\right\rangle
\over \langle J_0'J_0'\rangle} = g{a_0'\over a_s a_{\tilde s}}\int d^3\vec x
\lim_{z\to 0} z^{-2} B_{\Delta=2}^{(0,0)}(\vec x,z),
\fe
where $\langle J_0'J_0'\rangle=N$ is the coefficient of the two point function
of $J_0'$ (follows from (\ref{gxa})).
Recall that $a_s$ is the normalization constant associated to the spin-$s$ boundary-to-bulk propagators,
and $a_0'$ is that of the $\Delta=2$ scalar. We will determine in section 6.1 that $a_s/a_0=2^{-s} s!$, and that
$g/a_0 = -16/\pi$. We then see that (\ref{compttmp}) precisely holds provided the identification $a_0'={1\over 2}a_0$.
We conclude that our result for the integrated three-point function of two higher spin currents
with the $\Delta=2$ scalar from Vasiliev theory
indeed agrees with that of the critical $O(N)$ model.

\section{More Three-Point Functions}

In this section, we will compute the tree level three-point function coefficients of Vasiliev theory
via the boundary expectation value of a higher spin outcoming field at second order in perturbation theory.
These computations are more involved than the cases we considered previously, where the outcoming field
is a scalar. In particular, the computation of $C(0,s;s')$ for $s>s'$ in section 6.1, combined with earlier result on $C(s,s';0)$, will allow us to determine the relative normalization of the boundary-to-bulk
propagators in Vasiliev theory that correspond to the spin-$s$ currents with normalized two-point function,
and consequently fix the normalization of all three-point functions up to one overall constant, namely the
coupling constant $g$ of Vasiliev theory. We will find complete agreement with the correlation function
of one scalar and two higher spin currents in the free $O(N)$ theory.

\subsection{$C(0,s;s')$ with $s>s'$, and fixing the normalization}

Let us revisit the computation of the three-point function coefficients of one scalar operator with two higher spin fields, but take the scalar operator to be a boundary source, rather than the outcoming field.
There are two distinct cases, $C(0,s;s')$ for $s>s'$ and for $s\leq s'$. The former receives contribution from
the source $J^\Omega(y)$ only (recall (\ref{jomy})), whereas the latter comes entirely from $J'(y)$.
It turns out that the computation in the first case, $s>s'$, is easier, as we shall perform in this
subsection.

To compute the spin-$s'$ component of $J^{(s')}(y)=J^\Omega(y)|_{y^{2s'}}$,
we need the $(2s',0)$ and $(2s'+1,1)$ components of $J_{\A\db}$ in its $(y,\bar y)$ expansion,
\ie\label{jsptemp}
J^{(s')} &= \left.-{z\over 2}\partial_y({\slash\!\!\!\partial} - {s'+2\over z}\sigma^z){\slash\!\!\! J} y
-{2s'+1\over 2}\partial^\A\partial^\db J_{\A\db}\right|_{y^{2s'},\bar y=0}
\\
&=\left.-{z\over 2} \left[\partial_y({\slash\!\!\!\partial} - {s'+2\over z}\sigma^z)\right]^\db y^\A
\left[ \Omega_{\A\db}*\tilde B - \tilde B*\pi(\Omega_{\A\db}) \right]
\right.\\
&~~~~~\left.
+{2s'+1\over 2}\partial^\A\partial^\db \left[ \Omega_{\A\db}*\tilde B - \tilde B*\pi(\Omega_{\A\db}) \right]
\right|_{y^{2s'},\bar y=0}
\\
&=\left. -z \left[\partial_y({\slash\!\!\!\partial} - {s'+2\over z}\sigma^z)\right]^\db y^\A (\Omega_{\A\db}*\tilde B)
+ (2s'+1) \partial^\A\partial^\db (\Omega_{\A\db}*\tilde B)
\right|_{y^{2s'},\bar y=0}
\\
&=\left.\sum_{n=-s'}^{s'} \left\{  {z\over s^2-n^2} ({\slash\!\!\!\partial} - {s'+2\over z}\sigma^z)^{\C\db} \partial_\C \left[  y^\A(\partial_\A\partial_\db\Omega_{++}^n*\tilde B) \right]
+ {2s'+1\over s+n} \partial_\A\Omega_{++}^n*\partial^\A\tilde B
\right\} \right|_{y^{2s'},\bar y=0}
\\
&=-\sum_{n=-s'}^{s'}\int d^4u d^4v e^{uv+\bar u\bar v} \left\{ {z\over s^2-n^2} \partial_y ({\slash\!\!\!\partial} - {s'+2\over z}\sigma^z)\bar v
\left[  (y v) \Omega_{++}^n(x|y+u,\bar u) \tilde B(x|y+v,\bar v) \right]\right. \\
&~~~\left.\left.
+ (2s'+1){uv+2\over s+n} \Omega_{++}^n(x|y+u,\bar u) \tilde B(x|y+v,\bar v)
\right\} \right|_{y^{2s'}}
\fe
In the third line, we have made the replacement $-\tilde B*\pi(\Omega_{\A\db})|_{y^{2s'},\bar y=0}$
by $\Omega_{\A\db}*\tilde B|_{y^{2s'},\bar y=0}$. This is because the contribution comes from taking the star product of
$\Omega_{\A\db}^{(s-1+s'-n,s-1-s'+n)}$ and $B^{(s-1-s'+n,s-1-s'+n)}$, $|n|\leq s'$, where $s-s'-1$ pairs of $y$'s
and $s-1-s'+n$ pairs of $\bar y$'s are contracted to get a term of order $y^{2s'}$. The sign is such that
$-\tilde B*\pi(\Omega_{\A\db})|_{y^{2s'},\bar y=0} = \Omega_{\A\db}*\tilde B|_{y^{2s'},\bar y=0}$.
Similarly, for our gauge choice, $\Omega_{\A\db}^n\sim \partial_\A\partial_\db\Omega_{++}^n$,
and $\partial^\A\partial^\db (\Omega_{\A\db}*\tilde B) = \Omega_{\A\db}*\partial^\A\partial^\db \tilde B$,
and so the same argument can be applied to the second term in the third line of (\ref{jsptemp}).
In the fourth line, note that the sum is over $n=-s',\cdots s'$, as these are the only components among
the spin-$s$ field $\Omega_{++}^{(s-1+n,s-1-n)}$ to $J^{(s')}$.

Let us recall the formulae for the boundary-to-bulk propagators of the relevant master fields, $\Omega_{++}^n$ and
$\Omega_{++}^{-n}$ with $n\geq 0$ for the spin-$s$ field sourced at $\vec x=0$, and $\tilde B$ for the scalar field
sourced at $\vec x=\vec\delta$.
\ie
&\Omega^{n}_{++} = {2^{-n-2}\over (2s-1)!}{z^s\over (x^-)^{s+n}}
(y{\bf x}\sigma^{-z}{\bf x}y)^n \partial_+^{2s} {(y{\bf x}\bar y)^{s-n} \over x^2}\\
&~~~~= s2^{-s-1}{z^s\over (x^2)^{2s+1}} (y{\bf x}\sigma^{-z}{\bf x}y)^n
(y{\bf x}\sigma^-{\bf x}\bar y)^{s-n} , \\
&\Omega^{-n}_{++} = s2^{-s-1}{z^s\over (x^2)^{2s+1}} (\bar y{\bf x}\sigma^{-z}{\bf x}\bar y)^n
(y{\bf x}\sigma^-{\bf x}\bar y)^{s-n} , \\
&\tilde B = \tilde K e^{-y\tilde\Sigma \bar y}.
\fe
Here we have worked in the light cone coordinates in which the null polarization vector of the spin-$s$
current is $\varepsilon^+=1$, $\varepsilon^-=\varepsilon^\perp=0$.
Let us decompose $J^{(s')}$ according to the contributions from the components $\Omega_{++}^n$,
$J^{(s')}=\sum_{n=-s'}^{s'} J^{(s')}_n$. For each $n\geq 0$, we have
\ie
J^{(s')}_n
&= - s2^{-s-1}
\int d^4u d^4v \left[ {z\over s^2-n^2} \partial_y ({\slash\!\!\!\partial} - {s'+2\over z}\sigma^z)\bar v (y v)
+ (2s'+1){(u-y)(v-y)+2\over s+n}\right] \\
&~~~\left.\times (u{\bf x}\sigma^{-z}{\bf x}u)^n
(u{\bf x}\sigma^-{\bf x}\bar u)^{s-n} e^{-v\tilde\Sigma\bar v}
e^{(u-y)(v-y)+\bar u\bar v} {z^{s+1}\over (x^2)^{2s+1}\tilde x^2}  \right|_{y^{2s'}},
\\
J^{(s')}_{-n} &= - s2^{-s-1}
\int d^4u d^4v \left[ {z\over s^2-n^2} \partial_y ({\slash\!\!\!\partial} - {s'+2\over z}\sigma^z)\bar v (y v)
+ (2s'+1){(u-y)(v-y)+2\over s-n}\right] \\
&~~~\left.\times (\bar u{\bf x}\sigma^{-z}{\bf x}\bar u)^n
(u{\bf x}\sigma^-{\bf x}\bar u)^{s-n} e^{-v\tilde\Sigma\bar v}
e^{(u-y)(v-y)+\bar u\bar v} {z^{s+1}\over (x^2)^{2s+1}\tilde x^2}  \right|_{y^{2s'}},
\fe
where we have shifted the integration variables $u,v$ in comparison to (\ref{jsptemp}).
We will calculate (\ref{jstmpa}) from a generating function. As before, we introduce a polarization spinor $\lambda=\sigma^z\bar\lambda$,
with $\sigma^{-z}_{\da\db} = \bar\lambda_\da\bar\lambda_\db$, so that $(u{\bf x}\sigma^{-z}{\bf x} u) = (u{\bf x}\bar\lambda)^2$.
We will make use of the generating function
\ie\label{jstmpa}
I(j_u,j_v,\bar j_{\bar u},\bar j_{\bar v})&=\int d^4u d^4v e^{\tau uv+\bar u\bar v} e^{-v\tilde\Sigma\bar v + \eta u{\bf x}\sigma^-{\bf x}\bar u
+ j_u u+ j_v v+\bar j_{\bar u}\bar u+\bar j_{\bar v}\bar v}\\
&={1\over \Xi(\tau)} \exp\left[
\left( \begin{matrix}  j_v, & \bar j_{\bar u}\end{matrix}\right)
\left( \begin{matrix} \tau & \eta {\bf x}\sigma^-{\bf x} \\ -\tilde\Sigma & -1 \end{matrix} \right)^{-1}
{j_u\choose \bar j_{\bar v}}
\right]
\\
&={1\over \Xi(\tau)} \exp\left[{1\over\Xi(\tau)}
\left( \begin{matrix}  j_v, & \bar j_{\bar u}\end{matrix}\right)
\left( \begin{matrix} 1 & \eta {\bf x}\sigma^-{\bf x} \\ -\tilde\Sigma & -\tau \end{matrix} \right)
{(\tau-\eta\tilde\Sigma{\bf x}\sigma^-{\bf x})j_u\choose (\tau-\eta{\bf x}\sigma^-{\bf x}\tilde\Sigma)\bar j_{\bar v}}
\right]
\fe
where $\Xi(\tau)\equiv \det(\tau -\eta{\bf x}\sigma^-{\bf x}\tilde\Sigma)=\tau^2-\tau\eta {\rm Tr}({\bf x}\sigma^-{\bf x}\tilde\Sigma)$.
Now for $n\geq 0$, (\ref{jstmpa}) can be rewritten as
\ie
J_n^{(s')} &= -(2n)!(s-n)! s2^{-s-1} \left.\left[ {z\over s^2-n^2}
\partial_y ({\slash\!\!\!\partial} - {s'+2\over z}\sigma^z)\partial_{\bar j}
(y\partial_j)+(2s'+1){2+\partial_\tau\over s+n} \right]\right|_{\tau=1,j=\bar j=0}\\
&~~~~~~\left.\times
I(\tau y-\zeta{\bf x}\bar\lambda, -\tau y+j, 0,\bar j)\right|_{y^{2s'}\zeta^{2n}\eta^{s-n}}{z^{s+1}\over (x^2)^{2s+1}\tilde x^2} ,\\
J_{-n}^{(s')} &=- (2n)!(s-n)! s2^{-s-1} \left.\left[ {z\over s^2-n^2}
\partial_y ({\slash\!\!\!\partial} - {s'+2\over z}\sigma^z)\partial_{\bar j}
(y\partial_j)+(2s'+1){2+\partial_\tau\over s-n} \right]\right|_{\tau=1,j=\bar j=0}\\
&~~~~~~\left.\times
I(\tau y, -\tau y+j,-\zeta{\bf x}\lambda,\bar j)\right|_{y^{2s'}\zeta^{2n}\eta^{s-n}}{z^{s+1}\over (x^2)^{2s+1}\tilde x^2}.
\fe
To proceed in the $n\geq 0$ case, we shall collect some useful formulae involving
special cases of $I(j_u,j_v,\bar j_{\bar u},\bar j_{\bar v})$,
\ie
& I(\tau y-\zeta{\bf x}\bar\lambda, -\tau y+j, 0,\bar j)
\\
&~~~={1\over \Xi(\tau)} \exp\left\{ {1\over\Xi(\tau)}
(j-\tau y)\left[
(\tau-\eta\tilde\Sigma{\bf x}\sigma^-{\bf x})(\tau y-\zeta{\bf x}\bar\lambda)
+\eta{\bf x}\sigma^-{\bf x} (\tau-\eta{\bf x}\sigma^-{\bf x}\tilde\Sigma)\bar j \right]
\right\}
\\
&~~~={1\over \Xi(\tau)} \exp\left\{ {\tau\over\Xi(\tau)}
(j-\tau y)\left[
(\tau-\eta\tilde\Sigma{\bf x}\sigma^-{\bf x})y - \zeta{\bf x}\bar\lambda
+\eta{\bf x}\sigma^-{\bf x} \bar j \right]
\right\},
\\
& I(\tau y-\zeta{\bf x}\bar\lambda, -\tau y, 0,0)
= {1\over \Xi(\tau)} \exp\left\{ -{\tau^2\over\Xi(\tau)}
y\left[
(\tau-\eta\tilde\Sigma{\bf x}\sigma^-{\bf x})y - \zeta{\bf x}\bar\lambda \right]
\right\}
\\
&~~~\to {1\over \Xi(\tau)} \exp\left[ {\tau^2\kappa\over\Xi(\tau)}
\lambda(
\kappa\eta\tilde\Sigma{\bf x}\sigma^-{\bf x} + \zeta{\bf x}\sigma^z ) \lambda
\right]
\\
&~~~= {1\over \Xi(\tau)} \exp\left[ {2\tau^2\kappa x^-\over\Xi(\tau)}
\left(\kappa\eta {\rm Tr}_+(\tilde\Sigma{\bf x}\sigma^{-z})-\zeta\right)
\right],
\fe
and its derivatives,
\ie
&\left. (\partial_\tau+2) I(\tau y-\zeta{\bf x}\bar\lambda, -\tau y, 0,0) \right|_{\tau=1}
\to {\Xi-1\over \Xi^2} \left[ 1+{2\kappa x^-(\kappa\eta {\rm Tr}_+(\tilde\Sigma{\bf x}\sigma^{-z})-\zeta)\over\Xi} \right] e^{{2\kappa x^-(\kappa\eta {\rm Tr}_+(\tilde\Sigma{\bf x}\sigma^{-z})-\zeta)\over\Xi} }
\\
&~~~\to {\Xi-1\over \Xi^2} \left( 1+{\Lambda\over\Xi} \right) e^{\Lambda\over\Xi },
\fe
\begin{equation}\nonumber\begin{aligned}
& \left.(\partial_y \sigma^z \partial_{\bar j})(y\partial_j) I(\tau y-\zeta{\bf x}\bar\lambda, -\tau y+j, 0,\bar j)\right|_{\tau=1,j=\bar j=0}
\\
&= -{1\over \Xi^2} (\partial_y \sigma^z \partial_{\bar j}) y\left(
\eta\tilde\Sigma{\bf x}\sigma^-{\bf x} y+\zeta{\bf x}\bar\lambda
-\eta{\bf x}\sigma^-{\bf x}\bar j \right)
\exp\left\{ {1 \over\Xi }
y\left(
\eta\tilde\Sigma{\bf x}\sigma^-{\bf x} y+\zeta{\bf x}\bar\lambda
-\eta{\bf x}\sigma^-{\bf x}\bar j \right)
\right\}
\\
&= {\eta\over \Xi^2} \left(\partial_y \sigma^z {\bf x}\sigma^-{\bf x} y\right)
\left[1+{y\left(
\eta\tilde\Sigma{\bf x}\sigma^-{\bf x} y+\zeta{\bf x}\bar\lambda
\right)\over\Xi} \right]
e^{y\left(
\eta\tilde\Sigma{\bf x}\sigma^-{\bf x} y+\zeta{\bf x}\bar\lambda
\right)\over\Xi}
\end{aligned}\end{equation}
\ie
&= -{4\eta z x^-\over \Xi^2}
\left(1+{\Lambda\over\Xi} \right) e^{\Lambda\over\Xi}
-{\eta\over \Xi^3} \left\{ y{\bf x}\sigma^-{\bf x} \sigma^z \left(
\eta [\tilde\Sigma,{\bf x}\sigma^-{\bf x}]y+\zeta {\bf x}\bar\lambda \right) \right\}
\left(2+{\Lambda\over\Xi} \right) e^{\Lambda\over\Xi}
\\
&\to -{4\eta z x^-\over \Xi^2}
\left(1+{\Lambda\over\Xi} \right) e^{\Lambda\over\Xi}
-{\kappa\eta\over \Xi^3} {\rm Tr}_+\left\{{\bf x}\sigma^-{\bf x} \sigma^z \left(
\kappa\eta [\tilde\Sigma,{\bf x}\sigma^-{\bf x}]+\zeta {\bf x}\sigma^z \right)\sigma^{-z} \right\}
\left(2+{\Lambda\over\Xi} \right) e^{\Lambda\over\Xi}
\\
&= -{4\eta z x^-\over \Xi^2}
\left[1+{\Lambda\over\Xi}
+{2\over\Xi} \left(\kappa^2x^-{\Xi-1\over 2z}+\kappa\zeta x^- + \Lambda \right)
\left(2+{\Lambda\over\Xi} \right) \right]  e^{\Lambda\over\Xi},
\fe
where we wrote $\Xi\equiv \Xi(1) = 1-\eta{\rm Tr}({\bf x}\sigma^-{\bf x}\tilde\Sigma)$. $\Lambda$ is defined
as
\ie
\Lambda&= y(\eta\tilde\Sigma{\bf x}\sigma^-{\bf x}y+\zeta {\bf x}\bar\lambda)
\\
&\to 2\kappa x^-(\kappa\eta {\rm Tr}_+(\tilde\Sigma{\bf x}\sigma^{-z})-\zeta),
\fe
In the second step, we restrict ourselves to the case that the null polarization vector of the spin-$s$ current source
is the same as the polarization of the outcoming spin-$s'$ field. This is also the special case we considered in the computation of
$C(s,\tilde s;0)$ in the previous sections. ``$\to$" here stands for identifying $y=\kappa\lambda$,
with $\lambda=\sigma^z
\bar\lambda$, $\lambda_\A\lambda_\B=\sigma^{-z}_{\A\B}$; and we will extract the coefficient of $\kappa^{2s'}$ in the end. Note that we can only make this substitution after taking the derivative on $y$, as in $J^{(s')}(y)$, and
not in $J^\Omega_{\A\db}$. Now we shall compute the integral (\ref{ssint}), for the $n>0$ component of $J^{(s')}$.
It will be useful to consider a more general expression,
\ie
&I_n^{(s')}(\hat\delta,y;\epsilon) = \int d^3\vec xdz\,z^{s'-2+\epsilon} J_n^{(s')}(\vec x,z;0,\hat\delta|y)\\
&=-(2n)!(s-n)!s2^{-s-1} \int d^3\vec xdz{z^{s+s'-1+\epsilon}\over (x^2)^{2s+1}\tilde x^2} \left[\left.-{2s'+1+\epsilon\over s^2-n^2}(\partial_y\sigma^z\partial_{\bar j})(y\partial_j)\right|_{j=\bar j=0}\right.\\
&
~~~\left.\times I(y-\zeta{\bf x}\bar\lambda, -y+j, 0,\bar j)+\left.(2s'+1){\partial_\tau+2\over s+n}I(\tau y-\zeta{\bf x}\bar\lambda, -\tau y, 0,0) \right|_{\tau=1}  \right]
\\
&=-(2n)!(s-n)!s2^{-s-1} \int d^3\vec xdz{z^{s+s'-1+\epsilon}\over (x^2)^{2s+1}\tilde x^2} e^{\Lambda\over\Xi} \left\{
{2s'+1\over s+n} {\Xi-1\over \Xi^2} \left( 1+{\Lambda\over\Xi} \right) \right.\\
&~~~~~ \left.\left.+{2s'+1+\epsilon\over s^2-n^2} {4\eta z x^-\over \Xi^2}
\left[1+{\Lambda\over\Xi}
+{2\over\Xi} \left(\kappa^2x^-{\Xi-1\over 2z}+\kappa\zeta x^- + \Lambda \right)
\left(2+{\Lambda\over\Xi} \right) \right] \right\}\right|_{\kappa^{2s'}\zeta^{2n}\eta^{s-n}},
\fe
where $\epsilon$ is not assumed to be an integer.
Recall that
\ie
&x^-=-{1\over 2}\lambda {\bf x}\sigma^z\lambda,\\
&\Lambda = \kappa (\lambda{\bf x}\sigma^z\lambda)\left[\kappa\eta \left(\lambda{\bf x}\sigma^z\lambda + {2z\over\tilde x^2}\lambda{\bf x}{\slash\!\!\!\delta}\lambda\right)+\zeta\right],\\
& \Xi = 1+\eta{2z\over \tilde x^2}(\lambda{\bf x}\sigma^z\lambda - x^2\lambda{\slash\!\!\!\delta}\sigma^z\lambda).
\fe
We have then
\ie
&I_n^{(s')}(\hat\delta,y;\epsilon)= -{(2n)!(s-n)!s2^{-s-1}\over (s'+n)!} \int d^3\vec xdz{z^{s+s'-1+\epsilon}\over (x^2)^{2s+1}\tilde x^2}
\left({\Lambda\over\Xi}\right)^{s'+n} \left\{
{(2s'+1)(s'+n+1) \over s+n} {\Xi-1\over \Xi^2} \right.\\
&~~~~~ \left.\left.+{2s'+1+\epsilon\over s^2-n^2} {4\eta z x^-\over \Xi^2}
\left[s'+n+1
+2 (s'+n)(s'+n+1)\left({\kappa^2x^-{\Xi-1\over 2z}+\kappa\zeta x^- \over \Lambda}+1 \right)
\right] \right\}\right|_{\kappa^{2s'}\zeta^{2n}\eta^{s-n}}
\\
&= -{(s-n)!s2^{-s-1}\over (s'-n)!} \int d^3\vec xdz{z^{s+s'-1+\epsilon}\over (x^2)^{2s+1}\tilde x^2}
{\left[\lambda{\bf x}\sigma^z\lambda(\lambda{\bf x}\sigma^z\lambda+{2z\over\tilde x^2}\lambda{\bf x}{\slash\!\!\!\delta}\lambda)\right]^{s'-n}(\lambda{\bf x}\sigma^z\lambda)^{2n}}
\left.{1\over\Xi^{s'+n+2}}\right|_{\eta^{s-s'-1}} \\
&~~~ \times\left\{
{(2s'+1)(s'+n+1) \over s+n}{2z\over \tilde x^2}(\lambda{\bf x}\sigma^z\lambda - x^2\lambda{\slash\!\!\!\delta}\sigma^z\lambda)
\right.\\
&~~~+{2s'+1+\epsilon\over s^2-n^2} {4 z x^-}\left[s'+n+1
+2 (s'+n)(s'+n+1)\right.
\\
&~~~~~\cdot \left.\left.\left({s'-n\over s'+n}{x^-\over \tilde x^2}{(\lambda{\bf x}\sigma^z\lambda - x^2\lambda{\slash\!\!\!\delta}\sigma^z\lambda)\over (\lambda{\bf x}\sigma^z\lambda)(\lambda{\bf x}\sigma^z\lambda+{2z\over\tilde x^2}\lambda{\bf x}{\slash\!\!\!\delta}\lambda) }+{2n\over s'+n}
{x^-\over \lambda{\bf x}\sigma^z\lambda} +1 \right)
\right] \right\}
\\
&=-{ s2^{-s'-1}(s-n)!(s+n-1)!\over (s-s'-1)!(s'-n)!(s'+n)!} \int d^3\vec xdz{z^{2s-1+\epsilon}\over (x^2)^{2s+1}(\tilde x^2)^{s-s'}}
\\
&~~~\times(\lambda{\bf x}\sigma^z\lambda)^{s'+n}(\lambda{\bf x}\sigma^z\lambda+{2z\over\tilde x^2}\lambda{\bf x}{\slash\!\!\!\delta}\lambda)^{s'-n}
(\lambda{\bf x}\sigma^z\lambda - x^2\lambda{\slash\!\!\!\delta}\sigma^z\lambda)^{s-s'-1}
\\
&~~~ \times\left\{
- (2s'+1) {\lambda{\bf x}\sigma^z\lambda - x^2\lambda{\slash\!\!\!\delta}\sigma^z\lambda\over \tilde x^2}
+{2s'+1+\epsilon\over s-n}(\lambda{\bf x}\sigma^z\lambda)
\left[ 2s'+  1
-{s'-n\over \tilde x^2}{(\lambda{\bf x}\sigma^z\lambda - x^2\lambda{\slash\!\!\!\delta}\sigma^z\lambda)\over (\lambda{\bf x}\sigma^z\lambda+{2z\over\tilde x^2}\lambda{\bf x}{\slash\!\!\!\delta}\lambda) }
\right] \right\}
\fe
The evaluation of $I_n^{(s')}(\hat\delta,y;\epsilon)$ is now straightforward using our integration formulae,
and a tedious one. The result is
\ie
I_n^{(s')}(\hat\delta,y;\epsilon) &=
 {2^{-2-2s-3s'-2\epsilon} \pi^{5\over 2} } {\Gamma(2s+\epsilon)\Gamma(s+s'+\epsilon)\Gamma(s+{1-\epsilon\over 2})
\over \Gamma(2s) \Gamma(s-s')\Gamma(s'+1+{\epsilon\over 2})}{(-)^n\over (s'+n)!(s'-n)!}\\
&~~~\times{\Gamma(s+n)\over\Gamma(s+n+1+\epsilon)}
\left[ 2 (s+n) (2s'+1) + (3s'+n+2) \epsilon \right].
\fe
In particular, we have at $\epsilon=0$
\ie
& I_n^{(s')}(\hat\delta,y;\epsilon=0) = {2^{-2s-3s'-1}\pi^{5\over 2}(2s'+1)\Gamma(s+s')\Gamma(s+{1\over 2})\over
s'!\Gamma(s-s')}{(-)^n\over (s'+n)!(s'-n)!}.
\fe
For the integral (\ref{ssint}), we will need $\left.\partial_\epsilon I_n^{(s')}(\hat\delta,y;\epsilon)\right|_{\epsilon=0}$,
whose expression is too tedious to write explicitly here.

Let us now turn to $I_{-n}^{(s')}(\hat\delta,y;\epsilon)$, with $n>0$. We have the analogous formulae for the generating function
\ie
&I(\tau y,-\tau y+j,-\zeta {\bf x}\lambda,\bar j)
\\
&~~~= {1\over\Xi(\tau)}\exp\left[{1\over\Xi(\tau)}
\left( \begin{matrix}  j-\tau y, & \zeta\lambda{\bf x}\end{matrix}\right)
\left( \begin{matrix} 1 & \eta {\bf x}\sigma^-{\bf x} \\ -\tilde\Sigma & -\tau \end{matrix} \right)
{(\tau-\eta\tilde\Sigma{\bf x}\sigma^-{\bf x})\tau y\choose (\tau-\eta{\bf x}\sigma^-{\bf x}\tilde\Sigma)\bar j}
\right],
\\
&I(\tau y,-\tau y,-\zeta {\bf x}\lambda,0) = {1\over\Xi(\tau)}\exp\left[{\tau\over\Xi(\tau)}
(-\tau y-\zeta\lambda{\bf x}\tilde\Sigma )
(\tau-\eta\tilde\Sigma{\bf x}\sigma^-{\bf x}) y
\right]
\\
&~~~={1\over\Xi(\tau)}\exp\left[{\tau^2\over\Xi(\tau)}
y\tilde\Sigma{\bf x} (\eta \sigma^-{\bf x}y + \zeta \lambda ) \right]
\\
&~~~\to {1\over\Xi(\tau)}\exp\left[{\tau^2\kappa\over\Xi(\tau)}
\lambda\tilde\Sigma{\bf x} (\kappa \eta \sigma^-{\bf x} + \zeta) \lambda \right]\\
&~~~= {1\over\Xi(\tau)}\exp\left[{\kappa\tau^2\over\Xi(\tau)}(2\kappa\eta x^-+\zeta){\rm Tr}_+(\tilde\Sigma{\bf x}\sigma^{-z}) \right],
\fe
as well as its derivatives
\ie\label{nntmpa}
&\left.(\partial_\tau+2)I(\tau y,-\tau y,-\zeta {\bf x}\lambda,0)\right|_{\tau=1}
\to {\Xi-1\over \Xi^2}\left(1+{\Lambda'\over\Xi}\right)e^{\Lambda'\over\Xi},
\fe
and
\ie\label{nntmpb}
&\left.(\partial_y \sigma^z\partial_{\bar j})(y\partial_j)
I(\tau y,-\tau y+j,-\zeta {\bf x}\lambda,\bar j)\right|_{\tau=1,j=\bar j=0}
\\
&= {\eta\over \Xi^2} (\partial_y \sigma^z\partial_{\bar j}) \left[y{\bf x}\sigma^-{\bf x}(\tilde\Sigma y+\bar j)\right]
\exp\left[{1\over\Xi(\tau)}
\left( \begin{matrix}  - y, & \zeta\lambda{\bf x}\end{matrix}\right)
\left( \begin{matrix} 1 & \eta {\bf x}\sigma^-{\bf x} \\ -\tilde\Sigma & -1 \end{matrix} \right)
{(1-\eta\tilde\Sigma{\bf x}\sigma^-{\bf x}) y\choose (1-\eta{\bf x}\sigma^-{\bf x}\tilde\Sigma)\bar j}
\right]
\\
&= {\eta\over \Xi^2} \left[\partial_y \sigma^z {\bf x}\sigma^-{\bf x} y
-{1\over\Xi} \partial_y\sigma^z {\bf x}(\eta\sigma^-{\bf x}y+\zeta \lambda) (y{\bf x}\sigma^-{\bf x}\tilde\Sigma y)
\right]
e^{{1\over \Xi}y\tilde\Sigma{\bf x}(\eta\sigma^-{\bf x} y+\zeta\lambda)}
\\
&\to -{4\eta zx^-\over \Xi^2}\left( 1
-\kappa^2\eta{\lambda{\bf x}\sigma^-{\bf x}\tilde\Sigma \lambda\over\Xi} \right)
e^{\Lambda'\over \Xi}+{\kappa \eta\over \Xi^3} \left\{ 4\zeta zx^- \lambda{\bf x}\tilde\Sigma \lambda\right.
\\
&~~
\left.+ \left[{2\kappa\eta x^- +\zeta \over \Xi}\kappa(\lambda{\bf x}\sigma^-{\bf x}\tilde\Sigma \lambda)-4x^-\right] \lambda{\bf x}\sigma^z
(\kappa\eta [\tilde\Sigma,{\bf x}\sigma^-{\bf x}]
+\zeta \tilde\Sigma{\bf x})\lambda
\right\}
e^{\Lambda'\over \Xi}
\\
&= -{4\eta zx^-\over \Xi^2}\left( 1
+{2\kappa^2\eta x^-{\rm Tr}_+(\tilde\Sigma {\bf x} \sigma^{-z})\over\Xi} \right)
e^{\Lambda'\over \Xi}-{\kappa\eta\over \Xi^3} \left\{ 4\zeta zx^- {\rm Tr}_+(\tilde\Sigma{\bf x} \sigma^{-z})\right.
\\
&\left.
~~+ 2x^-\left({\Lambda' \over \Xi}+2\right)
\left[ {2z\Lambda'\over\kappa}+(2\kappa\eta x^-+\zeta){\Xi-1\over\eta}+4\kappa\eta zx^- {\rm Tr}_+(\tilde\Sigma{\bf x}\sigma^{-z}) \right]
\right\}
e^{\Lambda'\over \Xi}
\\
&= -{4z\eta x^-\over \Xi^2} \left\{ 1+{\Lambda'\over\Xi}
+ {1\over\Xi}\left({\Lambda' \over \Xi}+2\right)
\left[ 2\Lambda'+(2\kappa^2\eta x^- + \kappa\zeta){\Xi-1\over 2z\eta} - \kappa\zeta {\rm Tr}_+(\tilde\Sigma{\bf x}\sigma^{-z}) \right]
\right\}
e^{\Lambda'\over \Xi}.
\fe
Here we are writing $\Lambda'=y\tilde\Sigma{\bf x} (\eta \sigma^-{\bf x}y + \zeta \lambda )
\to \kappa(2\kappa\eta x^-+\zeta){\rm Tr}_+(\tilde\Sigma{\bf x}\sigma^{-z}) $.

The integral $I_{-n}^{(s')}(\hat\delta,y;\epsilon)$ can now be computed as ($n>0$)
\ie
&I_{-n}^{(s')}(\hat\delta,y;\epsilon) = \int d^3\vec xdz\,z^{s'-2+\epsilon} J_{-n}^{(s')}\\
&=-(2n)!(s-n)!s2^{-s-1} \int d^3\vec xdz{z^{s+s'-1}\over (x^2)^{2s+1}\tilde x^2} \left[\left.-{2s'+1+\epsilon\over s^2-n^2}(\partial_y\sigma^z\partial_{\bar j})(y\partial_j)\right|_{j=\bar j=0}I(y,- y+j,-\zeta {\bf x}\lambda,\bar j)\right.\\
&
~~~\left.+\left.(2s'+1){\partial_\tau+2\over s-n}I(\tau y,-\tau y,-\zeta {\bf x}\lambda,0) \right|_{\tau=1}  \right]
\fe
Using the formulae (\ref{nntmpa}) and (\ref{nntmpa}), we find
\ie
&I_{-n}^{(s')}(\hat\delta,y;\epsilon)=-(2n)!(s-n)!s2^{-s-1} \int d^3\vec xdz{z^{s+s'-1+\epsilon}\over (x^2)^{2s+1}\tilde x^2} e^{\Lambda'\over\Xi} \left\{
{2s'+1\over s-n} {\Xi-1\over \Xi^2} \left( 1+{\Lambda'\over\Xi} \right) \right.\\
&~~~\left.\left.+{2s'+1+\epsilon\over s^2-n^2} {4\eta z x^-\over \Xi^2}
\left[1+{\Lambda'\over\Xi}
+{1\over\Xi} \left((2\kappa^2\eta x^-+\kappa\zeta){\Xi-1\over 2z \eta}
\right.\right.\right.\right.\\
&~~~\left.\left.\left.\left.-\kappa\zeta {\rm Tr}_+(\tilde\Sigma{\bf x}\sigma^{-z}) + 2\Lambda' \right)
\left(2+{\Lambda'\over\Xi} \right) \right] \right\}\right|_{\kappa^{2s'}\zeta^{2n}\eta^{s-n}}
\\
&=-{(2n)!(s-n)!s2^{-s-1}(s'+n+1)\over (s'+n)!(s-n)} \int d^3\vec xdz{z^{s+s'-1+\epsilon}\over (x^2)^{2s+1}\tilde x^2} \left(\Lambda'\over\Xi\right)^{s'+n}
\left\{ (2s'+1){\Xi-1\over \Xi^2} \right.\\
&~~~\left.\left.+{2s'+1+\epsilon\over s+n} {4\eta z x^-\over \Xi^2}
\left[1
+(s'+n) \left({(2\kappa^2\eta x^-+\kappa\zeta){\Xi-1\over 2z \eta}-\kappa\zeta {\rm Tr}_+(\tilde\Sigma{\bf x}\sigma^{-z})\over\Lambda'} + 2 \right) \right] \right\}\right|_{\kappa^{2s'}\zeta^{2n}\eta^{s-n}}
\\
&=-{(s-n-1)!s2^{-s-1}(s'+n+1)\over (s'-n)!} \int d^3\vec xdz{z^{s+s'-1+\epsilon}\over (x^2)^{2s+1}\tilde x^2} {\left[{\rm Tr}_+(\tilde\Sigma{\bf x}\sigma^{-z})\right]^{s'+n}(2 x^-)^{s'-n}}
\\
&~~\times\left.\left\{ (2s'+1) {\Xi-1\over\eta} +{2s'+1+\epsilon\over s+n} {4 z x^-}
\left[2s'+1
+(s'+n) {\Xi-1\over 2z \eta{\rm Tr}_+(\tilde\Sigma{\bf x}\sigma^{-z})}  \right] \right\}{1\over\Xi^{s'+n+2}}\right|_{\eta^{s-s'-1}}
\\
&={(s-n-1)!(s+n)!s2^{-s'-1}\over (s-s'-1)!(s'-n)!(s'+n)!} \int d^3\vec xdz{z^{2s-1+\epsilon}\over (x^2)^{2s+1}
(\tilde x^2)^{s-s'+1}}
\\
&~~\times (\lambda{\bf x}\sigma^z\lambda)^{s'-n}\left(\lambda{\bf x}\sigma^z\lambda + {2z\over\tilde x^2}\lambda{\bf x}{\slash\!\!\!\delta}\lambda\right)^{s'+n}\left(\lambda{\bf x}\sigma^z\lambda - x^2\lambda{\slash\!\!\!\delta}\sigma^z\lambda\right)^{s-s'}
\\
&~~\times \left\{ (2s'+1) -{2s'+1+\epsilon\over s+n} (\lambda{\bf x}\sigma^z\lambda)
\left[ \tilde x^2{2s'+1\over \lambda{\bf x}\sigma^z\lambda - x^2\lambda{\slash\!\!\!\delta}\sigma^z\lambda}
-{s'+n\over \lambda{\bf x}\sigma^z\lambda + {2z\over\tilde x^2}\lambda{\bf x}{\slash\!\!\!\delta}\lambda}  \right] \right\}
\fe
This is precisely the same analytic expression as $I_{n}^{(s')}(\hat\delta,y;\epsilon)$ derived earlier for $n>0$, with $n$ now replaced
by $-n$.

Now $I^{(s')}(\hat\delta,\lambda;\epsilon)$ is given by
\ie
I^{(s')}(\hat\delta,\lambda;\epsilon) = \sum_{n=-s'}^{s'} I_{n}^{(s')}(\hat\delta,\lambda;\epsilon)
\fe
In particular, one can show that at integer values of $\epsilon$, $I^{(s')}(\hat\delta,\lambda;\epsilon)$ has the property
\ie
I^{(s')}(\hat\delta,\lambda; \epsilon ) = 0,~~~~~\epsilon =-2s'+1, -2s'+2,\cdots,-1,0.
\fe
This means that the first branch of the small $z$ expansion of ${\cal K}_{(s')}(\vec x,z|y,\lambda)$ (\ref{kzser}), involving integers powers of $z$,
starting at ${\cal O}(z^{2-s'})$ up to ${\cal O}(z^{s'+1})$, do not contribute to the integral with $J^{(s')}$.
Therefore, only the term in ${\cal K}_{(s')}(\vec x,z|y,\lambda)$ will contribute, as claimed. We have
\ie
I_{s'}(\hat\delta, \lambda) &= \left.\partial_\epsilon I^{(s')}(\hat\delta,y; \epsilon )\right|_{\epsilon=0}
\\
&= { 2^{-2-2s-3s'} \pi^{5/2}(s'+1) \Gamma(s+{1\over 2}) \over s' s'!} (\lambda{\slash\!\!\!\delta}\sigma^z\lambda)^{s+s'}
\\
&= { 2^{-2-s'} \pi^{5/2} (s'+1)\Gamma(s+{1\over 2}) \over s' s'!} (\hat\delta\cdot \vec\varepsilon)^{s+s'}.
\fe
Let us consider the boundary expectation value of the spin-$s'$ component of $B$ field, in the limit $\delta/|\vec x|\to 0$,
and for the special polarization vector $\varepsilon$ such that $\vec\varepsilon\cdot\vec x=0$.
The latter implies that $\lambda$ is an eigen-spinor of ${\bf \hat x}\sigma^z$, with ${\bf \hat x}\sigma^z\lambda
=i|\vec x|\lambda$. We then have
\ie
\lim_{z\to 0} z^{-s'-1} B^{(2s',0)}_{h=s'}(\vec x,z|y=\lambda) &\to -{\cal N}_{s'} {2\delta^{s'-s-1}\over (s'+1) (\vec x^2)^{2s'+1}}
I_{s'}(\hat\delta, y={\bf\hat x} \sigma^z \lambda;\lambda) \\
&=
-{ 2^{s'-2} \pi^{1\over 2} \over s'!}  \Gamma(s+{1\over 2})
{(\vec\delta\cdot \vec\varepsilon)^{s+s'}\over  (\vec x^2)^{s'+1}\delta^{2s+1}}.
\fe
Taking into account the normalization factor from the $z\to 0$ limit of the boundary-to-bulk propagator of $B^{(2s',0)}(\vec x,z|y)$ with one source,\footnote{In the $z\to 0$ limit and upon replacing $yy\rightarrow \lambda \lambda =\sigma^{-z}$, the boundary-to-bulk propagator (\ref{B-btb}) of $B^{(2s',0)}(\vec x,z|y)$ goes to $2^{3s'-1}z^{s'+1} \frac{(\vec{\epsilon}\cdot \vec{x})^{2s'}}{(\vec{x}^2)^{2s'+1}}$ (assuming $\vec{x}$ is away from the origin).} 
we conclude that
\ie
C(0,s;s') =
-{ 2^{-2s'-1} \pi^{1\over 2} \over s'!}  \Gamma(s+{1\over 2}).
\fe
Recall our earlier result
\ie
C(s,s';0) =- {\pi^{1\over 2}\over 2} \Gamma(s+s' + {1\over 2}).
\fe
Let ${\cal O}_s$ be the operator dual to the spin-$s$ field in the boundary CFT, and
denote by $\langle{\cal O}_s{\cal O}_s\rangle$ the two point function coefficient, after we strip off the 
polarization dependent factor.\footnote{In our conventions, we define the polarization dependent factor to be $c_s (\vec{\epsilon}\cdot \vec{x})^{2s}/(x^2)^{2s+1}$, namely $\langle{\cal O}_s(\vec{x},\vec{\epsilon}){\cal O}_s(0,\vec{\epsilon})\rangle \equiv \langle{\cal O}_s{\cal O}_s\rangle \cdot c_s \frac{(\vec{\epsilon}\cdot \vec{x})^{2s}}{(x^2)^{2s+1}}$. Here $c_s=1$ for $s\ge 2$ and $c_0=2$, see eq. (\ref{JJe1e2}).}
In the $\delta/|\vec x|\to 0$ limit, we have
\ie
&{\langle{\cal O}_0(0) {\cal O}_s(\vec\delta,\vec\varepsilon) {\cal O}_{s'}(\vec x,\vec\varepsilon)\rangle\over
\langle {\cal O}_{s'} {\cal O}_{s'}\rangle} \to
g C(0,s;s'){(\vec\delta\cdot \vec\varepsilon)^{s+s'}\over  (\vec x^2)^{s'+1}\delta^{2s+1}},\\
&{\langle{\cal O}_s(0,\vec\varepsilon) {\cal O}_{s'}(\vec\delta,\vec\varepsilon) {\cal O}_{0}(\vec x)\rangle\over
\langle {\cal O}_{0} {\cal O}_{0}\rangle} \to
g C(s,s';0){(\vec\delta\cdot \vec\varepsilon)^{s+s'}\over  (\vec x^2)\delta^{2s+2s'+1}}.
\fe
where $g$ is the coupling constant of Vasiliev theory.
Under the identification ${\cal O}_s=a_s J_s$, the three-point functions of $J_s$'s completely agree with that of
the free field theory computed in section 4.5, provided
\ie
{a_s\over a_0} = 2^{-s} s!.
\fe
Comparison with free field theory fixes the relation between the normalization factor $a_0$ and the coupling constant $g$ in terms of $N$,
\ie
a_0={1\over\sqrt{N}},~~~~{g}=-{16\over \pi\sqrt{N}}.
\fe

\subsection{$C(0,0;s)$, and a puzzle}

Now let us turn to the computation of the three-point function coefficient sewing
two scalar sources into one outcoming spin-$s$ field, $C(0,0;s)$. Unlike all the computations we have explicitly
so far, which involved only the contribution from $J^\Omega(y)$,
$C(0,0;s)$ receives contribution from $J'(y)$ alone.

Recall that we have derived in section 4.1 the expression for $y^\A J'_{\A\db}|_{\bar y=0}$,
which for a pair of scalar sources takes the form
\ie
&\left.y^\A J'^{(s)}_{\A\db}\right|_{\bar y=0} =
{z\over 2x^2\tilde x^2} \left.\int_0^1 dt\,t(1-t) y^\A \left[ (\Sigma\bar y)_\A \bar y_\db  e^{ty\Sigma\bar y},
e^{-y\tilde\Sigma\bar y} \right]_*\right|_{y^{2s+1},\bar y=0} + (x\leftrightarrow\tilde x)\\
&= {z\over x^2\tilde x^2} \int_0^1 dt\,t(1-t) \int d^4 u d^4 v
\,e^{uv+\bar u\bar v}  (y\Sigma\bar u) \bar u_\db  e^{t(y+u)\Sigma\bar u -(y+v)\tilde\Sigma\bar v}  + (x\leftrightarrow\tilde x)
\\
&= {z\over x^2\tilde x^2} \int_0^1 dt\,t(1-t) \int d^4 u d^4 v
\,e^{(u-y)(v-y)+\bar u\bar v} (y\Sigma\bar u) \bar u_\db  e^{t u\Sigma\bar u - v \tilde\Sigma\bar v}  + (x\leftrightarrow\tilde x).
\fe
where in the second line, we used the fact that $s$ is even. The integration over $u,v$ can be performed as
\ie\label{jzztmpx}
&\left.y^\A J'^{(s)}_{\A\db}\right|_{\bar y=0} = {z\over x^2\tilde x^2}\int_0^1 dt\,{t(1-t)} \left.(y\Sigma\partial_{\bar j}) \partial_{\bar j^\db}\right|_{\bar j=0} \\
&~~~\times\int d^4 u d^4 v  \exp\left[
\left( \begin{matrix} u, & \bar v\end{matrix}\right) \left( \begin{smallmatrix} 1&t\Sigma \\ -\tilde\Sigma &-1 \end{smallmatrix} \right)
{v\choose \bar u}
\right] \,e^{y(u-v)+\bar j\bar u}+ (x\leftrightarrow\tilde x)
\\
&={z\over x^2\tilde x^2}\int_0^1 dt\,{t(1-t)\over \det(1-t\Sigma\tilde\Sigma)} \left.(y\Sigma\partial_{\bar j}) \partial_{\bar j^\db}\right|_{\bar j=0} \exp\left[
\left( \begin{matrix} -y, & \bar j\end{matrix}\right) \left( \begin{smallmatrix} 1&t\Sigma \\ -\tilde\Sigma &-1 \end{smallmatrix} \right)^{-1}
{y\choose 0}
\right]+ (x\leftrightarrow\tilde x)
\\
&={z\over x^2\tilde x^2}\int_0^1 dt\,{t(1-t)\over \det(1-t\Sigma\tilde\Sigma)} \left.(y\Sigma\partial_{\bar j}) \partial_{\bar j^\db}\right|_{\bar j=0} \exp\left[-{(y+\bar j\tilde\Sigma) (1-t\tilde\Sigma\Sigma)y
\over\det(1-t\Sigma\tilde\Sigma)}
\right]+ (x\leftrightarrow\tilde x)
\\
&=- {z\over x^2\tilde x^2}\int_0^1 dt\,{t(1-t)\over \Omega(t)^3} (y\tilde\Sigma\Sigma y)
((\tilde\Sigma-t\Sigma)y)_\db \exp\left[{t
\over\Omega(t)}y\tilde\Sigma\Sigma y
\right]+ (x\leftrightarrow\tilde x)
\\
&\to -{z\over (s-1)! x^2\tilde x^2}\int_0^1 dt\,{t^s(1-t)^2\over \Omega(t)^{s+2}}
((\Sigma+\tilde\Sigma)y)_\db \left( y\tilde\Sigma\Sigma y
\right)^s.
\fe
In the above, we have defined
\ie
&\Omega(t)\equiv \det(1-t\Sigma\tilde\Sigma) = (1-t)^2+ 4t {z^2\over x^2\tilde x^2}.
\fe
Here we have used $\tilde x=x-\hat\delta$, $\hat\delta^2=1$.
In the last line of (\ref{jzztmpx}), we have restrict to the ${\cal O}(y^{2s+1})$, which
contribute to the outcoming spin-$s$ components of the $B$ field.
Also recall that
\ie
\left.\partial^\A\partial^\db J'_{\A\db}\right|_{\bar y=0}=0,
\fe
and that $J^{(s)}(y)$ is given by
\ie
J^{(s)}(y) &= -{z\over 2} \partial_y \left({\slash\!\!\!\partial} - {s+2\over z}\sigma^z\right)
{\slash\!\!\! J}^{(2s,0)} y.
\fe
Let us now consider the generalized integral of (\ref{ssint}),
\ie\label{itmpx}
I_s(\hat\delta,y;\epsilon) &= \int d^3\vec x dz\,z^{s-2+\epsilon} J^{(s)}(\vec x,z;0,\hat\delta|y)
\\
&={2s+1+\epsilon\over 2} \int d^3\vec x dz\,z^{s-2+\epsilon} \partial_y \sigma^z {\slash\!\!\! J}^{(s)}(x,z;0,\hat\delta|y) y \\
&= {2s+1+\epsilon\over 2(s-1)!} \int d^3\vec x dz{z^{s-1+\epsilon}\over x^2\tilde x^2} \left[\partial_y \sigma^z (\Sigma+\tilde\Sigma)y \right]
(y\tilde\Sigma\Sigma y)^s\int_0^1 dt \,{ t^s(1-t)^2 \over \Omega(t)^{s+2}}
\\
&= -{2s+1+\epsilon\over 2(s-1)!} \int d^3\vec x dz{z^{s-1+\epsilon}\over x^2\tilde x^2}\int_0^1 dt \,{ t^s(1-t)^2 \over \Omega(t)^{s+2}}\\
&~~~\times \left[{\rm Tr}\left(\sigma^z (\Sigma+\tilde\Sigma)\right) (y\tilde\Sigma\Sigma y)^s
+s \left(y(\Sigma+\tilde\Sigma)\sigma^z[\tilde\Sigma,\Sigma] y\right) (y\tilde\Sigma\Sigma y)^{s-1} \right]
.
\fe
We will make use of the formulae
\ie
& {\rm Tr}\left(\sigma^z (\Sigma+\tilde\Sigma)\right) = 4\left( 1-{z^2\over x^2} - {z^2\over\tilde x^2} \right),\\
& y\tilde\Sigma\Sigma y = 2z({1\over x^2}-{1\over \tilde x^2})y{\bf\hat x}\sigma^z y + {2z\over \tilde x^2} y{\slash\!\!\!\delta}\sigma^z y + {4z^2\over x^2\tilde x^2} y {\bf x}{\slash\!\!\!\delta} y.
\fe
and
\ie
& y \Sigma \sigma^z [\tilde\Sigma,\Sigma]y = 4(1-{2z^2\over x^2}) y\tilde\Sigma\Sigma y - 2 y(\tilde\Sigma-\Sigma)\sigma^z y
-{4z^2\over x^2\tilde x^2} y\Sigma\sigma^z y,\\
& y \tilde\Sigma \sigma^z [\tilde\Sigma,\Sigma]y = 4(1-{2z^2\over \tilde x^2}) y\tilde\Sigma\Sigma y - 2 y(\tilde\Sigma-\Sigma)\sigma^z y
+{4z^2\over x^2\tilde x^2} y\tilde \Sigma\sigma^z y,\\
& y (\Sigma+\tilde\Sigma) \sigma^z [\tilde\Sigma,\Sigma]y = 8(1-{z^2\over x^2}-{z^2\over \tilde x^2}) y\tilde\Sigma\Sigma y -
4(1-{z^2\over x^2\tilde x^2})y(\tilde\Sigma-\Sigma)\sigma^z y.\\
\fe
The integral over $t$ in the last line of (\ref{itmpx}) can formally be expanded in powers of $z$ as
\ie
\int_0^1 dt \,{ t^s(1-t)^2 \over \Omega(t)^{s+2}} &= \sum_{n=0}^\infty (-)^n {s+n+1\choose n}
B(s+n+1,-2s-2n-1) \left( {4z^2\over x^2\tilde x^2} \right)^n\\
&= - {1\over 2}\sum_{n=0}^\infty {s+n+1\choose n}
B(s+n+1,s+n+1) \left( {4z^2\over x^2\tilde x^2} \right)^n,
\fe
where we have performed analytic interpolation in $s$. We may now write
\ie
& I_s(\hat\delta,y;\epsilon) = {2s+1+\epsilon\over 4(s-1)!}
\sum_{n=0}^\infty {s+n+1\choose n}
B(s+n+1,s+n+1) 4^n A_n(\hat\delta,y;\epsilon),
\fe
with $A_n$ given by the integral
\ie\label{anint}
A_n(\hat\delta,y;\epsilon) &= \int d^3\vec x dz {z^{s+2n-1+\epsilon}\over (x^2)^{n+1}(\tilde x^2)^{n+1}}
\left[{\rm Tr}\left(\sigma^z (\Sigma+\tilde\Sigma)\right) (y\tilde\Sigma\Sigma y)^s
\right.\\
&~~~~\left.+s \left(y(\Sigma+\tilde\Sigma)\sigma^z[\tilde\Sigma,\Sigma] y\right) (y\tilde\Sigma\Sigma y)^{s-1} \right].
\fe
Using our integration formulae, we find that $A_n(\hat\delta,y;\epsilon=0)=0$, and
\ie
\left.\partial_\epsilon A_n(\hat\delta,y;\epsilon)\right|_{\epsilon=0}
= {2^{1-s}\pi^2(s+1)\Gamma(s+{1\over 2})\Gamma(n+{1\over 2})
\over s(s+n)\Gamma(s+n+2)} (y{\slash\!\!\!\delta}\sigma^z y)^s.
\fe
After performing the sum over $n$, we have
\ie
I_s(\hat\delta,y)
&= \left.\partial_\epsilon I_s(\hat\delta,y;\epsilon)\right|_{\epsilon=0}
\\
&= {2^{-3s}\pi^{5\over 2}\Gamma(s+{3\over 2})\over s (s!)^2}
(y{\slash\!\!\!\delta}\sigma^z y)^s.
\fe
Finally, we arrive at the boundary expectation value of $B$, in the limit where the two scalar sources
collide, $\delta/|\vec x|\to 0$,
\ie
\lim_{z\to 0} z^{-s-1} B^{(2s,0)}_{h=s}(\vec x,z|y) &\to -{\cal N}_s {2\delta^{s-1}\over (s+1) (\vec x^2)^{2s+1}}
I_{s}(\hat\delta, {\bf\hat x} \sigma^z y) \\
&= -{\pi^{1\over 2} \Gamma(s+{3\over 2}) \over s! (s+1)!} {\left[ 2(\vec\delta\cdot \vec x)(\vec\varepsilon\cdot x)
-\vec x^2 (\vec\delta\cdot\vec\varepsilon) \right]^s
\over (\vec x^2)^{2s+1}\delta} .
\fe
This is however a different expression from with our previous result on $C(0,s';s)$, for $s'>s$. In particular, it
is {\sl not} consistent with the three-point function interpretation when the normalization leg factors are taken into account,
as opposed to our results on $C(s_1,s_2;0)$ and $C(0,s_1;s_2)$ ($s_1>s_2$) which are consistent.
We believe that this is
because the computation of $C(0,0;s)$ is singular, and one may need to examine the more general $C(s_1,s_2;s)$ case
for $s_1\not=s_2$ and
take the limit $s_{1,2}\to 0$ in the end to recover the regularized answer. This subtlety was already seen in
the computation of $C(s,s;0)$ earlier.
We hope to return to this issue in future works.

\subsection{$C(0,0;s)$ in the $\Delta=2$ case}

In this subsection, we consider the three-point function coefficient for a spin-$s$ outcoming field
with two scalar sources with $\Delta=2$ boundary conditions. Using the boundary-to-bulk propagator
for the $\Delta=2$ scalar, we can compute $J'^{(s)}$ as
\ie
&\left.y^\A J'^{(s)}_{\A\db}\right|_{\bar y=0} =
{z^3\over 2(x^2)^2(\tilde x^2)^2} \int_0^1 dt\,t(1-t)
(1+\partial_{\tau_1}|_{\tau_1=1})(1+\partial_{\tau_2}|_{\tau_2=1})
\\
&~~~~\times \left. y^\A \left[ (\tau_1\Sigma\bar y)_\A \bar y_\db  e^{\tau_1 ty\Sigma\bar y},
e^{-\tau_2 y\tilde\Sigma\bar y} \right]_*\right|_{y^{2s+1},\bar y=0} + (x\leftrightarrow\tilde x)\\
&={z^3\over 2(x^2)^2(\tilde x^2)^2} \int_0^1 dt\,(1-t)
(1+t\partial_t)t^{-1}(1+t\partial_t) t^2
\left. y^\A \left[ (\Sigma\bar y)_\A \bar y_\db  e^{ty\Sigma\bar y},
e^{-y\tilde\Sigma\bar y} \right]_*\right|_{y^{2s+1},\bar y=0}\\
&~~~~ + (x\leftrightarrow\tilde x)\\
&={z^3\over 2(x^2)^2(\tilde x^2)^2}
\left. y^\A \left[ (\Sigma\bar y)_\A \bar y_\db  e^{y\Sigma\bar y},
e^{-y\tilde\Sigma\bar y} \right]_*\right|_{y^{2s+1},\bar y=0} + (x\leftrightarrow\tilde x)
\fe
where in the last step we have integrated by part in $t$, and picked up only the boundary term at $t=1$.
Using the results from the previous subsection, we see that the two terms in the last line, related by exchanging $x$ and $\tilde x$,
in fact cancel. Therefore, the contribution from $J'$ to $C^{\Delta=2}(0,0;s)$ vanishes identically.
This is of course not the case in the critical $O(N)$ model, at leading order in the $1/N$ expansion.
For instance, the $s=2$ case gives the three point function of the stress-energy tensor with the scalar
operator, which is nonzero. Just like in the previous subsection, we suspect that our result $C^{\Delta=2}(0,0;s)=0$ is due to a singular behavior
of Vasiliev theory when the two sources are both scalars, and should be regularized in some way that
we do not understand.

\section{Discussion}

We have computed the three-point functions of Vasiliev theory that involve one scalar operator and two currents of
general spins, with the $\Delta=1$ boundary condition for the bulk scalar field,
and found complete agreement with the free $O(N)$ vector theory. To be precise, what we have computed is
$C(s, s';0)$ with $s\not=s'$, and $C(0,s;s')$ in the case $s>s'$. The results can be extrapolated to $s=s'$ by
a formal analytic continuation in the spins. The coefficient $C(0,s;s')$ with $s<s'$ will presumably give the same
answer, although it involves a qualitatively different computation (contribution from $J'$ rather than
$J^\Omega$), which we have not performed in this paper.

In the case of $\Delta=2$ boundary condition on the bulk scalar field, we have computed $C(s,s';0)$.
On the critical $O(N)$ vector model side, we considered the corresponding three-point function, with the simplification that we integrate out the position of the scalar operator. This is sufficient for extracting the coefficient of the three-point function.
We needed to extract the higher spin primary currents from the $SS$ OPE by analyzing the operator mixing
at leading order in $1/N$. After doing so, the result from the critical $O(N)$ vector model precisely
agrees with that of Vasiliev theory with $\Delta=2$ boundary condition.

In conclusion, we have found highly nontrivial agreement of the tree level three-point functions of Vasiliev theory
with free and critical $O(N)$ vector model, at leading order in their $1/N$ expansion.
We have also been able to identify the relation between the coupling constant of Vasiliev theory and $N$ of the dual CFT.

Our computation of tree level three-point functions is not yet complete, as we have not treated the most general
case $C(s_1,s_2;s_3)$. This case requires a lengthier calculation, which is left to future work.
We expect the general answer for $s_1\not=s_2$ to provide a way
to regularize the case $s_1=s_2$, when the computation in Vasiliev theory appears to be singular.

Let us emphasize that we only expect this duality to hold in the $O(N)$-singlet sector of the dual CFT, for either $\Delta=1$ or
$\Delta=2$ boundary conditions. In other words, the $O(N)$ symmetry of the boundary theory is gauged
(with zero gauge coupling). An interesting
generalization is to couple the $O(N)$ symmetry to a Chern-Simons gauge field at level $k$, and fine
tune the mass terms so that we obtain a family of CFTs parameterized by $N$ and $k$ (see \cite{Gaiotto:2007qi}
for discussions on supersymmetric versions of such theories). The duality discussed in this paper
would be obtained in the $k\to \infty$ limit. Vasiliev's minimal higher spin gauge theory in $AdS_4$ should then be a degenerate limit
of a more general dual bulk theory.

As pointed out in \cite{Sezgin:2002ru} and in the introduction, however, Vasiliev's minimal higher spin gauge theory is subject to an ambiguity
in its quartic and higher order interactions. These are encoded in the function $f(\Psi)=1+\Psi + ic\Psi*\Psi*\Psi+\cdots$, as in (\ref{veqna}). Demanding that Vasiliev theory is dual to the free $O(N)$ vector
theory should determine $f(\Psi)$ entirely. This requires analyzing higher point correlation functions.
We have so far been considering only the classical theory. In general, one may expect a nonlocal field theory, such as
Vasiliev theory which has arbitrarily high order derivatives at each given order in the fields,
to have poor UV behavior. However, on the other hand, the structure of Vasiliev theory appears to be highly constrained by the higher spin gauge symmetry. While we do not have a proof, it is conceivable that the loop corrections in Vasiliev theory can only modify the function $f(\Psi)$. The conjecture that one of such $f(\Psi)$ leads to a holography dual of the free $O(N)$ theory is remarkable in that, it implies that such a nonlocal gauge theory
in $AdS_4$ is UV complete and make sense as a full quantum theory of interacting higher spin gauge fields.

As shown in \cite{Mikhailov:2002bp}, once it is demonstrated that the three-point functions of the
higher spin currents have the same structure as in the free field theory, the $n$-point functions are
determined up to finitely many constants at given $n$. More concretely, the $n$-point function takes the form
\ie
\langle J_{s_1}(x_1)\cdots J_{s_n}(x_n)\rangle = \sum_{\sigma\in S_n} A_\sigma G_{free, cyclic}(x_{\sigma(1)},\cdots,
x_{\sigma(n)}),
\fe
where $G_{free, cyclic}(x_1,\cdots, x_n)$ is the term in the correlation function of the corresponding currents
in the free field theory, with the scalar fields $\phi$ in the $n$ currents contracted in a cyclic order.
$A_\sigma$ are undetermined constants. Working at tree level, one can in principle calculate $A_\sigma$
in the boundary CFT of the Vasiliev theory with higher order interactions specified by the function $f(\Psi)$.
By comparing this with the free field theory correlators, it should be possible to fix $f(\Psi)$ in the bulk theory dual to the free $O(N)$ vector theory, to leading nontrivial
order in $1/N$.

Another important aspect is the duality with the critical $O(N)$ model. The latter does not have exact higher spin
symmetry at higher order in its $1/N$ expansion, and hence the bulk dual should not have exact
higher spin gauge symmetry either. Classically, there is no known $AdS_4$ solution of Vasiliev theory
in which the scalar field acquires a vacuum expectation value and spontaneously break the high spin gauge symmetries.
However, it has been suggested \cite{Girardello:2002pp,Gubser:2002zh} that the loop corrections
in Vasiliev theory will generate an effective action, such that the bulk scalar field may condense in a new
$AdS_4$ vacuum, breaking all of the higher spin gauge symmetries. It is clearly essential to
understand this mechanism in detail. We hope to report on it in future works.

\subsection*{Acknowledgments}

We are grateful to D. Gaiotto, S. Hartnoll, I. Klebanov, L. Rastelli and P. Sundell for discussions and correspondences.
S.G. would like thank the 7th Simons Workshop in Mathematics and Physics,
and X.Y. would like to thank the Banff International Research Station, the Erwin Schroedinger
International Institute for Mathematical Physics, Aspen Center for Physics, and the Workshop on Holography and Universality of
Black Holes at McGill University, for their hospitality during the course of this work.
This work is supported by the Fundamental Laws Initiative Fund at Harvard University. S.G. is supported
in part by NSF Award DMS-0244464. X.Y. is supported in part by a Junior Fellowship from the Harvard Society of Fellows
and by NSF Award PHY-0847457.

\appendix

\section{Consistency of the correlation function computation: $B$ versus $W$}

Our approach to computing the three point function coefficient $C(s,\tilde s;s')$ is by solving the boundary
expectation value of the master $B={\cal B}|_{z=\bar z=0}$ at the second order, using the equation
$\tilde D_0 B = -\hat W*B+B*\pi(\hat W)$. At the linearized level, the spin-$s$ gauge field is contained in both $B$ and $\Omega=\hat W|_{z=\bar z=0}$ (for $s>0$). Their relation has been described in the previous section.
In computing the correlation function, we could a priori extract the answer from either $B$ or $\Omega$ at the second
order. As long as the sources (RHS of the perturbative equations (\ref{perteqn})) are localized in the bulk,
we expect the linearized relations among ${\cal B}, S$ and $W$ still hold near the boundary at nonlinear orders.
However, Vasiliev theory is highly nonlocal, and it is a priori not at all obvious that
the sources in (\ref{perteqn}) are localized away from the boundary in the appropriate sense.
In this appendix we will argue that this is indeed the case, by examining the equations in some more detail.

The equations of motion for the second order master fields are
\ie\label{pertsnd}
& \tilde D_0 {\cal B}^{(2)} = -W^{(1)}*B^{(1)}+B*\pi(W^{(1)}),\\
& d_Z {\cal B}^{(2)} = -S^{(1)}*B^{(1)}+B^{(1)}*\pi(S^{(1)}),\\
& d_Z S^{(2)} = -S^{(1)}*S^{(1)} + {\cal B}^{(2)}*(Kdz^2+\bar Kd\bar z^2),\\
& d_Z W'^{(2)} = -\{S^{(1)},W^{(1)}\}_* - D_0 S^{(2)},\\
& D_0 \Omega^{(2)} = -W^{(1)}*W^{(1)} - D_0 W'^{(2)}.
\fe
To begin, recall the relations among the fields at linear order
\ie
&S^{(1)} = -z_\A dz^\A \int_0^1 dt\,t B^{(1)}(-tz,\bar y) K(t) + c.c.,\\
&W'^{(1)} = z^\A \int_0^1 dt \left.[W_0, S^{(1)}_\A]_*\right|_{z\to tz} + c.c.
\fe
The $\hat z$-dependent part of ${\cal B}^{(2)}$ is solved from the second equation of (\ref{pertsnd}),
\ie
{\cal B}'^{(2)} &= -z^\A \int_0^1 dt\left[ S^{(1)}_\A*B^{(1)}-B^{(1)}*\bar\pi(S^{(1)}_\A) \right]_{z\to tz} + c.c.
\fe
As before we will use the notation $B^{(2)}={\cal B}^{(2)}|_{z=\bar z=0}$.
We now solve the third equation of (\ref{pertsnd}), and find
\ie
& S^{(2)} = dz^\A \left\{ z_\A \int_0^1 dt \,t \left.\left(-S^{(1)}_\B*S^{(1)\B}-{\cal B}^{(2)}*K
\right)\right|_{z\to tz}+{1\over 2}\bar z^\db \int_0^1 dt \,\left[S_\A^{(1)},S_\db^{(1)}\right]_*|_{\bar z\to t\bar z} \right\}
+ c.c.
\fe
The $\hat z$-independent part of $S^{(2)}$ is gauged away, as before.
Now using the fourth equation of (\ref{pertsnd}), we can solve for the $\hat z$-dependent
of $W^{(2)}$,
\ie
&W'^{(2)} = z^\A\int_0^1 dt \left\{-\left[S^{(1)}_\A,W^{(1)}\right]_*
+D_0\left[z_\A \int_0^1 du \,u \left.\left(-S^{(1)}_\B*S^{(1)\B}-{\cal B}^{(2)}*K
\right)\right|_{z\to uz}\right.\right.\\
&~~~\left.\left.\left.+{1\over 2}\bar z^\db \int_0^1 du \,\left[S_\A^{(1)},S_\db^{(1)}\right]_*|_{\bar z\to u\bar z}  \right]\right\}\right|_{z\to tz}+c.c.
\\
&= -z^\A\int_0^1 dt \left.\left\{\left[S^{(1)}_\A,W^{(1)}\right]_*
+{1\over 2}\left[{\partial W_0\over\partial \bar y^\db}, \int_0^1 du \,\left[S_\A^{(1)},S^{(1)\db}\right]_*|_{\bar z\to u\bar z}  \right]_*
\right.\right.\\
&~~~\left.\left.+\left[{\partial W_0\over\partial y^\A}, \int_0^1 du \,u \left.\left(S^{(1)}_\B*S^{(1)\B}+{\cal B}^{(2)}*K
\right)\right|_{z\to uz} \right]_*\right\}\right|_{z\to tz}+c.c.
\\
&\equiv z^\A w_{\A} + \bar z^\da \bar w_\da
\fe
We will shortly need the expressions for $w_\A$ and $\bar w_\da$ restrict to $\hat z=\bar{\hat z}=0$,
\ie
&w_{\A}|_{z=\bar z=0} = \left.-\left[S^{(1)}_\A,W^{(1)}\right]_*\right|_{z=\bar z=0}
-{1\over 2}\left[{\partial W_0\over\partial \bar y^\db}, \left.\left[S_\A^{(1)},S^{(1)\db}\right]_*\right|_{z=\bar z=0}  \right]_*
\\
&~~~-{1\over 2}\left[{\partial W_0\over\partial y^\A}, \left.S^{(1)}_\B*S^{(1)\B}\right|_{z=\bar z=0}+\left.{\cal B}^{(2)}\right|_{y\to 0,z\to -y,\bar z=0}
\right]_*
\\
&= \left.-\left[S^{(1)}_\A,W^{(1)}\right]_*\right|_{z=\bar z=0}
-{1\over 2}\left[{\partial W_0\over\partial \bar y^\db}, \left.\left[S_\A^{(1)},S^{(1)\db}\right]_*\right|_{z=\bar z=0}  \right]_*
\\
&~~~-{1\over 2}\left[{\partial W_0\over\partial y^\A}, \left.S^{(1)}_\B*S^{(1)\B}\right|_{z=\bar z=0}+ y^\B \int_0^1 dt\left.\left[ S^{(1)}_\B*B^{(1)}-B^{(1)}*\bar\pi(S^{(1)}_\B) \right]\right|_{y\to 0,z\to -ty,\bar z=0}
\right]_*\\
&~~~-{1\over 2}\left[{\partial W_0\over\partial y^\A}, B^{(2)}(0,\bar y)
\right]_*
\fe
In the end, we would like to consider the equation that relates $\Omega^{(2)}$
to $B^{(2)}$ as in the linearized relation, plus additional source terms that are expressed in terms of
first order fields,
\ie\label{ombsnd}
D_0\Omega^{(2)} &= \left.-W^{(1)}*W^{(1)}-D_0W'^{(2)}\right|_{z=\bar z=0}
\\
&=\left.-W^{(1)}*W^{(1)}-\{W_0,W'^{(2)}\}_*\right|_{z=\bar z=0}
\\
&=\left.-W^{(1)}*W^{(1)}\right|_{z=\bar z=0}+\left\{{\partial W_0\over\partial y^\A}, w^\A|_{z=\bar z=0}\right\}_*
+\left\{{\partial W_0\over\partial \bar y^\da}, \bar w^\da|_{z=\bar z=0}\right\}_*
\\
&=\Delta^{(2)}+ {1\over 2}\epsilon^{\A\B}\left\{{\partial W_0\over\partial y^\A}, \left[{\partial W_0\over\partial y^\B}, B^{(2)}(0,\bar y)
\right]_*\right\}_*
+{1\over 2}\epsilon^{\da\db}\left\{{\partial W_0\over\partial \bar y^\da},\left[{\partial W_0\over\partial \bar y^\db}, B^{(2)}(y,0)
\right]_*\right\}_*\\
\fe
where $\Delta^{(2)}$ represents the ``corrections" to the linearized relation between $\Omega^{(2)}$ and $B^{(2)}$.
As shown in the previous subsection, the three-point functions $C(s,\tilde s;s')$ are extracted using the spin-$s'$ component
of the boundary expectation value of $B^{(2)}|_{\bar y=0}$. The latter scales like $z^{s'+1}$ as the boundary coordinate $z$
goes to zero ($z$ is not to be confused with $\hat z$ which is the noncommutative variable in the master fields).
We could alternatively extract the three-point functions using $\Omega^{(2)}$, which scales like $z^s$ near the boundary. They would be consistent only
if $\Delta^{(2)}$ vanishes at this order in $z$, so that the linearized relation still holds between $\Omega^{(2)}$
and $B^{(2)}$ near the boundary.

To illustrate this, let us examine $\Delta^{(2)}$ explicitly in the case $s=\tilde s=0$, i.e. in computing
the three-point function $C(0,0;s')$.
Given any function $f(y,\bar y)$, we may write
\ie
&\left[ {\partial W_0\over \partial y^\A}, f\right]_*
= -{1\over 2z} \left[ d{\bf x}(\partial_{\bar y}+\sigma^z \partial_y) - dz\partial_y \right]_\A f,
\\
&\left[ {\partial W_0\over \partial \bar y^\db}, f\right]_*
= -{1\over 2z} \left[ d{\bf x}(\partial_y+\sigma^z \partial_{\bar y}) - dz\partial_{\bar y} \right]_\db f.
\fe
In the $(0,0;s)$ case, where both sources are scalars, the linearized fields have boundary-to-bulk
propagators
\ie\label{bswlin}
& B^{(1)} = K e^{-y\Sigma\bar y},\\
& S^{(1)} = -K z_\A dz^\A \int_0^1 dt \,t e^{tz(y+\Sigma\bar y)} + c.c.,\\
& W^{(1)} = W'^{(1)} = {K\over 2z}\int_0^1 dt\,t(1-t) (\hat z d{\bf x}(\Sigma-\sigma^z)\hat z) e^{t\hat z(y+\Sigma\bar y)} + c.c.
\fe
for the field sourced at $\vec x=0$, and similarly for the other source at $\vec x=\vec\delta$,
given by the same expression with $x$ replaced by $\tilde x=x-\vec\delta$, and $\Sigma=\sigma^z-{2z\over x^2}{\bf x}$ by $\tilde\Sigma$ etc.
Note that the scalar does not enter $\Omega^{(1)}$.
In the $z\to 0$ limit, the spin-$s'$ component of $B^{(2)}(y,0)$ scales like $z^{s'+1}$ to leading order in $z$. Correspondingly, from section 3.3, $\Omega_{++}^{n}$
are of order $z^{s'}$, whereas $D_0\Omega^{(2)}$ is of order $z^{s'-1}$. We are thus asking if there are
terms in $\Delta^{(2)}$ of order $z^{s'-1} y^{s'-1+n} \bar y^{s'-1-n}$, $|n|\leq s-1$.

Let us consider one of the terms in $\Delta^{(2)}$, of the form
$\left\{ \partial_\A W,\left[\partial^\A W, S^{(1)}_\B * \tilde S^{(1)\B}|_{\hat z=\bar{\hat z}=0}\right]_*\right\}_*$, coming from $w_\A$. Using the second line of (\ref{bswlin}), we have
\ie
& \left.S^{(1)}_\A * \tilde S^{(1)\A}\right|_{z=\bar z=0} = K\tilde K \int_0^1 dt\int_0^1 d\tilde t\,t\tilde t \left.\left(
z_\A e^{tz(y+\Sigma \bar y)}\right)*\left( z^\A e^{\tilde tz(y+\tilde\Sigma \bar y)} \right)\right|_{z=\bar z=0}\\
&~~~~~=
-K\tilde K \int_0^1 dt\int_0^1 d\tilde t\,t\tilde t \int d^4 u d^4 v e^{uv+\bar u\bar v}(uv)
e^{tu(y+\Sigma (\bar y+\bar u))} e^{-\tilde tv(y+\tilde\Sigma (\bar y+\bar v))}
\fe
where we used the integral representation of the star product. After performing the Gaussian integration over $u$ and $v$, we find
\ie\label{sstmp}
&\left.S^{(1)}_\A * \tilde S^{(1)\A}\right|_{z=\bar z=0} = -K\tilde K \int_0^1 dt\int_0^1 d\tilde t\,t\tilde t \partial_\tau|_{\tau=1}
{\exp\left[{t\tilde t\over \det(\tau-t\tilde t\Sigma\tilde\Sigma)}
(y-\bar y\tilde\Sigma)(\tau-t\tilde t\tilde\Sigma\Sigma)(y+\Sigma\bar y) \right]\over \det(\tau-t\tilde t\Sigma\tilde\Sigma)}
\\
&= -{z^2\over x^2\tilde x^2} \int_0^1 dt\int_0^1 d\tilde t\,t\tilde t \partial_\tau|_{\tau=1}
{\exp\left\{{t\tilde t\over \det(\tau-t\tilde t\Sigma\tilde\Sigma)}
\left[ -t\tilde t y\tilde\Sigma\Sigma y -\tau \bar y\tilde\Sigma\Sigma\bar y + (\tau+t\tilde t)y(\Sigma-\tilde\Sigma)\bar y \right] \right\}\over \det(\tau-t\tilde t\Sigma\tilde\Sigma)}
\\
&= -{z^2\over x^2\tilde x^2} \int_0^1 dt\int_0^1 d\tilde t\,t\tilde t \partial_\tau|_{\tau=1}
{\exp\left\{-{2zt\tilde t\over (\tau-t\tilde t)^2}
\left[ t\tilde t y{\bf\Delta}\sigma^z y +\tau \bar y{\bf\Delta}\sigma^z\bar y + (\tau+t\tilde t)y{\bf\Delta}\bar y \right] \right\}\over (\tau-t\tilde t)^2}\\
&~~~~+{\rm higher~order~in~}z.
\fe
where $\det(\tau-t\tilde t\Sigma\tilde\Sigma)$ is understood to be the determinant of a $2\times 2$ matrix.
We have defined
\ie
{\bf \Delta} = {{\bf x}\over x^2}-{{\bf\tilde x}\over\tilde x^2}.
\fe
Note that ${\bf x}$ and $x^2$ contain $z$ by definition, although they do not matter in the last line of
(\ref{sstmp}). There are potentially divergences from the $t$ and $\tilde t$ integral near $t=\tilde t=1$.
Such divergences, if present, will be regularized using Gamma function regularization, as discussed in section 6.

Now extracting the spin-$s'$ components of $\left\{ \partial_\A W,\left[\partial^\A W, S^{(1)}_\B * \tilde S^{(1)\B}_{\hat z=\bar{\hat z}=0}\right]_*\right\}_*$, we need the $y^{s'+n}\bar y^{s'-n}$ components
of (\ref{sstmp}). By expanding the exponential in the last line of (\ref{sstmp}), we see that such terms in
$S*\tilde S|_{\hat z=\bar{\hat z}=0}$ are of order $z^{2s'+2}$, for $\vec x$ away from $0$ and $\vec\delta$.
Therefore to leading order
in $z$, $\left\{ \partial_\A W,\left[\partial^\A W, S^{(1)}_\B * \tilde S^{(1)\B}_{\hat z=\bar{\hat z}=0}\right]_*\right\}_*$ scales like $z^{s'} y^{s'-1+n} \bar y^{s'-1-n}$, one power of $z$ higher than the terms
in $D_0\Omega^{(2)}$, and hence do not affect the computation of the three-point function.

Similarly, the other terms in $\Delta^{(2)}$ do not contribute at order $z^{s'-1} y^{s'-1+n} \bar y^{s'-1-n}$ either. We conclude
that to the leading nontrivial order in $z$, the linearized relation between $\Omega$ and $B$ holds for the second order fields $\Omega^{(2)}$ and $B^{(2)}$ near the boundary, therefore one would get the same answer for
the three-point function from the boundary expectation value of either field. In practice, it is simpler to consider
$B^{(2)}$, as we analyzed in the section 4.2.

\section{An integration formula}

In this section, we give some formulae for Feynman type integrals that we encounter repeatedly
in the computation of the three-point functions. These are integrals over $\vec x$ and $z$ that
arise in (\ref{scint}) and (\ref{ssint}).

In the following we use the notation ${\bf{\hat x}}=\vec x\cdot\vec \sigma$, as opposed to
${\bf x}=x^\mu\sigma_\mu$. We will also write ${\slash\!\!\!\delta}=\hat\delta\cdot\vec\sigma$.
$\hat\delta$ is a unit vector, with its norm factored out. We will need the integral
\ie
&I(k,m,n,a,b) \equiv \int d^3\vec x dz {z^k\over (x^2)^n (\tilde x^2)^m} (y{\bf \hat x}\sigma^zy)^a(y{\bf x}{\slash\!\!\!\delta}y)^b \\
&= \int d^3\vec x dz {z^k\over (\vec x^2+z^2)^n ((\vec x-\hat\delta)^2+z^2)^m} (y{\bf \hat x}\sigma^zy)^a(y{\bf x}{\slash\!\!\!\delta}y)^b
\\
&= {\Gamma(n+m)\over\Gamma(n)\Gamma(m)}\int_0^1 du\,u^{m-1}(1-u)^{n-1} \int d^3\vec x dz {z^k\over ((\vec x-u\hat\delta)^2+z^2+u(1-u))^{n+m}} (y{\bf \hat x}\sigma^zy)^a(y{\bf x}{\slash\!\!\!\delta}y)^b
\\
&= {\Gamma(n+m)\over\Gamma(n)\Gamma(m)}\int_0^1 du\,u^{m-1}(1-u)^{n-1} \int d^3\vec x dz {z^k\over (\vec x^2+z^2+u(1-u))^{n+m}} (y({\bf \hat x}+u{\slash\!\!\! \delta})\sigma^zy)^a(y{\bf x}{\slash\!\!\!\delta}y)^b
\fe
Here $a$ and $b$ are assumed to be positive integers. We will need to apply it to the case where $k$ is a non-integer, in order
to extract the integral with a $\log(z)$ factor in the integrand.
We observe that
\ie
\int d^3\vec x dz {z^k\over (\vec x^2+z^2+1)^n} (y{\bf\hat x}\sigma^z y)^a (y{\bf\hat x}{\slash\!\!\!\delta}y)^b = 0,~~~~{\rm for~}a\not=0.
\fe
Therefore
\ie\label{ietc}
I(k,m,n,a,b) = {\Gamma(n+m)\over\Gamma(n)\Gamma(m)}\int_0^1 du\,u^{m-1+a}(1-u)^{n-1} \int d^3\vec x dz {z^k(y{\slash\!\!\! \delta}\sigma^zy)^a(y{\bf x}{\slash\!\!\!\delta}y)^b\over (\vec x^2+z^2+u(1-u))^{n+m}}
\fe
It remains to compute
\ie
\int d^3\vec x dz {z^k\over (\vec x^2+z^2+1)^n} (y{\bf\hat x}{\slash\!\!\!\delta}y)^a = J(k,n,a) (y{\slash\!\!\! \delta}\sigma^z y)^a
\fe
where $J(k,n,a)$ is a numerical factor. It vanishes for odd $a$; for even $a$,
using
\ie
\int d^3\vec x dz {z^k x_1^{a}\over (\vec x^2+z^2+1)^n} = {\pi\Gamma({a+1\over 2})\Gamma({k+1\over 2})\Gamma(n-{a+k\over 2}-2)\over 2\Gamma(n)},
\fe
we have
\ie
J(k,n,a) =   (-)^{a\over 2}{\pi\Gamma({a+1\over 2})\Gamma({k+1\over 2})\Gamma(n-{a+k\over 2}-2)\over 2\Gamma(n)}.
\fe
Plugging these back in (\ref{ietc}), we arrive at
\ie\label{jfive}
&I(k,m,n,a,b) = {\Gamma(n+m)\over\Gamma(n)\Gamma(m)}\int_0^1 du\,u^{m-1+a}(1-u)^{n-1}\sum_{\ell=0}^b (-)^{b-\ell}{b\choose \ell}(y{\slash\!\!\! \delta}\sigma^zy)^{a+b-\ell}\\
&~~~\times \int d^3\vec x dz {z^{k+b-\ell}(y{\bf \hat x}{\slash\!\!\!\delta}y)^\ell \over (\vec x^2+z^2+u(1-u))^{n+m}}\\
&= (-)^b{\Gamma(n+m)\over\Gamma(n)\Gamma(m)}(y{\slash\!\!\! \delta}\sigma^zy)^{a+b}B(2-n+a+{k+b\over 2},2-m+{k+b\over 2})\\
&~~~\times
\sum_{\ell=0,even}^b {b\choose \ell} J(k+b-\ell,n+m,\ell)
\\
&=(-)^{b}k!{\pi^{2}B(2-n+a+{k+b\over 2},2-m+{k+b\over 2})\Gamma(m+n-{b+k\over 2}-2)\over 2^{k+1}\Gamma({k-b\over 2}+1)\Gamma(m)\Gamma(n)} (y{\slash\!\!\! \delta}\sigma^zy)^{a+b} \\
&\equiv J(k,m,n,a,b) (y{\slash\!\!\! \delta}\sigma^zy)^{a+b}.
\fe


\begin{thebibliography}{}

\bibitem{Maldacena:1997re}
  J.~M.~Maldacena,
  ``The large N limit of superconformal field theories and supergravity,''
  Adv.\ Theor.\ Math.\ Phys.\  {\bf 2}, 231 (1998)
  [Int.\ J.\ Theor.\ Phys.\  {\bf 38}, 1113 (1999)]
  [arXiv:hep-th/9711200].

\bibitem{Gubser:1998bc}
  S.~S.~Gubser, I.~R.~Klebanov and A.~M.~Polyakov,
  ``Gauge theory correlators from non-critical string theory,''
  Phys.\ Lett.\  B {\bf 428}, 105 (1998)
  [arXiv:hep-th/9802109].

\bibitem{Witten:1998qj}
  E.~Witten,
  ``Anti-de Sitter space and holography,''
  Adv.\ Theor.\ Math.\ Phys.\  {\bf 2}, 253 (1998)
  [arXiv:hep-th/9802150].

\bibitem{Aharony:1999ti}
  O.~Aharony, S.~S.~Gubser, J.~M.~Maldacena, H.~Ooguri and Y.~Oz,
  ``Large N field theories, string theory and gravity,''
  Phys.\ Rept.\  {\bf 323}, 183 (2000)
  [arXiv:hep-th/9905111].

\bibitem{Sundborg:2000wp}
  B.~Sundborg,
  ``Stringy gravity, interacting tensionless strings and massless higher
  spins,''
  Nucl.\ Phys.\ Proc.\ Suppl.\  {\bf 102}, 113 (2001)
  [arXiv:hep-th/0103247].

\bibitem{HaggiMani:2000ru}
  P.~Haggi-Mani and B.~Sundborg,
  ``Free large N supersymmetric Yang-Mills theory as a string theory,''
  JHEP {\bf 0004}, 031 (2000)
  [arXiv:hep-th/0002189].

\bibitem{Aharony:2003sx}
  O.~Aharony, J.~Marsano, S.~Minwalla, K.~Papadodimas and M.~Van Raamsdonk,
  ``The Hagedorn / deconfinement phase transition in weakly coupled large N
  gauge theories,''
  Adv.\ Theor.\ Math.\ Phys.\  {\bf 8}, 603 (2004)
  [arXiv:hep-th/0310285].

\bibitem{Beisert:2003te}
N.~Beisert, M.~Bianchi, J.~F.~Morales and H.~Samtleben,
``On the spectrum of AdS/CFT beyond supergravity,''
JHEP {\bf 0402}, 001 (2004)
[arXiv:hep-th/0310292].

\bibitem{Beisert:2004di}
N.~Beisert, M.~Bianchi, J.~F.~Morales and H.~Samtleben,
``Higher spin symmetry and N = 4 SYM,''
JHEP {\bf 0407}, 058 (2004)
[arXiv:hep-th/0405057].

\bibitem{Gopakumar:2003ns}
  R.~Gopakumar,
  ``From free fields to AdS,''
  Phys.\ Rev.\  D {\bf 70}, 025009 (2004)
  [arXiv:hep-th/0308184].

\bibitem{Gopakumar:2004qb}
  R.~Gopakumar,
  ``From free fields to AdS. II,''
  Phys.\ Rev.\  D {\bf 70}, 025010 (2004)
  [arXiv:hep-th/0402063].

\bibitem{Gopakumar:2005fx}
  R.~Gopakumar,
  ``From free fields to AdS. III,''
  Phys.\ Rev.\  D {\bf 72}, 066008 (2005)
  [arXiv:hep-th/0504229].

\bibitem{David:2006qc}
  J.~R.~David and R.~Gopakumar,
  ``From spacetime to worldsheet: Four point correlators,''
  JHEP {\bf 0701}, 063 (2007)
  [arXiv:hep-th/0606078].

\bibitem{Berkovits:2008qc}
  N.~Berkovits,
  ``Perturbative Super-Yang-Mills from the Topological $AdS_5\times S^5$ Sigma Model,''
  JHEP {\bf 0809}, 088 (2008)
  [arXiv:0806.1960 [hep-th]].

\bibitem{Berkovits:2008ga}
  N.~Berkovits,
  ``Simplifying and Extending the $AdS_5\times S^5$ Pure Spinor Formalism,''
  JHEP {\bf 0909}, 051 (2009)
  [arXiv:0812.5074 [hep-th]].

\bibitem{Isberg:1992ia}
  J.~Isberg, U.~Lindstrom and B.~Sundborg,
  ``Space-time symmetries of quantized tensionless strings,''
  Phys.\ Lett.\  B {\bf 293}, 321 (1992)
  [arXiv:hep-th/9207005].

\bibitem{Isberg:1993av}
  J.~Isberg, U.~Lindstrom, B.~Sundborg and G.~Theodoridis,
  ``Classical and quantized tensionless strings,''
  Nucl.\ Phys.\  B {\bf 411}, 122 (1994)
  [arXiv:hep-th/9307108].

\bibitem{Lindstrom:2003mg}
  U.~Lindstrom and M.~Zabzine,
  ``Tensionless strings, WZW models at critical level and massless higher  spin
  fields,''
  Phys.\ Lett.\  B {\bf 584}, 178 (2004)
  [arXiv:hep-th/0305098].

\bibitem{Bonelli:2003zu}
  G.~Bonelli,
  ``On the covariant quantization of tensionless bosonic strings in AdS
  spacetime,''
  JHEP {\bf 0311}, 028 (2003)
  [arXiv:hep-th/0309222].

\bibitem{Bakas:2004jq}
  I.~Bakas and C.~Sourdis,
  ``On the tensionless limit of gauged WZW models,''
  JHEP {\bf 0406}, 049 (2004)
  [arXiv:hep-th/0403165].
  
\bibitem{Sagnotti:2003qa}
  A.~Sagnotti and M.~Tsulaia,
  ``On higher spins and the tensionless limit of string theory,''
  Nucl.\ Phys.\  B {\bf 682}, 83 (2004)
  [arXiv:hep-th/0311257].

\bibitem{Klebanov:2002ja}
  I.~R.~Klebanov and A.~M.~Polyakov,
  ``AdS dual of the critical O(N) vector model,''
  Phys.\ Lett.\  B {\bf 550}, 213 (2002)
  [arXiv:hep-th/0210114].

\bibitem{Konstein:2000bi}
  S.~E.~Konstein, M.~A.~Vasiliev and V.~N.~Zaikin,
  ``Conformal higher spin currents in any dimension and AdS/CFT
  correspondence,''
  JHEP {\bf 0012}, 018 (2000)
  [arXiv:hep-th/0010239].

\bibitem{Shaynkman:2001ip}
  O.~V.~Shaynkman and M.~A.~Vasiliev,
  ``Higher spin conformal symmetry for matter fields in 2+1 dimensions,''
  Theor.\ Math.\ Phys.\  {\bf 128}, 1155 (2001)
  [Teor.\ Mat.\ Fiz.\  {\bf 128}, 378 (2001)]
  [arXiv:hep-th/0103208].

\bibitem{Sezgin:2001zs}
  E.~Sezgin and P.~Sundell,
  ``Doubletons and 5D higher spin gauge theory,''
  JHEP {\bf 0109}, 036 (2001)
  [arXiv:hep-th/0105001].

\bibitem{Vasiliev:2001zy}
  M.~A.~Vasiliev,
  ``Conformal higher spin symmetries of 4D massless supermultiplets and
  osp(L,2M) invariant equations in generalized (super)space,''
  Phys.\ Rev.\  D {\bf 66}, 066006 (2002)
  [arXiv:hep-th/0106149].

\bibitem{Mikhailov:2002bp}
  A.~Mikhailov,
  ``Notes on higher spin symmetries,''
  arXiv:hep-th/0201019.

\bibitem{Sezgin:2002rt}
  E.~Sezgin and P.~Sundell,
  ``Massless higher spins and holography,''
  Nucl.\ Phys.\  B {\bf 644}, 303 (2002)
  [Erratum-ibid.\  B {\bf 660}, 403 (2003)]
  [arXiv:hep-th/0205131].

\bibitem{Klebanov:1999tb}
  I.~R.~Klebanov and E.~Witten,
  ``AdS/CFT correspondence and symmetry breaking,''
  Nucl.\ Phys.\  B {\bf 556}, 89 (1999)
  [arXiv:hep-th/9905104].

\bibitem{Gubser:2002vv}
  S.~S.~Gubser and I.~R.~Klebanov,
  ``A universal result on central charges in the presence of double-trace
  deformations,''
  Nucl.\ Phys.\  B {\bf 656}, 23 (2003)
  [arXiv:hep-th/0212138].

\bibitem{Vasiliev:1995dn}
  M.~A.~Vasiliev,
  ``Higher-spin gauge theories in four, three and two dimensions,''
  Int.\ J.\ Mod.\ Phys.\  D {\bf 5}, 763 (1996)
  [arXiv:hep-th/9611024].

\bibitem{Vasiliev:1999ba}
  M.~A.~Vasiliev,
  ``Higher spin gauge theories: Star-product and AdS space,''
  arXiv:hep-th/9910096.

\bibitem{Vasiliev:2003ev}
  M.~A.~Vasiliev,
  ``Nonlinear equations for symmetric massless higher spin fields in
  (A)dS(d),''
  Phys.\ Lett.\  B {\bf 567}, 139 (2003)
  [arXiv:hep-th/0304049].


\bibitem{Bekaert:2005vh}
  X.~Bekaert, S.~Cnockaert, C.~Iazeolla and M.~A.~Vasiliev,
  ``Nonlinear higher spin theories in various dimensions,''
  arXiv:hep-th/0503128.

\bibitem{Sezgin:2002ru}
  E.~Sezgin and P.~Sundell,
  ``Analysis of higher spin field equations in four dimensions,''
  JHEP {\bf 0207}, 055 (2002)
  [arXiv:hep-th/0205132].
  
\bibitem{Osborn:1993cr}
  H.~Osborn and A.~C.~Petkou,
  ``Implications of Conformal Invariance in Field Theories for General
  Dimensions,''
  Annals Phys.\  {\bf 231}, 311 (1994)
  [arXiv:hep-th/9307010].

\bibitem{Petkou:2003zz}
  A.~C.~Petkou,
  ``Evaluating the AdS dual of the critical O(N) vector model,''
  JHEP {\bf 0303}, 049 (2003)
  [arXiv:hep-th/0302063].

\bibitem{Leigh:2003gk}
  R.~G.~Leigh and A.~C.~Petkou,
  ``Holography of the N = 1 higher-spin theory on AdS(4),''
  JHEP {\bf 0306}, 011 (2003)
  [arXiv:hep-th/0304217].

\bibitem{Sezgin:2003pt}
  E.~Sezgin and P.~Sundell,
  ``Holography in 4D (super) higher spin theories and a test via cubic  scalar
  couplings,''
  JHEP {\bf 0507}, 044 (2005)
  [arXiv:hep-th/0305040].

\bibitem{Fotopoulos:2008ka}
  A.~Fotopoulos and M.~Tsulaia,
  ``Gauge Invariant Lagrangians for Free and Interacting Higher Spin Fields. A
  Review of the BRST formulation,''
  Int.\ J.\ Mod.\ Phys.\  A {\bf 24}, 1 (2009)
  [arXiv:0805.1346 [hep-th]].


\bibitem{D'Hoker:1999ea}
  E.~D'Hoker, D.~Z.~Freedman, S.~D.~Mathur, A.~Matusis and L.~Rastelli,
  ``Extremal correlators in the AdS/CFT correspondence,''
  arXiv:hep-th/9908160.

\bibitem{Diaz:2006nm}
  D.~E.~Diaz and H.~Dorn,
  ``On the AdS higher spin / O(N) vector model correspondence: Degeneracy  of
  the holographic image,''
  JHEP {\bf 0607}, 022 (2006)
  [arXiv:hep-th/0603084].

\bibitem{Girardello:2002pp}
  L.~Girardello, M.~Porrati and A.~Zaffaroni,
  ``3-D interacting CFTs and generalized Higgs phenomenon in higher spin
  theories on AdS,''
  Phys.\ Lett.\  B {\bf 561}, 289 (2003)
  [arXiv:hep-th/0212181].

\bibitem{Gubser:2002zh}
  S.~S.~Gubser and I.~Mitra,
  ``Double-trace operators and one-loop vacuum energy in AdS/CFT,''
  Phys.\ Rev.\  D {\bf 67}, 064018 (2003)
  [arXiv:hep-th/0210093].

\bibitem{Witten:1985cc}
  E.~Witten,
  ``Noncommutative Geometry And String Field Theory,''
  Nucl.\ Phys.\  B {\bf 268}, 253 (1986).

\bibitem{Lang:1990re}
  K.~Lang and W.~Ruhl,
  ``Anomalous dimensions of tensor fields of arbitrary rank for critical
  nonlinear O(N) sigma models at $2 < d < 4$ to first order in $1/N$,''
  Z.\ Phys.\  C {\bf 51}, 127 (1991).

\bibitem{Lang:1990ni}
  K.~Lang and W.~Ruhl,
  ``Field algebra for critical O(N) vector nonlinear sigma models at $2 < d <
  4$,''
  Z.\ Phys.\  C {\bf 50}, 285 (1991).

\bibitem{Lang:1991kp}
  K.~Lang and W.~Ruhl,
  ``The Critical O(N) Sigma Model At Dimension $2 < D < 4$ And Order $1/N^2$:
  Operator Product Expansions And Renormalization,''
  Nucl.\ Phys.\  B {\bf 377}, 371 (1992).

\bibitem{Lang:1991jm}
  K.~Lang and W.~Ruhl,
  ``The Scalar ancestor of the energy momentum field in critical sigma models
  at $2 < d < 4$,''
  Phys.\ Lett.\  B {\bf 275}, 93 (1992).

\bibitem{Lang:1992zw}
  K.~Lang and W.~Ruhl,
  ``The Critical O(N) sigma model at dimensions $2 < d < 4$: Fusion coefficients
  and anomalous dimensions,''
  Nucl.\ Phys.\  B {\bf 400}, 597 (1993).

\bibitem{Kazakov:1983ns}
  D.~I.~Kazakov,
  ``Calculation Of Feynman Integrals By The Method Of 'Uniqueness',''
  Theor.\ Math.\ Phys.\  {\bf 58}, 223 (1984)
  [Teor.\ Mat.\ Fiz.\  {\bf 58}, 343 (1984)].

\bibitem{Gaiotto:2007qi}
  D.~Gaiotto and X.~Yin,
  ``Notes on superconformal Chern-Simons-matter theories,''
  JHEP {\bf 0708}, 056 (2007)
  [arXiv:0704.3740 [hep-th]].

\end{thebibliography}
\end{document}